\documentclass[a4paper,structabstract]{aa}  
\usepackage{graphicx}
\usepackage{txfonts}
\usepackage{natbib}
\usepackage{soul}

\begin{document}

\title{Detectability of giant planets in protoplanetary disks \\ by CO emission lines}
\titlerunning{Detectability of giant planets in disks}


\author{
	Zs. Reg\'aly\inst{1}
    \and
	Zs. S\'andor\inst{2}
	\and
	C. P. Dullemond\inst{2}
	\and
	R. van Boekel\inst{3}
}

\institute{
	Konkoly Observatory of the Hungarian Academy of Sciences, P.O. Box 67, H-1525 Budapest, Hungary\\
	\email{regaly@konkoly.hu}
	\and
	Junior Research Group, Max-Planck Institut f\"ur Astronomie, K\"onigstuhl 17, D-69117 Heidelberg, Germany
	\and
	Max-Planck-Institut f\"ur Astronomie, K\"onigstuhl 17, D-69117 Heidelberg, Germany
}

\date{Received March 15, 2010; accepted July 6, 2010}

\abstract
{Planets are thought to form in protoplanetary accretion disks around young stars. Detecting a giant planet still embedded in a protoplanetary disk would be very important and give observational constraints on the planet-formation process. However, detecting these planets with the radial velocity technique is problematic owing to the strong stellar activity of these young objects.}
	{We intend to provide an indirect method to detect Jovian planets by studying near infrared emission spectra originating in the protoplanetary disks around T\,Tauri stars. Our idea is to investigate whether a massive planet could induce any observable effect on the spectral lines emerging in the disks atmosphere. As a tracer molecule we propose CO, which is excited in the ro-vibrational fundamental band in the disk atmosphere to a distance of $\sim 2-3\,\mathrm{AU}$ (depending on the stellar mass) where terrestrial planets are thought to form.}
	{We developed a semi-analytical model to calculate synthetic molecular spectral line profiles in a protoplanetary disk using a double layer disk model heated on the outside by irradiation by the central star and in the midplane by viscous dissipation due to accretion. 2D gas dynamics were incorporated in the calculation of synthetic spectral lines. The motions of gas parcels were calculated by the publicly available hydrodynamical code FARGO which was developed to study planet-disk interactions.}
	{We demonstrate that a massive planet embedded in a protoplanetary disk strongly influences the originally circular Keplerian gas dynamics. The perturbed motion of the gas can be detected by comparing the CO line profiles in emission, which emerge from planet-bearing to those of planet-free disk models. The planet signal has two major characteristics: a permanent line profile asymmetry, and short timescale variability correlated with the orbital phase of the giant planet. We have found that the strength of the asymmetry depends on the physical parameters of the star-planet-disk system, such as the disk inclination angle, the planetary and stellar masses, the orbital distance, and the size of the disk inner cavity. The permanent line profile asymmetry is caused by a disk in an eccentric state in the gap opened by the giant planet. However, the variable component is a consequence of the local dynamical perturbation by the orbiting giant planet. We show that a forming giant planet, still embedded in the protoplanetary disk, can be detected using contemporary or future high-resolution near-IR spectrographs like VLT/CRIRES and ELT/METIS.}
	{}


\maketitle

\section{Introduction}

According to the general consensus, planets and planetary systems form in circumstellar disks. These disks consist of gaseous and solid dust material. There are two major mechanisms proposed for the formation of giant planets. The first is the disk instability, which requires a gravitationally unstable disk in which giant planets form by direct collapse of the gas as a consequence of its self-gravity, originally suggested by \citet{Kuiper1951,Cameron1978} and more recently by \citet{Boss2001}. Later on, the gas giant may collect dust, which after settling to its center, forms a solid core. The second mechanism is the core-accretion process \citep{BodenheimerPollack1986, Pollacketal1996}, which is the final stage of planet formation in the planetesimal hypothesis \citep{Safronov1972}. In this hypothesis dust coagulates first and forms planetesimals (meter to kilometer-sized objects). The subsequent growth of planetesimals consists of two stages, the runaway \citep{WetherillStewart1989} and the oligarchic growth \citep{KokuboIda1998}, in which the formed bodies are massive enough to increase their masses by gravity-assisted collisional accretion processes. In this way planetary embryos and, through their consecutive collisions, protoplanets, terrestrial planets, and planetary cores of giant planets can be formed. During the gas accretion phase the precursor of the proto giant planet reaches a critical mass (where the planetary envelope contains more material than its core), and at about the order of $10\,M_{\oplus}$ starts to contract, causing an increased accretion rate, which in turn raises radiative energy losses resulting in a runaway gas accretion. In this model the planetary core rapidly builds up a massive envelope of gas from its surrounding disk, in less than $10\,\mathrm{Myr}$, and eventually a gas giant forms. Because the disk dispersal time is presumably about $5\,\mathrm{Myr}$ \citep{Haischetal2001,Hillenbrand2005}, this model should also incorporate planetary migration to explain giant planets observed well inside the snow line, see \citet{Alibertetal2004}.

Both planetary formation processes have their weak points. Disk instability assumes a very massive disk that is gravitationally unstable or only marginally stable. Another as yet unsolved problem is that during the gravitational collapse an effective cooling mechanism should work to ensure the formation of gravitationally bound clumps. While the radiative cooling is not effective enough and the recently invoked thermal convection is also ineffective, the disk instability as a common planet-formation process is questionable \citep{Klahr2008}.

Regarding the giant planet formation in the planetesimal hypothesis, \citet{Mordasinietal2009} convincingly presented the success of the core accretion model to produce giant planets within the disk lifetime in a planet population synthesis model, but the artificial slow-down of the type I migration was necessary. Type I migration has such a short time scale \citep{Ward1997} that the planet inevitably will be engulfed by the host star within the disk lifetime. Another unsolved issue of the planetesimal hypothesis is the so-called ``meter-sized barrier problem'': i.e. the formation of meter-sized bodies is impeded by their quick inward drift to the star \citep{Weidenschilling1977}, and by mutual disruptive collisions due to their high relative velocities above $1\,\mathrm{m/s}$ \citep{BlumandWurm2008}. If no other physical processes are taking place, these two mechanisms would impede the formation of the meter-sized and consequently the larger planetesimals.

A wealth of information about the planet formation process would clearly be provided if newly formed planets, or at least their signatures, could be observed in still gas-rich protoplanetary disks. There are already attempts to discover signatures of giant planet-formation assuming the disk instability mechanism in the works of \citet{Narayananetal2006} and  \citet{Jang-CondellBoss2007}. In the first study, for instance, the authors suggest observations of $\mathrm{HCO}^{+}$ emission lines as a tracer of the accumulation of large clumps of cold material, arguing that with high-resolution millimeter and sub-millimeter interferometers the cold clumps can be directly imaged. The observability of an already formed planet embedded in a circumstellar disk has also been studied recently by \citet{WolfDAngelo2005} and \citet{Wolfetal2007}. In these works the authors investigated whether the influence of a forming planet on the protoplanetary disk can be detected by analyzing the spectral energy distribution (SED) coming from the star-disk system. It was found that by studying only the SED it is impossible to infer the presence of an embedded planet. On the other hand, it was also concluded that the dust re-emission in the hot regions near the gap could certainly be detected and mapped by ALMA for the nearby (less than $100\,\mathrm{pc}$ in distance) and approximately face-on protoplanetary disks. 

Another interesting attempt was presented by \citet{ClarkeArmitage2003}, where the authors have investigated the possibility of detection of giant planets by excess CO overtone emission originating from the planetary accretion inflow in FU\,Ori type objects. Nevertheless, the periodic line profile distortions are observable only in tight systems in which the giant planet is orbiting within $0.25\,\mathrm{AU}$. Another possibility to detect embedded giant planets is based on the radial velocity measurements, which favor high-mass, short-period companions. The detectability limit of an embedded giant planet by radial velocity measurements is $\sim10\,M_\mathrm{J}$ or $\sim 2\,M_\mathrm{J}$ ($M_\mathrm{J}$ is the mass of Jupiter) orbiting at 1\,AU or 0.5\,AU, respectively, due to heavy variability in the optical spectra of young ($<10\,\mathrm{Myr}$) stars (e.g., \citet{PaulsonYelda2006,Pratoetal2008}). Note that although there were attempts to detect planets by radial velocity measurements in young systems, no firm detection has yet been repeated see, e.g., \citet{Setiawanetal2008} for TW\,Hya\,b, debated later by \citet{Huelamoetal2008}. The discovery of planetary systems by direct imaging nowadays is restricted to large separations $>10-100\,\mathrm{AU}$ \citep{Verasetal2009} and favors higher mass planets. As of today nine exoplanetary systems have been directly imaged in the optical or near-infrared band, but only five of them are still embedded: GQ Lup \citep{Neuhauseretal2005} and CT Cha \citep{Schmidtetal2008}, but they are rather hosting a stellar companion as the companion masses are, because high as $21.5\,\mathrm{M_{J}}$ and $17\,\mathrm{M_{J}}$, respectively; 2M1207  \citep{Chauvinetal2004,Chauvinetal2005a} hosting a $4\,\mathrm{M_{J}}$ mass planet; UScoCTIO 108 \citep{Kashyapetal2008} hosting a $14\,\mathrm{M_{J}}$ mass planet; $\beta\,\mathrm{Pic}$ \citep{Lagrangeetal2009a,Lagrangeetal2009b} hosting an $8\,\mathrm{M_{J}}$ mass planet. AB\,Pic \citep{Chauvinetal2005b}, HR 8799 \citep{Maroisetal2008} (triple system), SR 1845 \citep{Billeretal2006} and Fomalhaut \citep{Kalasetal2008} are mature planetary systems with ages of about 30\,Myr, 60\,Myr, 100\,Myr, and 200\,Myr. Note that centimeter wavelength radio observations of a very young ($<0.1\,\mathrm{Myr}$) disk around HL\,Tau, presented by \citet{Greavesetal2008}, revealed the possibility of a 12 Jupiter mass giant planet being in formation at $\sim75\,\mathrm{AU}$.

We investigate a new possibility to detect young giant planets embedded in protoplanetary disks of T\,Tauri stars via line profile distortions in its near-IR spectra. Contrary to the studies mentioned above, instead of dealing with the disk's density perturbations, we investigate the gas \emph{dynamics} in the disk, particularly near the gap opened by the planet. We have found that there is a reasonable possibility to discover giant planets embedded in the protoplanetary disk of young stars with our approach.

Our idea has been inspired by recent findings of \citet{KleyDirksen2006}, who showed that a massive giant planet embedded in a protoplanetary disk produces a transition of the disk from the nearly circular state into an eccentric state. The mass limit for this transition is $3\,M_{\mathrm{J}}$ for a disk with $\nu=10^{-5}$ uniform kinematic viscosity (measured in dimensionless units, see details in Sect. \ref{sect:Hydro}). The most visible manifestation of the disk's eccentric state is that the outer rim of the gap opened by the giant planet has an elliptic shape. After performing a series of hydrodynamical simulations with the code FARGO \citep{Masset2000}, we can also confirm the results of \citet{KleyDirksen2006} even using $\alpha$-type viscosity. Note that beside the elliptic geometry that is clearly visible in the disk surface-density distribution, the giant planet also disturbs the motion of the gas parcels near the gap. Using CO as tracer molecule of gas, which is excited to $\sim3-5\,\mathrm{AU}$ (depending on the stellar mass) in the disk atmosphere \citep{Najitaetal2007}, we investigate how the CO emission line profiles at $4.7\,\mathrm{ \mu m}$ are distorted by the gas dynamics. With our semi-analytical synthetic spectral model we explore the influence of the mass of the giant planet, the disk inclination angle, the mass of the hosting star, the orbital distance of planet, and finally, the size of inner cavity on the line profile distortions.

The paper is structured as follows: in Sect. 2 we review the basic physics of our synthetic spectral line models. In Sect. 3 we present disk models and details of numerical simulations. Our results of the CO ro-vibrational line profile distortions, the detectability of an embedded giant planet, and its observability constraints are presented in Sect. 4. The paper closes with the discussion of the results and concluding remarks.

\section{Synthetic spectral line model}

\begin{figure}
	\centering
	\includegraphics[width=\columnwidth]{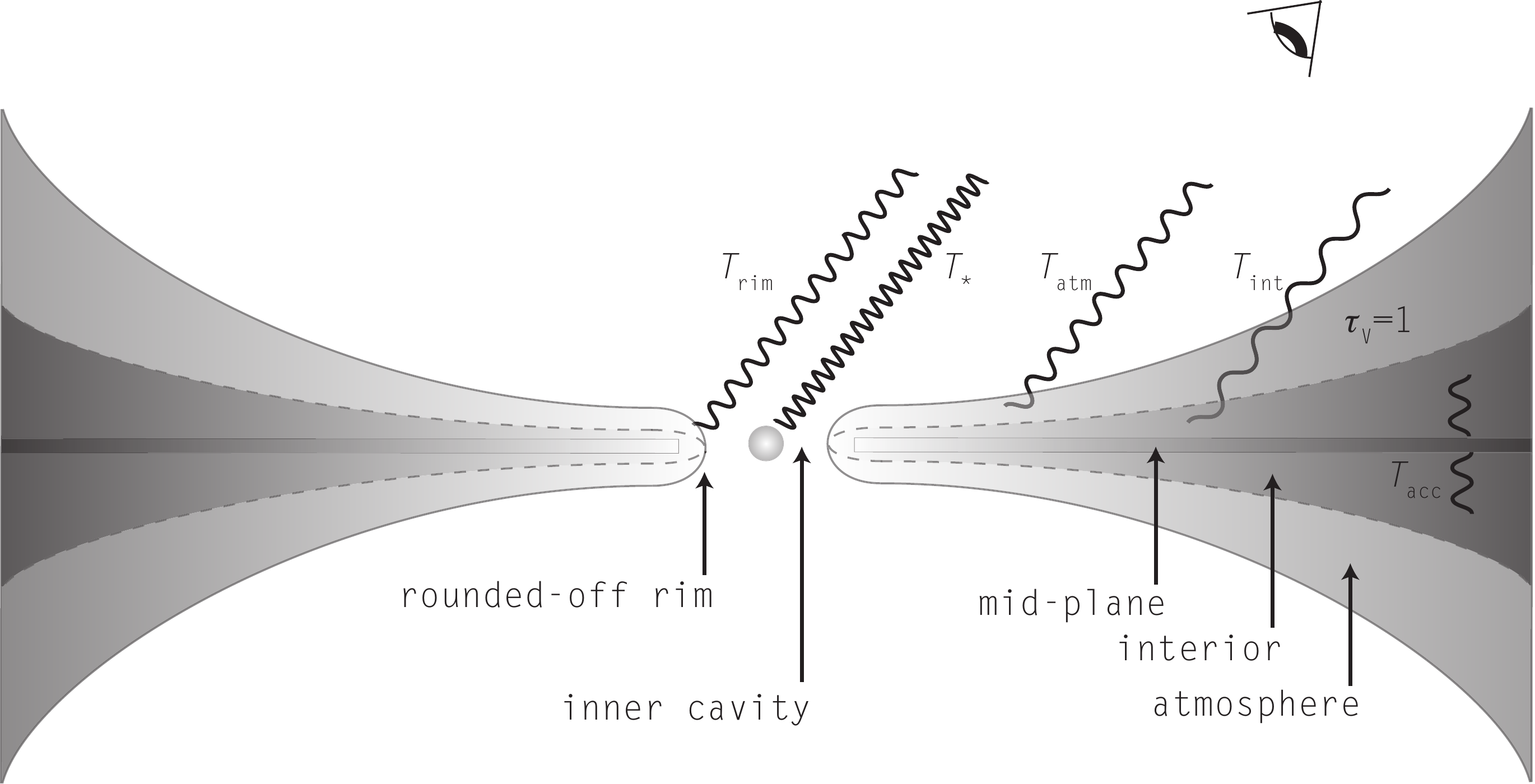}
	\caption{Double-layer flared disk model used in our simulations, after \citet{ChiangGoldreich1997}. In our model the following emission components are included: the optically thin atmosphere emission, with a temperature $T_\mathrm{atm}$ above the optically thick disk interior, with the temperature $T_\mathrm{int}$; the continuum emission of the rounded-off disk inner rim, with a   temperature $T_\mathrm{rim}$ and the stellar continuum assumed to be a blackbody with a temperature $T_*$. The accretion heating of disk interior is also accounted for.}
	\label{fig:disk-model}
\end{figure}

To calculate the CO ro-vibrational spectra emerging from protoplanetary disks we developed a semi-analytical line spectral model. The thermodynamical model of the disk is based on the two-layer flared disk model, originally proposed by \citet{ChiangGoldreich1997}, which describes the heating by stellar irradiation in a disk with a flared geometry. In our model we use the flared disk approximation, and assume the disk to consist of two layers: an optically thick interior, producing continuum radiation and an optically thin atmosphere, producing line emission or absorption in the spectra.\footnote{We only consider the inner part of disk ($R\leq 5\,\mathrm{AU}$). A significant part of the fundamental band CO ro-vibrational emission arises in $R= 2-3\,\mathrm{AU}$, however the line-to-continuum flux at $4.7\,\mathrm{\mu m}$ is sensitive to material that lies up to 5\,AU in our models. In this way the outer part of disk, where the disk interior becomes optically thin to its own and even to the radiation from superheated atmosphere, is not considered.} The boundary between the two layers lies where the dust becomes optically thick in the visual wavelengths, i.e. at optical depth $\tau_\mathrm{V}=1$ along the line of the incident stellar irradiation. The irradiation from the boundary layer formed at the inner edge of the disk \citep{Pophametal1993} is neglected though because its existence had not yet been confirmed conclusively by observations yet. Nevertheless, the emission from the inner disk rim is substantial, resulting in strong near-IR bump in Herbig Ae/Be \citep{Nattaetal2001} and weaker one in T\,Tauri SEDs \citep{Muzerolleetal2003}. In addition it was found by \citet{Monnieretal2006} that the simple vertical wall assumption for the inner rim is inconsistent with the interferometric measurements. Indeed, the shape of the disk inner rim is presumably rounded-off, resulting in less inclination-angle-dependent near-IR excess emission \citep{IsellaNatta2005}. A schematic representation of the disk model is shown in Fig.\ref{fig:disk-model}. If the disk atmosphere is superheated by the stellar irradiation with respect to its interior, we may expect emission lines. Conversely, if the disk interior is heated above the temperature of the disk atmosphere by viscous dissipation, for example due to an abrupt increase in accretion rate, spectral lines are expected in absorption.

The vertical structure of a geometrically thin disk can be derived by considering vertical hydrostatic equilibrium \citep{ShakuraSunyaev1973}. Ignoring any contribution from the gravitational force of the disk itself, the vertical density profile would be set by the equilibrium of gas pressure, the stellar gravitation and centrifugal force owing to orbital motion, resulting in a Gaussian vertical density profile. Because the thickness of the disk atmosphere in the double layer model is principally determined by the grazing angle of the stellar irradiation, which is narrow ($\delta(R)\ll 1$) in our computational domain ($R\leq 5\,\mathrm{AU}$), we assume that the disk atmosphere has uniform vertical density distribution. Note that the surface density in the disk atmosphere according to Eq. (\ref{eq:dens-surf}) does not depend on the surface density ($\Sigma(R)$) distribution by definition.

The expected flux emerging from the double-layer disk is the result of the radiation of gas emission from the optically thin disk atmosphere above the continuum emission of dust in the optically thick disk interior and inner rim. Because the stellar continuum contributes considerably to the line flux in the near-IR band compared to the disk continuum, especially in disks with an inner cavity, we have to also take into account the stellar continuum in the total flux. In our model the stellar radiation is taken to be blackbody radiation of the stellar effective surface temperature. Because the gas temperature is regulated by collisions with dust grains and stellar X-ray heating (which is significant for a young star), the gas and dust components may be thermally uncoupled in the tenuous disk atmosphere below a critical density of $n_\mathrm{cr}\sim 10^{13}-–10^{14}\,\mathrm{cm^{-3}}$ \citep{ChiangGoldreich1997}. As a result the gas temperature can be as high as $\sim 4000-5000\,\mathrm{K}$ in a region where the column density is below $\sim 10^{21}\,\mathrm{cm^{-2}}$ \citep{Glassgoldetal2004}. Because the gas column density is $\sim 10^{22}\,\mathrm{cm^{-2}}$ in our model disk atmosphere, we adopt thermal coupling of dust and gas, i.e. $T_\mathrm{g}(R)=T_\mathrm{d}(R)=T_\mathrm{atm,irr}(R)$, where $T_\mathrm{atm,irr}(R)$ is the atmospheric temperature defined by Eq. (\ref{eq:temp-surf}). Although \citet{Glassgoldetal2004} and \citet{KampDullemond2004} suggest the existence of an overheated layer above the superheated disk atmosphere, we did not consider this for simplicity's sake. In this model, assuming local thermodynamic equilibrium, the monochromatic intensity at frequency $\nu$ emitted by gas parcels at a distance $R$ to the star and an azimuthal angle $\phi$ observed at inclination angle $i$ ($i=0^{\circ}$ meaning face on disk) can be given by
\begin{eqnarray}
	\label{eq:intensity}
	I(\nu,R,\phi,i)&=&B(\nu,T_{\mathrm{int}}(R))e^{-\tau(\nu,R,\phi,i)}+ \nonumber \\
	&&B(\nu,T_{\mathrm{ atm}}(R))\left(1-e^{-\tau(\nu,R,\phi,i)}\right),
\end{eqnarray}
where $\tau(\nu,R,\phi,i)$ is the monochromatic optical depth of the disk atmosphere at a frequency $\nu$ along the line of sight, $B(\nu,T_\mathrm{int}(R))$ and $B(\nu,T_\mathrm{atm}(R))$ are the Planck functions of the dust temperature $T_\mathrm{int}(R)$ in the disk interior, and the gas temperature $T_\mathrm{atm}(R)$ in disk atmosphere, respectively. Regarding the disk inner edge, which has a temperature $T_\mathrm{rim}(R_0)$, it is assumed that it radiates as a blackbody ($I_\mathrm{rim}(\nu)=B(\nu,T_\mathrm{rim}(R_0))$), see Appendix \ref{apx:sinner-rim} for details. The total flux emerging from the protoplanetary disk seen by inclination angle $i$ at frequency $\nu$ can thus be given as
\begin{eqnarray}
	\label{eq:flux-disk}
	F_{\mathrm{disk}}(\nu,i)&=&\int _{R_{0}}^{R_{1}} \int _{0}^{2\pi} I(\nu,R,\phi,i)\frac{R dR d\phi}{D^2}\cos(i)+\nonumber \\
	&&\frac{I_\mathrm{rim}(\nu,R)}{D^2}A_\mathrm{rim}(i)\cos(90-i),
\end{eqnarray}
where $R_0$ and $R_1$ are the disk inner and outer radii and $D$ is the distance of the source to the observer. In Eq. \ref{eq:flux-disk}  $A_\mathrm{rim}\cos(90-i)$ is the visible area of the disk inner rim. If the disk inner rim were a perfect vertical wall, its surface area would be $A_\mathrm{rim}(i)=4\pi h(R_0)R_0^2$, where $h(R_0)$ is the disk aspect ratio at the inner edge, which is taken to be 0.05. In order to be consistent with the observations regarding the inclination-angle dependence of the rim flux \citep{IsellaNatta2005}, the rim is taken to be a vertical wall seen under a  constant 60 degree inclination angle.

To calculate the synthetic spectra numerically, we first set up the two-dimensional computational domain with $N_\mathrm{R}$ logarithmically distributed radial and $N_\phi$ equidistantly distributed azimuthal grid cells. For a given set of stellar parameters, shown in Table \ref{table:1}, first the grazing angle of the incident stellar irradiation is determined according to Eq. (\ref{eq:grazing-angle-flaring}). The temperature distributions in the disk atmosphere and  interior are calculated according to Eq. (\ref{eq:temp-surf}) and Eqs. (\ref{eq:temp-int-irr}, \ref{eq:acc-temp}, \ref{eq:temp-rim}, \ref{eq:temp-int-mod}), respectively. After determining the dust surface-density of the disk atmosphere with Eq. (\ref{eq:dens-surf}), the optical depth along the line of sight is calculated by Eq. (\ref{eq:tau_nu_2}) using the gas opacity given by Eq. (\ref{eq:gas-kappa}) and considering Eqs. (\ref{eq:ilp})-(\ref{eq:Doppler-shift}) to calculate appropriate intrinsic line profiles and Doppler shifts. In order to determine the mass absorption coefficient of the $^{12}\mathrm{C}^{16}\mathrm{O}$ molecule we used the transition data provided by \citet{Goorvitch1994}, such as transition probability $A$, lower and upper state energy $E_\mathrm{l},\,E_\mathrm{u}$ and statistical weight $g_\mathrm{u}$. Knowing the optical depth ($\tau(\nu,R,\phi,i)$), the temperature in disk atmosphere and interior in the 2D computational domain the expected flux along the line of sight can be calculated by applying Eq. (\ref{eq:intensity}) and Eq. (\ref{eq:flux-disk}) at each frequency in the vicinity of the investigated transition.

\section{Disk models}

In this section we describe in detail our disk models that fed into the synthetic spectral line model. As a first step, we calculated the disk surface-density and velocity distribution perturbed by a massive embedded planet with the publicly available hydrodynamical code FARGO of \citet{Masset2000}. A very important difference to the planet-free case is that the disk surface-density distribution is very heavily modified. It shows a clear elliptic character of the gap opened by the giant planet. According to \citet{KleyDirksen2006}, \citet{DAngeloetal2006}, and also to our calculations (see below), the disk becomes eccentric after several hundred orbits of the embedded planet. With regard to our synthetic spectral line model, the most relevant are those effects that are originating from the gas dynamics caused by the giant planet. It is well known that a pure Keplerian motion of the gas parcels on circular orbits results in symmetric double-peaked emission line profiles \citep{HorneMarsh1986}. As the CO emission is strongly depressed $2-3\,\mathrm{AU}$ in our models, the origin of the double-peaked profiles is clear. Any deviation of the gas dynamics from the pure rotation will break this symmetry, resulting in the distortion of line profiles. If the distorting effect is strong enough, the line profile distortion could be significant to be detected by high-resolution spectroscopic observations.

Indeed several T\,Tauri stars show symmetric but centrally peaked CO line profiles that can be explained by radially extent CO emission overriding the double peaks \citep{Najitaetal2003,Brittainetal2009}. Stronger CO emission can be expected if UV fluorescence plays a role \citep{Krotkovetal1980}, or the gas temperature is not well coupled to the dust \citep{KampDullemond2004,Glassgoldetal2004}. In the tenuous region above the disk atmosphere the gas can be significantly hotter than  is predicted by the double-layer model. As a consequence the slowly rotating distant disk parcels could produce a substantial contribution to the low-velocity part of the line profile, resulting in a centrally peaked profile. Nevertheless, for simplicity's sake here we do not consider any of the above mentioned effects.

\subsection{Hydrodynamical setup}
\label{sect:Hydro}

We adopt dimensionless units in hydrodynamical simulations, for which the unit of length and mass is taken to be the orbital distance of the planet, and the mass of the central star, respectively. The unit of time, $t_0$, is taken to be the reciprocal of the orbital frequency of the planet, resulting in $t_0=1/2\pi$, setting the gravitational constant to unity. In each disk model we assigned four planetary masses to the embedded planet, which are in increasing order $q=m_\mathrm{pl}/m_{*} = 0.001$, $0.003$, $0.005$, and $0.008$, where $m_\mathrm{pl}$ and $m_*$ are the planetary and stellar masses. Regarding the disks geometry, the flat thin disk approximation was assumed with an aspect ratio of $H(R)/R=0.05$. The disk extends between $0.2-5$ dimensionless units. This computational domain was covered by 256 radial and 500 azimuthal grid cells. The radial spacing was logarithmic, while the azimuthal spacing was equidistant. This results in practically quadratic grid cells, because the approximation $\Delta R\sim R\Delta\phi$ is valid at each radius. The disks are driven by $\alpha$-type viscosity \citep{ShakuraSunyaev1973}, which is consistent with our thermal disk model with an intermediate value of $\alpha$, which was taken to be $1\times10^{-3}$. Note that assuming a cold disk (therefore virtually non-ionized) the viscosity generated by magnetohydrodynamical turbulence is hard to explain, although, according to \citet{Mukhopadhyay2008}, the transient growth of two or three-dimensional pure hydrodynamic elliptic-type perturbations could result in $\alpha\simeq 10^{-1}-10^{-5}$, depending on the disk scale height (decreasing $\alpha$ with increasing scale height). The disk's initial surface density profile is given by a power law $\Sigma(R)=\Sigma_0 R^{-1/2}$, where the surface density at 1 distance unit $\Sigma_0$ is taken to be $2.15\times 10^{-5}$ in dimensionless units. For simplicity, a locally isothermal equation of state is applied for the gas, and the disk self-gravity is neglected. While the inner boundary of the disk is taken to be open, allowing the disk material to leave the disk on accretion timescale, the outer boundary is closed, i.e. no mass supply is allowed.

Below we assign physical units to the dimensionless quantities. If the stellar mass is assumed to be $m_*=1M_{\sun}$, the planetary masses in our models correspond to $1M_{\mathrm{J}}$, $3M_{\mathrm{J}}$, $5M_{\mathrm{J}}$, and $8 M_{\mathrm{J}}$. Thanks to the dimensionless calculations it is possible to scale the results to different stellar masses. For different stellar masses, the mass of the giant planet scales with the stellar mass, i.e $m_{\mathrm{pl}}= 1\times 10^{-3}\,m_{*}$, $3\times 10^{-3}\,m_{*}$, $5\times 10^{-3}\,m_{*}$, $8\times 10^{-3}\,m_{*}$, and the stellar masses are taken to be $m_*=0.5\,M_{\sun}$, $1\,M_{\sun}$, $1.5\,M_{\sun}$. Furthermore, the distance unit is set to the distance of the planet to the star, i.e., the planet orbits at 1\,AU. Below we extended our models to tight and wide systems, where the giant planets are orbiting at 0.5 and 2\,AU, respectively. In $1\,M_{\sun}$ stellar mass disk models the surface density at 1\,AU is $\Sigma_0=2.15\times 10^{-5} M_{\sun}\,\mathrm{AU^{-2}}\simeq 191\,\mathrm{g/cm^2}$ at the beginning of simulation, which is also scaled with the mass of the central star. But note that the solution to the hydrodynamical system of equations is independent of $\Sigma_0$, if there is no back-reaction to the planet, see the vertically integrated Navier-Stokes equations in \citet{Kley1999}. For $0.5\,M_{\sun}$, $1\,M_{\sun}$ and $1.5\,M_{\sun}$ stellar mass model the disk mass in the computational domain ($0.2\,\mathrm{AU}\leq R \leq 5\,\mathrm{AU}$) is $5\times10^{-4}\,M_{\sun}$, $1\times10^{-3}\,M_{\sun}$ and $1.5\times10^{-3}\,M_{\sun}$. These values refer to the mass of the inner disk only, as the whole disk may extend to several tens or hundreds of AU, and the total disk mass is in the conventional range of $0.01-0.1\,M_{\sun}$. More details about the models can be found in Table \ref{table:1}.

\begin{table}
\begin{minipage}[t]{\columnwidth}
\caption{Stellar and disk parameters in our hydrodynamical models. The stellar parameters were taken from a publicly available tool \citep{Siessetal2000}. The disk extends between $0.2-5\,\mathrm{AU}$ and the orbital distance of the massive planets is $1\,\mathrm{AU}$.}
\label{table:1}
\centering
\renewcommand{\footnoterule}{}
\begin{tabular}{c c c c c c}
\hline\hline

Model No. & $m_*\,(M_{\sun})$ & $T_{*}\,(\mathrm{K})$ & $R_*\,(R_{\sun})$ & $i\,(^\circ)$ & $m_\mathrm{p}\,(M_\mathrm{J})$\\

\hline

\#1  & 0.5 & 3760 & 1.4 &  20,40,60 & 0.5 \\
\#2  & 0.5 & 3760 & 1.4 &  20,40,60 & 1.5 \\
\#3  & 0.5 & 3760 & 1.4 &  20,40,60 & 2.5 \\
\#4  & 0.5 & 3760 & 1.4 &  20,40,60 & 4   \\
\hline                         
\#5  & 1   & 4266 & 1.83 & 20,40,60 & 1   \\
\#6  & 1   & 4266 & 1.83 & 20,40,60 & 3   \\
\#7  & 1   & 4266 & 1.83 & 20,40,60 & 5   \\
\#8  & 1   & 4266 & 1.83 & 20,40,60 & 8   \\
\hline
\#9  & 1.5 & 4584 & 2.22 & 20,40,60 & 1.5 \\
\#10 & 1.5 & 4584 & 2.22 & 20,40,60 & 4.5 \\
\#11 & 1.5 & 4584 & 2.22 & 20,40,60 & 7.5 \\
\#12 & 1.5 & 4584 & 2.22 & 20,40,60 & 12  \\

\hline
\end{tabular}
\end{minipage}
\end{table}

\subsection{Synthetic spectral calculation setup}
\label{sec:spectra-calc-setup}

We calculate the spectral line of a fundamental band CO transition ($\mathrm{V}=1\rightarrow0\mathrm{P}(10)$, $\lambda_0=4.7545\,\mathrm{\mu m}$), which is not blended by the higher excitation V=2$\rightarrow$1 and V=3$\rightarrow$2 lines, although in our thermal model the temperature is not high enough to excite the higher vibrational levels at all.\footnote{In our thermal model the temperature stayed below 1500\,K everywhere in the computational domain. At this temperature the $V\geq 2$ vibrational levels are not thermally excited.} The CO rotational and vibrational level populations were calculated in local thermodynamical equilibrium. The line profiles were calculated in 200 wavelengths and in the same numerical domain that was used in hydrodynamical simulations. According to our preliminary tests on ``circularly Keplerian''\footnote{We note that the term ``Keplerian'', widely used in the literature, here means  ``circularly Keplerian'' motion. On the other hand, elliptic motion is also Keplerian, but below we will use the widely accepted nomenclature for circular motion, though we think it is not entirely correct.} disks, the CO fundamental band spectra did not show significant changes with increasing outer disk boundary. This can be explained by the fact that a substantial part of the CO fundamental band emission is arising in the inner disk, consequently the outer part of the disk ($R>3-5\,\mathrm{AU}$, depending on the stellar mass) does not contribute to the disk emission at 4.7 micron. Contrary to this, the distance of the inner boundary of the disk to the central star heavily influences the CO fundamental band spectra, i.e. the closer the disk inner boundary to the star the line-over-continuum is weaker and broader. This is because the disk innermost regions give stronger contribution to the continuum, which in turn weakens the CO emission normalized to continuum, although the CO emission itself is also getting stronger due to an increased amount of hot CO. The line broadening is a natural consequence of the larger orbital velocity of gas orbiting closer to the star.

Because we investigate disks with already formed planets, it is reasonable to assume that a cavity is formed at the very inner disk like the ones found in disks of several T\,Tauri stars \citep{Akesonetal2005,Eisneretal2005,Eisneretal2007,Salyketal2009}. If dust evaporation is takes place close to the star, the CO will be depleted too, due to the disappearance of the dust which protects the CO molecules against UV photo dissociation. In this way, the observed inner radius of CO ($R_\mathrm{CO}$) should intuitively be similar or slightly larger than the dust sublimation radius $R_\mathrm{sub}$. Indeed, $R_\mathrm{CO}$ was found to be larger than $R_\mathrm{sub}$ by a factor of a few in several disks, e.g., see Fig.\,5. of \citet{Eisneretal2007} or Fig.\,10 in \citet{Salyketal2009}. Contrary to this, \citet{Carr2007} found that the CO inner radius is $\sim 0.7$ that of the dust. If the gas disk extends as far as 0.05\,AU, there would be an additional broad component to the emission that is not modeled here. Taking these argumentations into account it is reasonable to assume that the CO inner radius is set by the dust sublimation. The dust sublimation radius $R_\mathrm{sub}$ is given by
\begin{equation}
	\label{eq:sublimation-radius}
	R_{\mathrm{sub}}\simeq0.03\left(\frac{T_*}{T_{\mathrm{sub}}}\right)^{5/2}\frac{R_{*}}{R_{\sun}}\mathrm{ AU},
\end{equation}
where Eq. (\ref{eq:temp-surf}) with standard dust opacity law, $\beta=1$ \citep{Rodmannetal2006} was used. Assuming $T_\mathrm{sub}=1500\,\mathrm{K}$ for dust sublimation temperature, $T_*\sim4000\,\mathrm{K}$ and  $R_{*}=1.8\,R_{\sun}$ for stellar surface effective temperature and radius, appropriate for a 2.5\,Myr solar mass star, we obtain $R_\mathrm{sub}=0.05\,\mathrm{AU}$. To keep this simple, we set the disk inner boundary (namely the CO inner radius) fixed to $R_\mathrm{CO}\equiv4R_\mathrm{sub}=0.2\,\mathrm{AU}$, which is an acceptable value in disks hosted by T\,Tauri stars with 3500-5000\,K surface temperature and $1.4-2.2\,R_{\sun}$ radius. Note that $R_\mathrm{sub}$ was fixed throughout our models (except the ones where we investigated the effect of the size of the inner cavity) to make them comparable, neglecting the effect of the change in stellar luminosity on the $R_\mathrm{sub}$.

\begin{figure}
	\centering
	\includegraphics[width=\columnwidth]{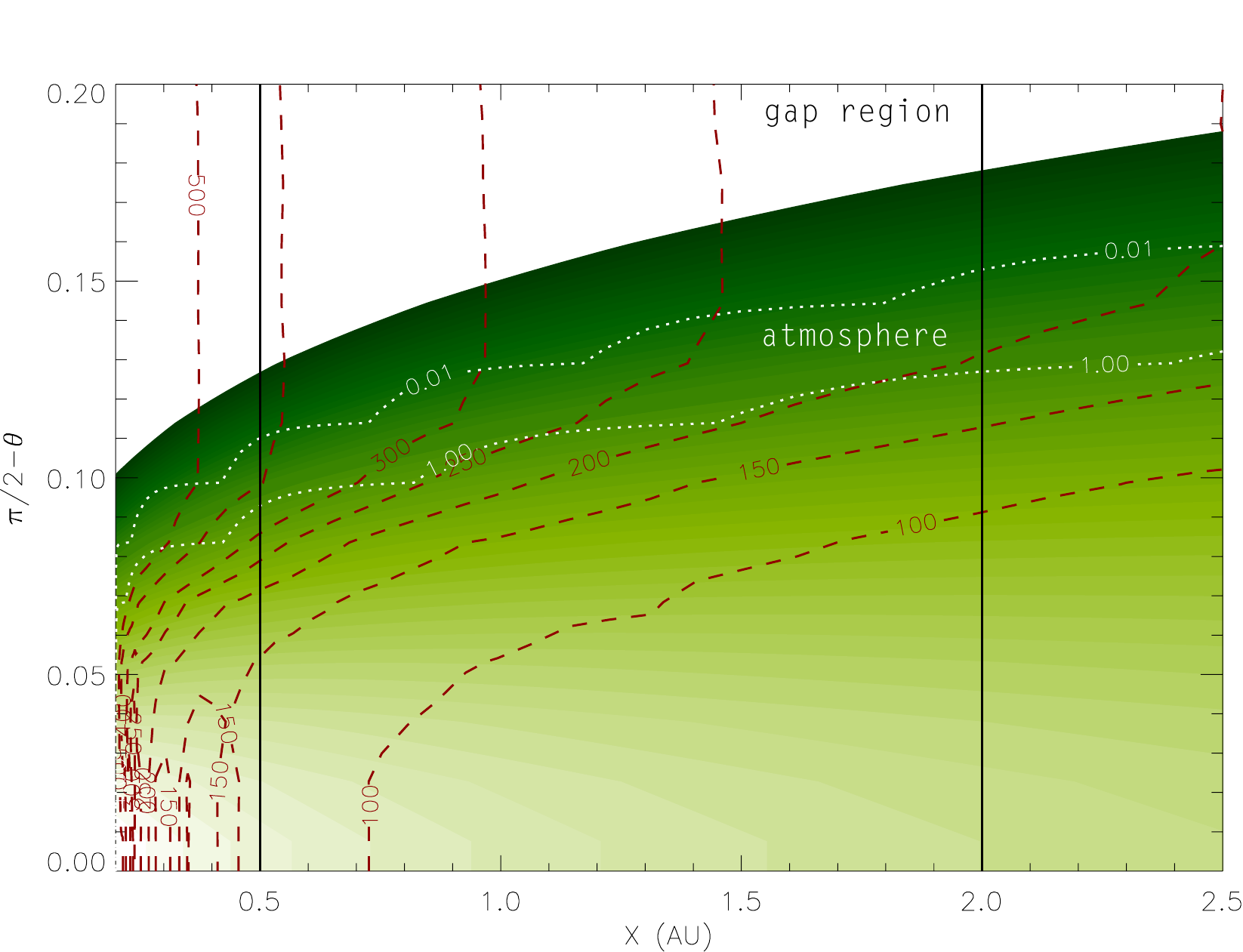}\\
	\includegraphics[width=\columnwidth]{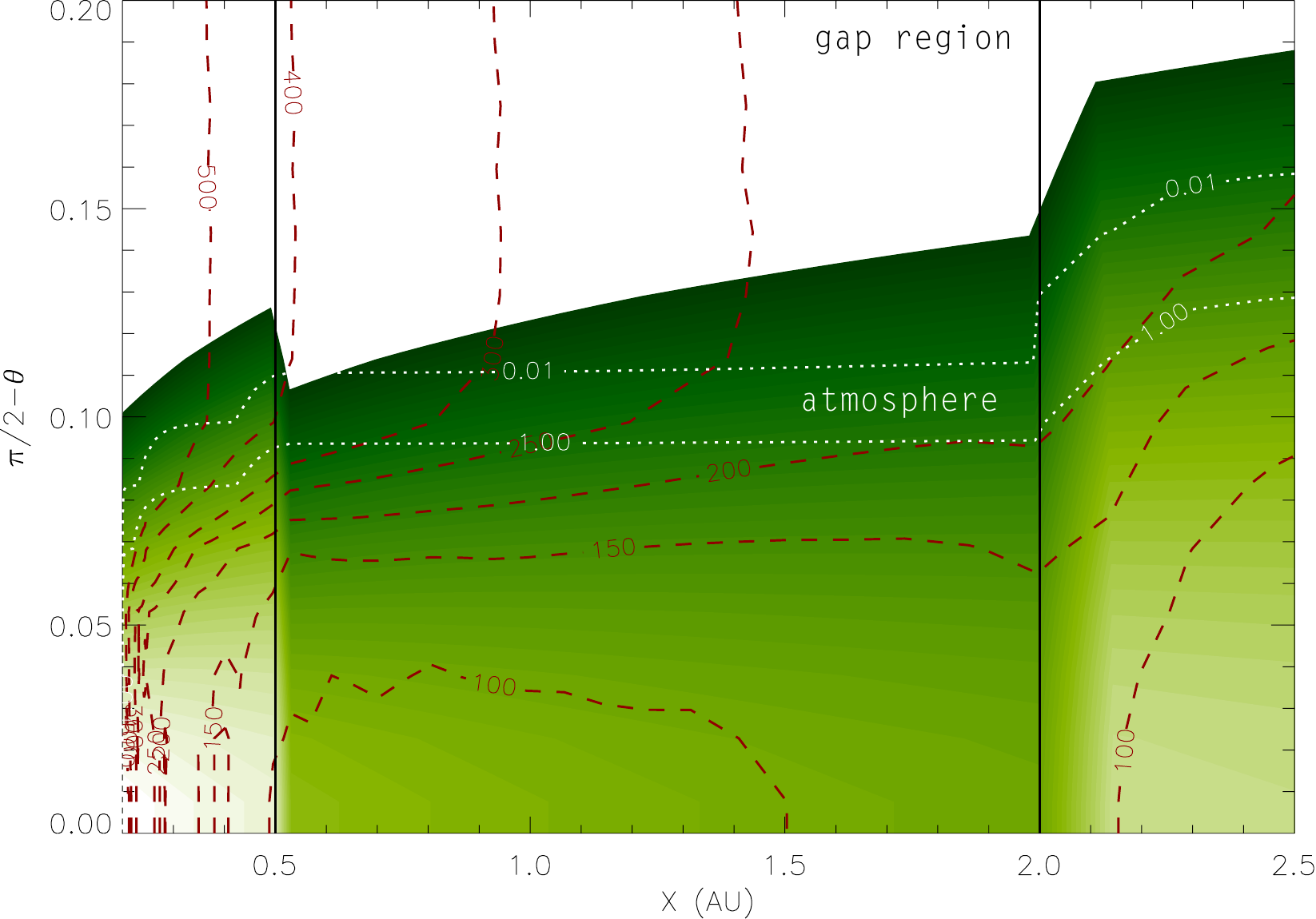}
	\caption{Logarithmic density (\emph{green shaded} in $g/cm^3$) and temperature contours (\emph{red dashed contours} in K) of the disk. The density contours are stopped at a gas density $\rho=10^{-34}\,\mathrm{g/cm^3}$ to avoid color crowding, while the maximum is $\rho=10^{-11}\,\mathrm{g/cm^3}$. The height of the disk atmosphere, i.e the range where the radial optical depth is less than 1 is shown with \emph{white dotted lines} in the disk inner region. Comparing the unperturbed case (\emph{top}) to the disk in which a gap exists at $0.5\,\mathrm{AU}<R<2\,\mathrm{AU}$, after depleting the density by 1/1000 (\emph{bottom}), it is appreciable that however the disk atmosphere is decreased in height, the temperature distribution is not considerably changed.}
	\label{fig:RADMC-output}
\end{figure}

It is reasonable to assume that a disk that is $2.5\,\mathrm{Myr}$ old contains a considerable amount of gas and tracer CO also in the inner disk, hence the stellar age is taken to be $2.5\,\mathrm{Myr}$ in all models. Note that according to \citet{Haischetal2001}, half of the observed young stars in nearby embedded clusters (NGC\,2024, Trapezium, IC\,348, NGC\,2264, NGC\,2362, NGC\,1960), with ages of about $3\,\mathrm{Myr}$, show near-IR  excess, which is related to the excess emission of dust in disks. Furthermore, \citet{Hillenbrand2005} also concluded that the median lifetime of the optically thick inner disk is between $2-3\,\mathrm{Myr}$. 

In order to simplify the models it is assumed that the dust consists of pure silicates with $0.1\,\mathrm{\mu m}$ grain size \citep{DrainLee1984}. According to this the mass absorption coefficient of the dust is taken to be $2320\,\mathrm{cm^2/g}$ at visual wavelength. Note that neither the effect of the coagulation nor the settling out of tenuous atmosphere of grains with a large size are taken into account. The size distribution and therefore the overall mass absorption coefficient of the dust is taken to be the same in the interior and the atmosphere. In this way the dust opacity is slightly overestimated. According to Eq. (\ref{eq:tau_nu_2}) the atmospheric gas density and thus gas line emission strength is slightly underestimated. The dust-to-gas and CO-to-gas mass ratios are assumed to be constant throughout the disk and are taken to be $10^{-2}$ and $4\times10^{-4}$ (measured per gram of gas+dust mass) in the disk atmosphere and interior, respectively.

For the planet-free disks circular Keplerian velocity distribution is applied to determine the observed line center shift, given by Eq. (\ref{eq:Doppler-shift}). For a massive planet-bearing disk observed under inclination angle $i$, the Doppler shift of the emission from a single patch of gas in the disk compared to its fundamental frequency $\nu_0$ emerging from a given $R,\phi$ point is calculated by
%
%
\begin{eqnarray}
	\Delta\nu(R,\phi,i)&=&\frac{\nu_{0}}{c}\left\{u_\mathrm{R}(R,\phi)\left[\sin(\phi)+\cos(\phi)\right]\right. +\nonumber\\
	&&\left. u_\mathrm{\phi}(R,\phi)\left[\cos(\phi)-\sin(\phi)\right]\right\}\sin(i),
	\label{eq:Doppler-shift-pertdisk}
\end{eqnarray}
where the radial $u_{\mathrm{R}}(R,\phi)$ and azimuthal $u_{\mathrm{\phi}}(R,\phi)$ velocity components of gas parcels are provided by the hydrodynamic simulations.

The perturbations in the surface density distribution of the disk interior and atmosphere were not taken into account, because the disk atmosphere density given by Eq. (\ref{eq:dens-surf}) is independent of the density distribution of the disk even in the gap, as long as there is enough dust in the gap to keep it optically thick. To test the optical thick assumptions, let us for a moment neglect the effect of disk shelf-shadowing that occurres in the inner edge of the gap. Applying the dust density in the disk $\Sigma_\mathrm{d}(R=1\,\mathrm{AU})\simeq\Sigma_{0}X_d$, where $X_d=0.01$ is the dust-to-gas ratio and the dust opacity $\kappa_\mathrm{V}=2320\,\mathrm{cm^2/g}$, in the gap, depleted to 0.1\% of the surrounding density, the optical depth at visual wavelengths along the incident stellar irradiation is  $\tau_\mathrm{V}=\Sigma_\mathrm{d}(R=1\,\mathrm{AU})\kappa_\mathrm{V}/\delta(R=1\,\mathrm{AU})\simeq1000$. To argue that the shelf shadowing can be neglected we calculated the height of the disk atmosphere and temperature distribution in a planet-free and planet-bearing disk by a 2-D Monte-Carlo code RADMC for dust continuum radiative transfer \citep{DullemondDominik2004} with the same dust prescription as in our synthetic spectral model. In the planet-bearing disk model the gap is taken to be extended from 0.5\,AU to 2\,AU and artificially depleted in density to 0.1\% that of the planet-free case. The amount of gap depletion is an average value measured in our hydrodynamic calculations. The results are shown in Fig. \ref{fig:RADMC-output}. It is evident that although the height of the disk atmosphere ($\tau_\mathrm{V}=1$ along the line of sight of stellar irradiation) above the mid-plane is decreased in the gap, the temperature in the disk atmosphere is not substantially changed. Because the temperature is not significantly decreased we neglect the surface density perturbations. Note that the effect of the distance of the gap from the central star on the shelf shadowing is not considered. The dust and gas temperature decoupling was also not taken into account. Because the gas has a temperature well in excess of the dust in depleted regions, where the gas column density is $\ll 10^{22}$ \citep{Glassgoldetal2004,KampDullemond2004}, we can expect increased contribution to the CO line flux originating from the gap. Contrary to this, if the base of the gap is shadowed completely by the inner wall, the gap does not contribute to the CO emission. Thus a gap could cause a strong permanent line profile asymmetry due to its asymmetric geometry (see Fig. \ref{fig:FARGO-output}), which requires further investigation.

Because the accretion heating has been taken into account according to Eq. (\ref{eq:acc-temp}), we had to set an appropriate accretion rate. The accretion rate measured in hydrodynamical simulations gives a value of about $2\times10^{-8}\,M_{\sun}/\mathrm{yr}$. Note that according to \citet{Fangetal2009} the accretion rate inferred from $\mathrm{H}\alpha$ emission luminosity in the young star population of Lynds\,1630N and 1641 clouds in the Orion\,GMC with an age about $2-3\,\mathrm{Myr}$, is between $10^{-10}-10^{-8}\,M_{\sun}/\mathrm{yr}$. In order to be consistent with the latter, the accretion rate is taken to be $5\times 10^{-9}\,M_{\sun}/\mathrm{yr}$ for all our models. However, note that at this accretion rate the contribution of the viscous dissipation to the disk continuum is weak in the near-IR band.

\section{Results}

\subsection{Effect of massive planets on the disk structure}

\begin{figure*}
	\centering
	\includegraphics[width=\columnwidth]{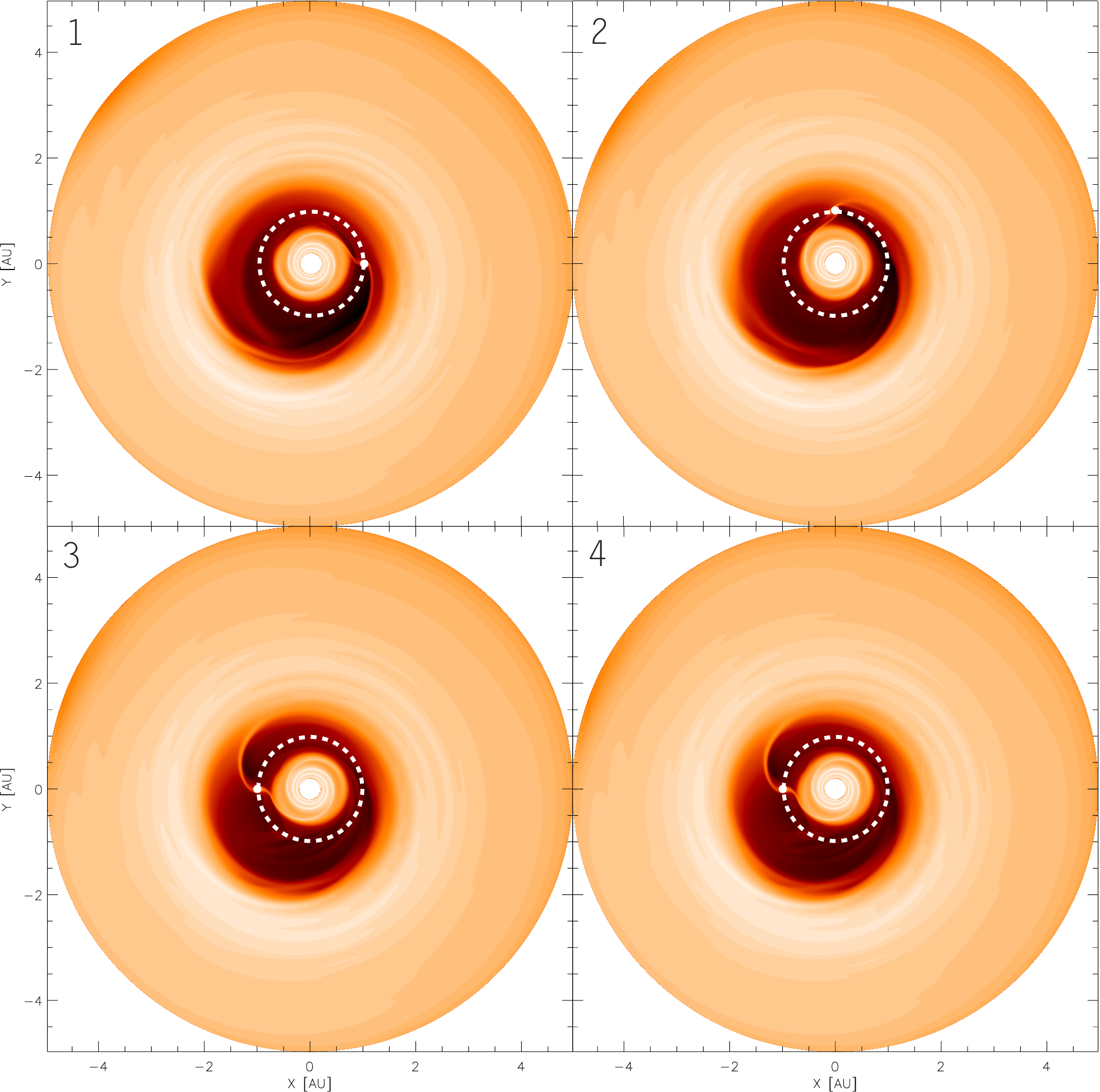}
	\includegraphics[width=\columnwidth]{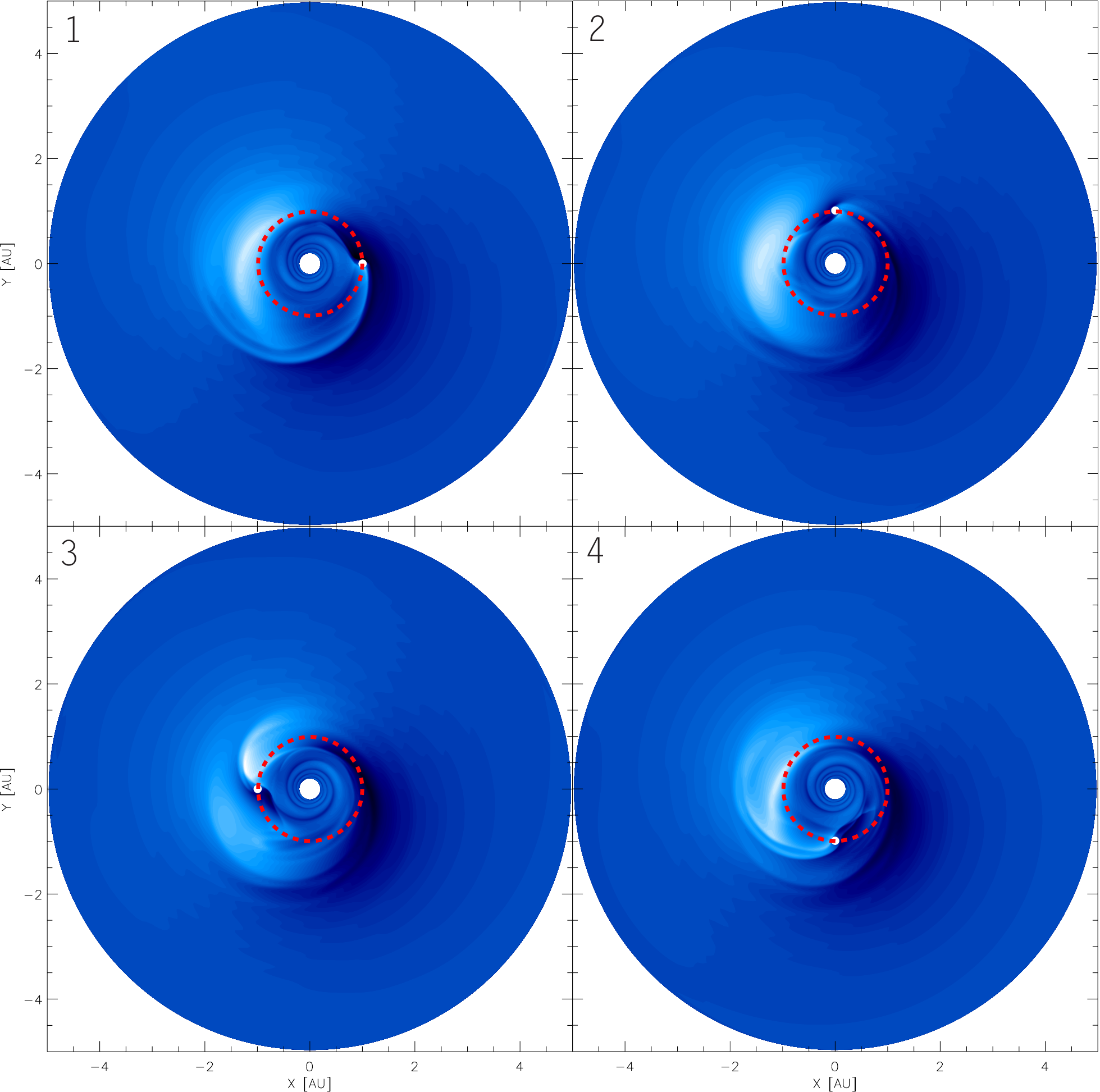}
	\caption{Surface density (\emph{left four figures, orange colors} in the range of $0.2-250\,\mathrm{g/cm^2}$) and the radial velocity component of the orbital velocity in the disk plane (\emph{right four figures, blue colors} in the range of $-10.86-+10.96\,\mathrm{km/s}$) distributions of the disks gaseous material taken from four different azimuthal positions of the planet's 2000th orbit. In our model the total disk mass is $0.05\,M_{\sun}$, the planet mass is $8\,M_\mathrm{J}$, while the stellar mass is $1\,M_{\sun}$. The planetary orbit is shown with a white and red dashed circle in the density and velocity plots. The gap (\emph{in dark}) and the elliptic shape of its outer rim is clearly visible in the surface density distribution. In the vicinity of the gap the disk material shows a strong deviation from the circular Keplerian rotation, see the white clumps in the radial velocity component distribution.}
	\label{fig:FARGO-output}
\end{figure*}

Three groups of models were computed through 2000 planetary orbits, with stellar and planetary parameters listed in Table \ref{table:1}. In each run the planet was kept fixed on circular orbit during the first 1000 orbits. After the 1000th orbit the planet was released. It thus felt the (gravitational) backreaction of the disk, resulting in its inward migration. In a good accordance with the expectations, we found that a more massive planet opened a broader and deeper gap. The depletion is 0.25\%-0.1\% for planets with mass in the range of $1\,M_\mathrm{J}-8\,M_\mathrm{J}$. After the first 1000 orbits a quasi steady state disk structure has developed in all simulations, which slightly changed during the following 1000 orbits, when the planet orbit was not fixed anymore.

To shed light on how the giant planet distorts the originally circularly Keplerian gas flow, several snapshots of density and velocity distributions were taken during the 2000th planetary orbit. Figure\,\ref{fig:FARGO-output} shows four snapshots of the density and the radial velocity component of the orbital velocity of gas for an $8\,M_\mathrm{J}$ mass planet orbiting an $1\,M_{\sun}$ star (model \#8) during one orbit. It is evident that in a disk with an embedded massive planet the overall orbits of gas parcels are non–circularly Keplerian because the radial components of their orbital velocity distribution are strongly departing from zero. Moreover, we found that the elliptic gap, while preserving its shape, precesses slowly retrograde with a period of about $150$ planetary orbits.

As we already mentioned, \citet{KleyDirksen2006} found that an originally circular disk with an embedded giant planet can reach an \emph{eccentric} equilibrium state if the mass of the embedded planet is larger than a certain limit ($3\,M_{\mathrm{J}}$ for a $1\,M_{\sun}$ star). Kley \& Dirksen found that for sufficiently wide gaps the growth of the eccentricity is induced by the interaction of the planet's gravitational potential with the disks material at the radial location of the 1:3 (outer) Lindblad resonance. For smaller planetary masses, this effect is damped mainly by the co-orbital and the 1:2 Lindblad resonances. If the gap is deep and wide enough, which is the case for a giant planet, the above resonances cannot damp the eccentricity-exciting effect appearing at the 1:3 Lindblad resonance, and the disk becomes eccentric. For a more detailed explanation of disk eccentricity growth see \citet{Lubow1991} and \citet{KleyDirksen2006}. The eccentric state of the disk in our simulations can also be clearly seen in Fig. \ref{fig:FARGO-output} where the shape of the outer rim of the gap becomes elliptic in the surface density distribution.

\begin{figure}
	\centering

	\includegraphics[width=\columnwidth]{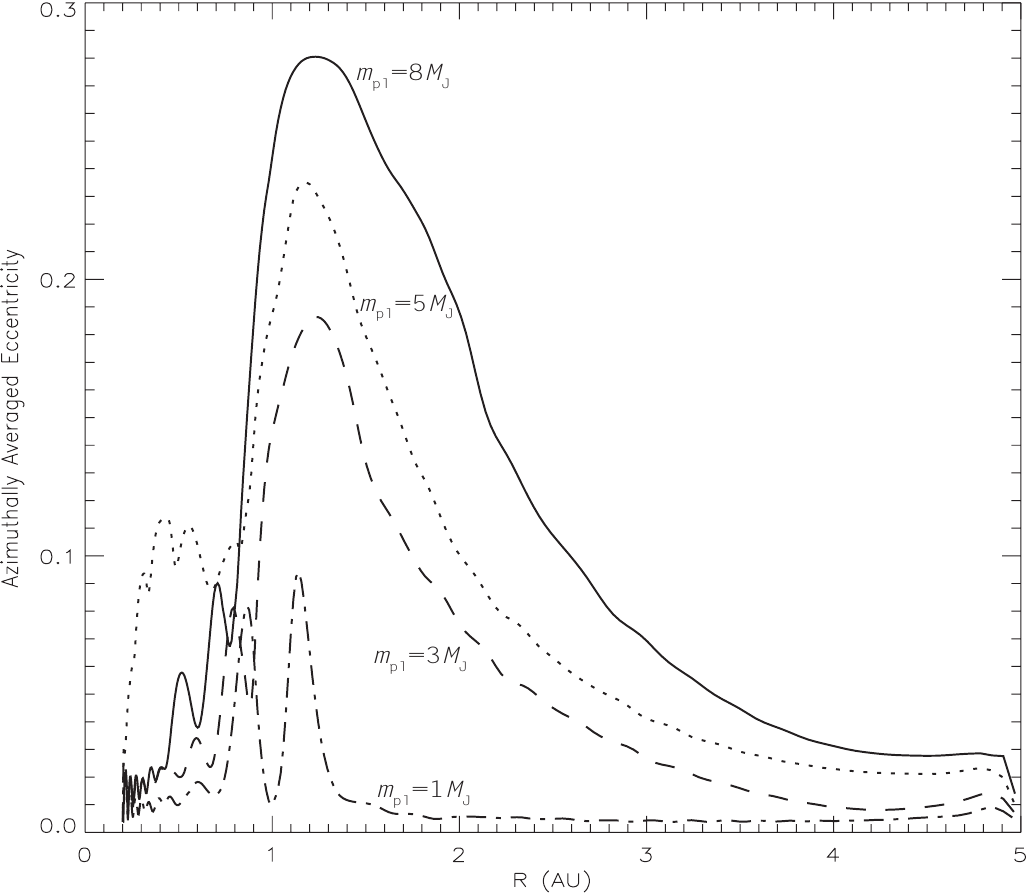}


	\caption{Azimuthally averaged eccentricities as the functions of radii after 2000 orbits of the giant planet. The eccentricity curves in models \#5-\#8 (where the planetary masses are $1\,M_\mathrm{J}$, $3\,M_\mathrm{J}$, $5\,M_\mathrm{J}$ and $8\,M_\mathrm{J}$ for $1\,M_{\sun}$ stellar mass) are shown with dot-dashed, dashed, dotted, and solid lines, respectively.}
	\label{fig:FARGO-ecc}
\end{figure}

The departure of the velocities of the gas parcels from the pure circularly Keplerian circular revolution can be characterized by calculating their eccentricities. Considering the eccentric equilibrium state of the disk, we expect that each gas parcel would move on a non-circular orbit, which may be characterized most conveniently by an average eccentricity value. Therefore, for each disk radius between 0.2\,AU and 5\,AU, the azimuthally averaged eccentricities of the orbits of the gas parcels were calculated in the following way
\begin{equation}
	\label{eq:ecc}
	e(R)=\int_0^{2\pi}\sqrt{1+2h(R,\phi) c(R,\phi)^2}d\phi.
\end{equation}
In Eq. (\ref{eq:ecc}) $c(R,\phi)$ and $h(R,\phi)$ stand for
\begin{equation}
	c(R,\phi)=x(R,\phi)v_\mathrm{y}(R,\phi)-y(R,\phi)v_\mathrm{x}(R,\phi)
\end{equation}
and
\begin{equation}
	h(R,\phi)=\frac{v_\mathrm{x}(R,\phi)^2+v_\mathrm{y}(R,\phi)^2}{2}-\frac{1}{\sqrt{x(R,\phi)^2+y(R,\phi)^2}},
\end{equation}
where $v_\mathrm{x}(R,\phi)$, $v_\mathrm{y}(R,\phi)$ and $x(R,\phi)$, $y(R,\phi)$ are the Cartesian velocity components and coordinates at point $R,\phi$. We found that the averaged eccentricities differ considerably from zero, and each eccentricity curve reaches its maximum near the outer boundary of the gap (Fig. \ref{fig:FARGO-ecc}). A priori one would expect that the averaged eccentricity values are higher for more massive planets. Indeed we found that the maximum of the azimuthally averaged eccentricity is monotonically increasing with a planetary mass in the range of $1\,M_\mathrm{J}\le m_\mathrm{pl} \le 8\,M_\mathrm{J}$. The eccentricity curve for the $8\,M_\mathrm{J}$ mass planet peaks about $e_{\mathrm{max}}\sim 0.3$, while the peak stays well bellow 0.1 for a $1\,M_\mathrm{J}$ mass planet. Note that a very similar behavior of the disk eccentricity has already been found by \citet{KleyDirksen2006}. In their cases, however, the maximum of the eccentricity curves is somewhat lower than in our cases, and at least $3\,M_\mathrm{J}$ is required to set the disk into eccentric state. This can be the consequence because contrary to \citet{KleyDirksen2006}, we used an $\alpha$-type viscosity in our simulations. The higher eccentricities we found are plausible inasmuch as \citet{KleyDirksen2006} found that the eccentricity of the disk is increasing with decreasing viscosity and in our approach the kinematic viscosity $\nu(R)=\alpha H^2 \Omega_\mathrm{K}(R)$ measured in dimensionless units is $2.5\times10^{-6}$ at $R=1$, which is smaller than the $\nu=1\times 10^{-5}$ used by \citet{KleyDirksen2006}.

\begin{figure}
	\includegraphics[width=\columnwidth]{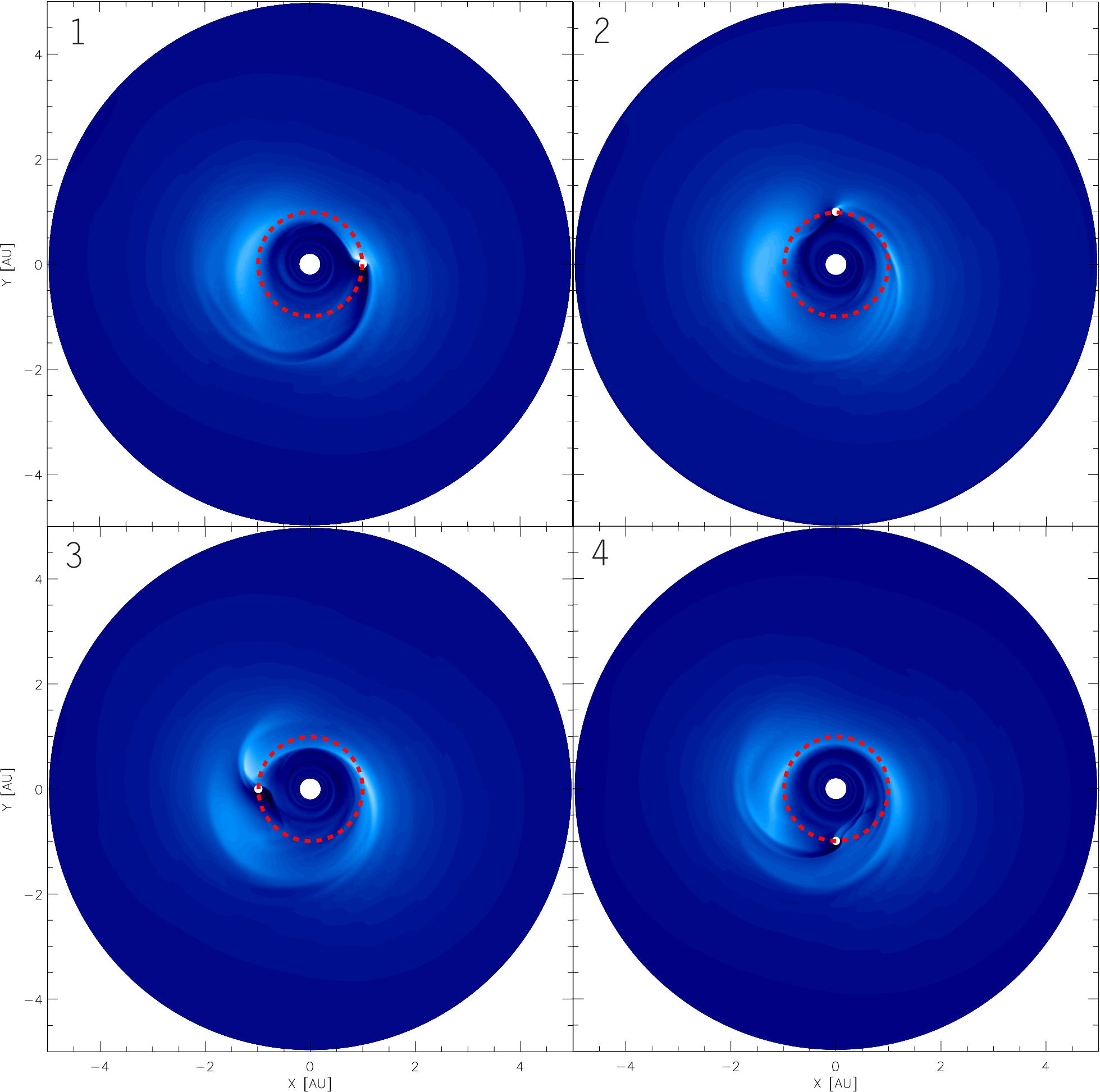}

	\caption{Offset of the line of sight component of the velocity distribution from circularly Keplerian values on the disk surface in the hydrodynamic simulations, in the same model presented in Fig. \ref{fig:FARGO-output}. Snapshots are calculated for four different azimuthal positions of the planet of the 2000th orbit. The velocity difference is in the range of $-7.92-4.88\,\mathrm{km/s}$. Here the inclination angle is taken to be $i=40^{\circ}$ and the viewing angle is taken to be $-90^{\circ}$, i.e the disk is seen from the bottom of Fig. \ref{fig:FARGO-output}. The planetary orbit is shown with a red dashed circle. Two permanent high-velocity regions can be seen at position angle $0^\circ$ and $180^\circ$, and a variable pattern in the vicinity of the planet orbits. The strength of the variable component can reach that of the permanent one, e.g, in the panel at the bottom left (planet is at position angle $180^\circ$).}
	\label{fig:FARGO-output-lsvel}
\end{figure}

As one can see, a giant planet has substantial impact on the density and velocity distributions of its host disk. It is an essential question, whether the velocity perturbations appear in the line-of-sight velocity with substantial strength. Figure\,\ref{fig:FARGO-output-lsvel} shows the subtracted distribution of the line-of-sight velocity in the perturbed and circularly Keplerian case. It is evident that the line-of-sight velocities show a significant departure from the circularly Keplerian fashion because the difference is non-vanishing. The radial velocity component of the orbital velocity and more importantly the line-of-sight velocity distributions have variable patterns following the planet on the top of a permanent excess seen at $0^\circ$ and $180^\circ$ position angle. Note that the the permanent pattern precesses slowly (with $\sim 150$ orbital periods), retrograde to the planet. The disk inclination angle $i$ is taken to be $40^{\circ}$ in the calculation of Fig. \ref{fig:FARGO-output-lsvel} and the disk was rotated to the line-of-sight in a way that the line-of-sight velocity component of dynamically perturbed gas parcels is maximized. Consequently we had expected that not only significant distortions appear in the line profiles, but that they vary in time within the orbital time scale of the giant planet. Because the deviation from circularly Keplerian velocity is in the range of $-7.92-4.88\,\mathrm{km/s}$, the width of variable component in the line profile should be $\sim10\,\mathrm{km/s}$, depending on the inclination angle.

\subsection{Distortion of CO lines}

Below we address the following questions: (i) Does the line-of-sight component of the non-circularly Keplerian velocity distribution (Fig. \ref{fig:FARGO-output-lsvel}) result in significant distortions in the CO spectral line profiles? (ii) Are these distortions varying in the planets orbital timescale? (iii) How do the distortions depend on the inclination angle, the planetary and stellar mass, the orbital distance of the planet, the size of the inner cavity and the disk geometry (if the disk is flared or not)? (iv) Are these distortions strong enough to be detected by high-resolution near-IR spectral measurements? A complete set of answers to the above questions may give us a novel method to detect massive planets embedded in protoplanetary accretion disks.

\begin{figure}
	\centering
	\includegraphics[width=\columnwidth]{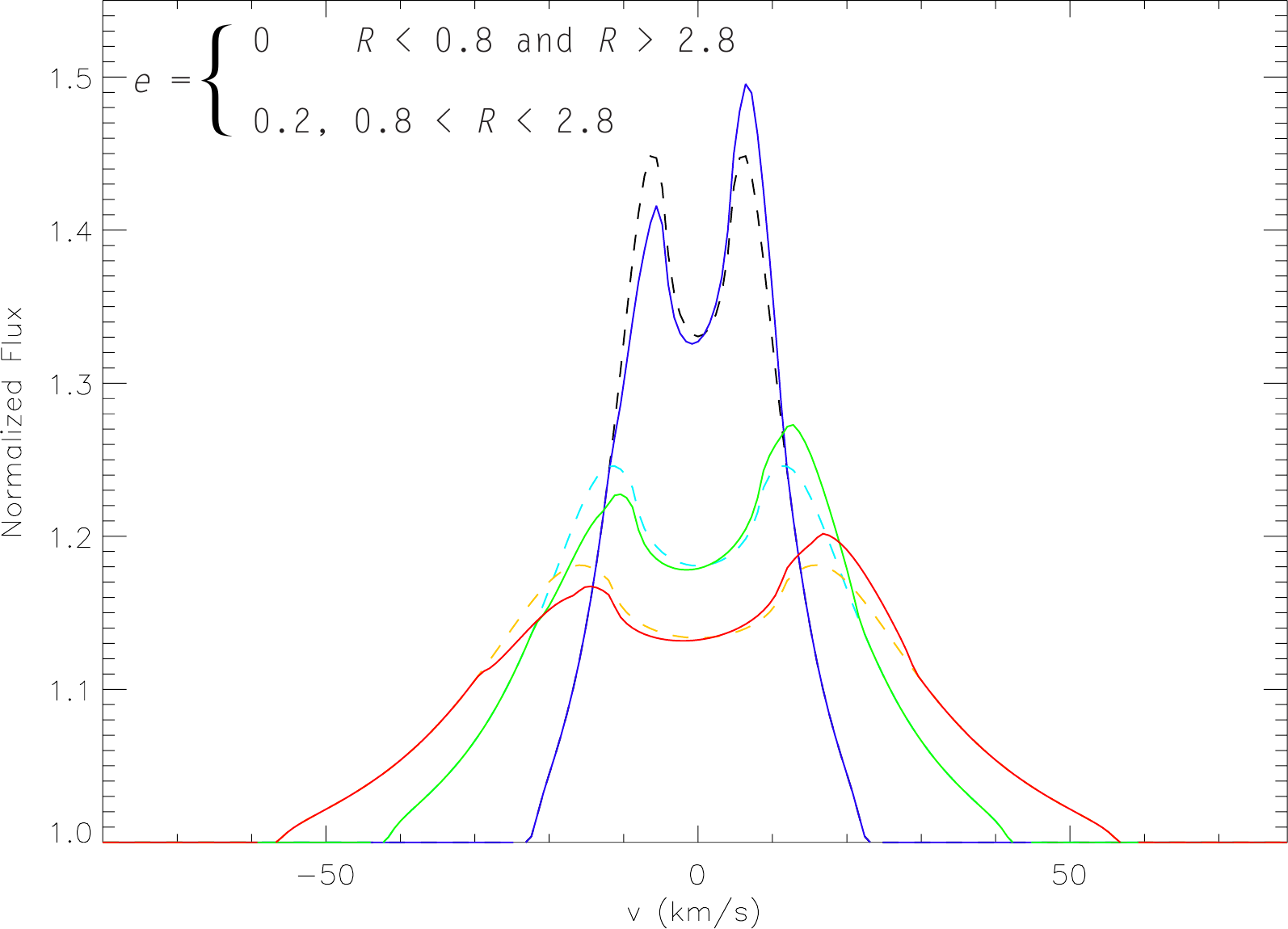}
	\caption{Asymmetric CO ro-vibrational line profile in a circularly Keplerian disk with a gap between $0.8-2.8\,\mathrm{AU}$ (similar to model \#8), where the gas is flowing in eccentric orbit with $e=0.2$. The gas outside the gap is in Keplerian circular orbit. It is evident that the line profiles are asymmetric. For comparison, we display the symmetric double-peaked line profiles (dashed lines) emerging from a planet-free disk as well.}
	\label{fig:Ms1-e02(08-28)}
\end{figure}

As was shown in Fig. \ref{fig:FARGO-ecc} the disk eccentricity is considerable near the gap in all planet-bearing disk models. If the gas parcels move on pure elliptic orbits in the gap, the CO line profile becomes permanently asymmetric. This effect is illustrated in Fig. \ref{fig:Ms1-e02(08-28)}, where we have calculated the emerging line profile of a disk with an eccentricity $e=0.2$ in the gap between $0.8\,\mathrm{AU}-2.8\,\mathrm{AU}$ hosted by a $1\,M_{\sun}$ star viewed $20^{\circ}$, $40^{\circ}$ and $60^{\circ}$ degree of inclination angles. Thus we can expect that the non-circularly Keplerian property of the velocity field shown in Figs. \ref{fig:FARGO-output} and \ref{fig:FARGO-output-lsvel} will break the symmetry of the spectral lines. Although the orbits of the gas parcels can be characterized by average eccentricities, their orbits are more complicated than those of a pure elliptic motion, because of the perturbations induced by the planet.

First we examined the CO emission line profile distortions from a disk in which a 8$M_\mathrm{J}$ planet orbits a 1$M_{\sun}$ star (model \#8). As expected, the strongly non-circularly Keplerian velocity flow significantly modified the CO line profiles. In Fig. \ref{fig:V1-0P10-profiles}(a) we display the V=1-0\,P(10) line profiles for three different inclination angles. The line profiles show a strongly asymmetric double-peaked shape. Studying Fig. \ref{fig:V1-0P10-profiles} (a) two distortion patterns of the spectral line profiles can be recognized. An excess can be seen near the red peak of the lines at about $11\,\mathrm{km/s}$, $21\,\mathrm{km/s}$ and $29\,\mathrm{km/s}$, for inclinations $i=20^{\circ}$, $40^{\circ}$ and $60^{\circ}$, respectively. Moreover, a deficiency appears around the blue peak of each spectral line at slightly lower velocities. The above velocities correspond to the orbital velocity of the gas revolving around a $1\,M_{\sun}$ star at 0.8-0.9\,AU, i.e. near the inner boundary of the gap, taking into account the appropriate inclination angles. Smaller-scale distortions also appear in the red and blue peaks as well, corresponding to the 0.8-2.9\,AU region, i.e. the whole gap.

Calculating the line profiles at different orbital phases during one orbit, we found that the shape of the already distorted line profile varies in time, yielding a clear dependence on the orbital position of the planet, see results in Fig. \ref{fig:variation-Ms1-Mb8-V1-0P10} for model \#8. The permanent asymmetry in lines is caused by the permanent velocity pattern seen at PA $0^\circ$ and $180^\circ$ in Fig. \ref{fig:FARGO-output-lsvel}. Note that the expression ``permanent'' is not entirely correct because the velocity pattern slowly ($\sim 150$ orbital periods of planet) precesses retrograde to the planet, but we still use it henceforward. The variable component indicated with arrows is moving in regions between $\sim\pm 25\,\mathrm{km/s}$ due to the variable pattern following the planet. Note that the maximum width of variable component is $\sim 10\,\mathrm{km/s}$, as is expected, which is resolvable by CRIRES, whose maximum resolution is $3\,\mathrm{km/s}$. In this particular model the time scale of variations is on the order of weeks, because the orbital period of the planet is one year, thus the time elapsed between snapshots is approximately 18 days.

\subsection{Planetary and stellar masses}

To gain further insights, it is useful to study the influence of the masses on the strength of the spectral line distortions. First we have investigated the cases in which a $1\,M_{\sun}$ star hosts a $1\,M_\mathrm{J}$, $3\,M_\mathrm{J}$, $5\,M_\mathrm{J}$, and $8\,M_\mathrm{J}$ mass planet at 1 AU, corresponding to models \#5-\#8. Parts of the results are presented in Fig. \ref{fig:V1-0P10-profiles}(b) for model \#5 ($m_\mathrm{pl}=1\,M_\mathrm{J}$, $m_*=1\,M_{\sun}$) and in Fig. \ref{fig:V1-0P10-profiles}(a) and Fig. \ref{fig:variation-Ms1-Mb8-V1-0P10} for model \#8. One can conclude that for a less massive planetary companion the line profile distortions are weaker. The same conclusion can be obtained by the growing influence of the increasing planetary mass on the disk eccentricity, see Fig. \ref{fig:FARGO-ecc}.

A natural way to study the effect of the mass of the central star on the line profile distortions would be that the planetary mass is kept fixed, and the stellar mass is increased. On the other hand, we recall that dimensionless units were used in hydrodynamical simulations thus the mass of the planet is expressed in stellar mass units. Thus changing either the planet mass or the stellar mass is equivalent to changing the ratio between the planetary and the stellar mass. Therefore, one could conclude from the previous results that the larger the stellar mass, the stronger the line profile distortions. The situation is  however a bit more complicated; if we change, for instance, the stellar mass, the luminosity of the star changes, which has an influence on the temperature distribution in the disk atmosphere and interior as well. The effect of the stellar mass can be therefore studied if the planetary-to-stellar mass ratio is kept fixed, while the mass along with the surface temperature and the stellar radius is changed in the synthetic spectral model. Due to the effect of the increased/decreased flux of stellar irradiation in case of the larger/smaller stellar mass, the outer boundary of the CO emitting region will be moved farther/closer to the star, and increased/decreased in size. To investigate the dependence of the observable line profile distortions on the stellar mass, we have recalculated the above presented line profiles for  a $0.5\,M_{\sun}$ (models \#1-\#4) and a $1.5\,M_{\sun}$ (models \#9-\#12) central stars. Part of the results are shown in Fig. \ref{fig:V1-0P10-profiles}(c) (for model \#4) and Fig. \ref{fig:V1-0P10-profiles}(d) (for model \#12), respectively. Comparing the line profiles to those presented in Fig. \ref{fig:V1-0P10-profiles}(a) (for model \#8), it is evident that the profiles are considerable narrower for $0.5\,M_{\sun}$ and broader for $1.5\,M_{\sun}$ models, due to the change in the Keplerian angular velocity ($\Omega_\mathrm{K}(R)=(Gm_{*}/R^3)^{1/2}$). Moreover, the line-to-continuum ratio at the maximum is slightly decreased and increased about a same amount for disks with 0.5 and $1.5\,M_{\sun}$ star. The change in line-to-continuum ratio can be easily explained: for a smaller mass star, the disk temperature is also decreased due to the decrease in stellar luminosity in a way that the ratio of the CO line to the disk plus star continuum is decreased too. The opposite is true for larger stellar masses. Finally, we can conclude that for a given planet-to-star mass ratio the larger the central stellar mass, the larger the observable CO line profile distortion caused by the giant planet. Here we have to note that although the observational data on T\,Tauri stars \citep{Akesonetal2005} show that the size of the disk inner cavity is increasing with increasing stellar luminosity (which can be the consequence of the increased dust sublimation radius, see Eq. (\ref{eq:sublimation-radius})), resulting in smaller continuum and stronger relative line strength, in our cases the size of inner cavity stayed fixed in order to make models comparable.

Analyzing  Fig. \ref{fig:V1-0P10-profiles}(b) we conclude that the line profile distortions in model \#5 ($m_\mathrm{pl}=1\,M_{\mathrm{J}}$, $m_*=1\,M_{\sun}$,  1\,AU orbital distance) are so small that the giant planet signal below $1\,M_{\mathrm{J}}$ is strongly suppressed. Note that a disk with a $1\,M_\mathrm{J}$ mass planet embedded into it will not get into the eccentricity state at all, see Fig. \ref{fig:FARGO-ecc}. Planets orbiting the host star closer than 1\,AU, however, can induce stronger distortions, as shown in the following section.

\subsection{Orbital distance of the planet}

\begin{figure*}
	\centering
	
	\includegraphics[width=8cm]{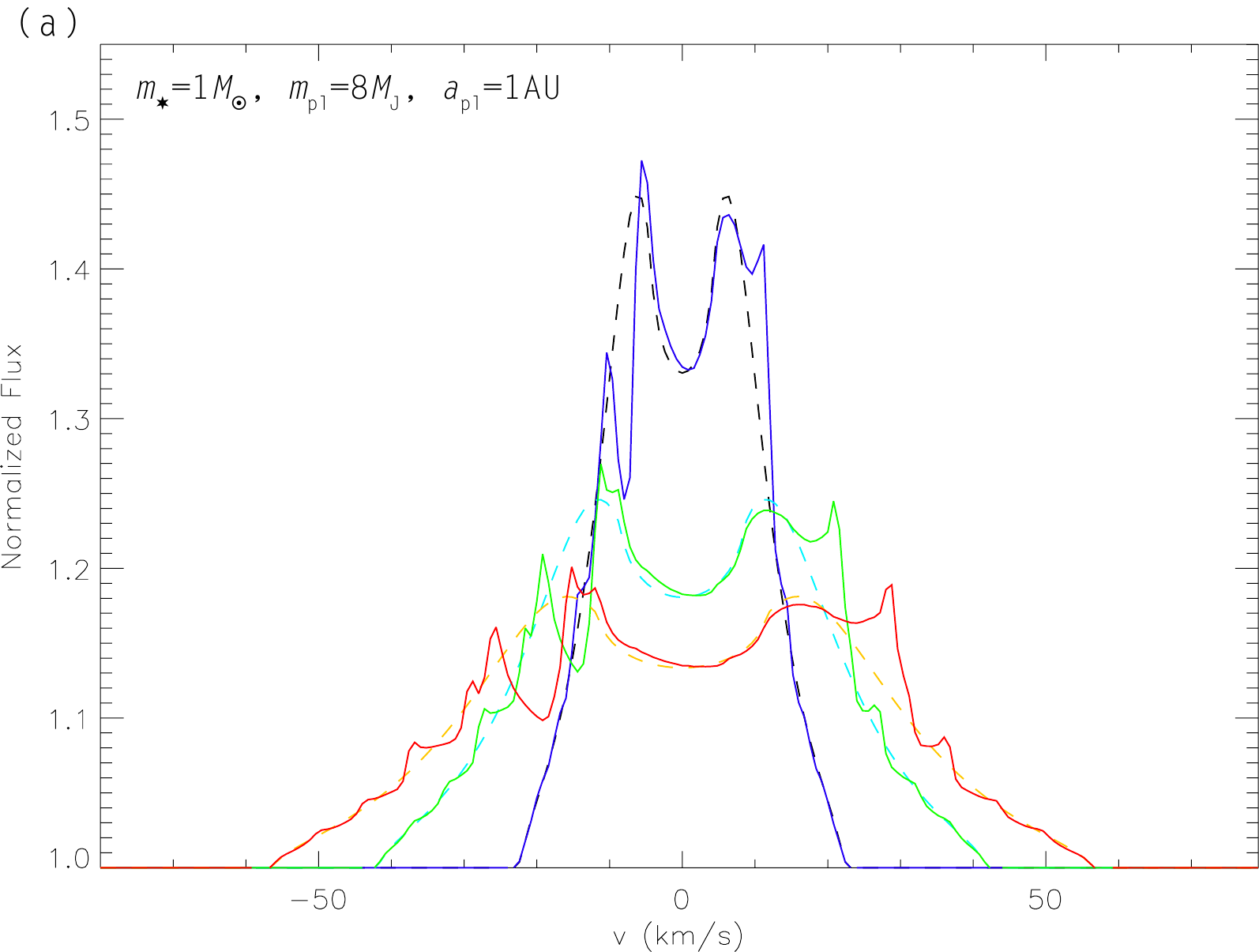}
	\includegraphics[width=8cm]{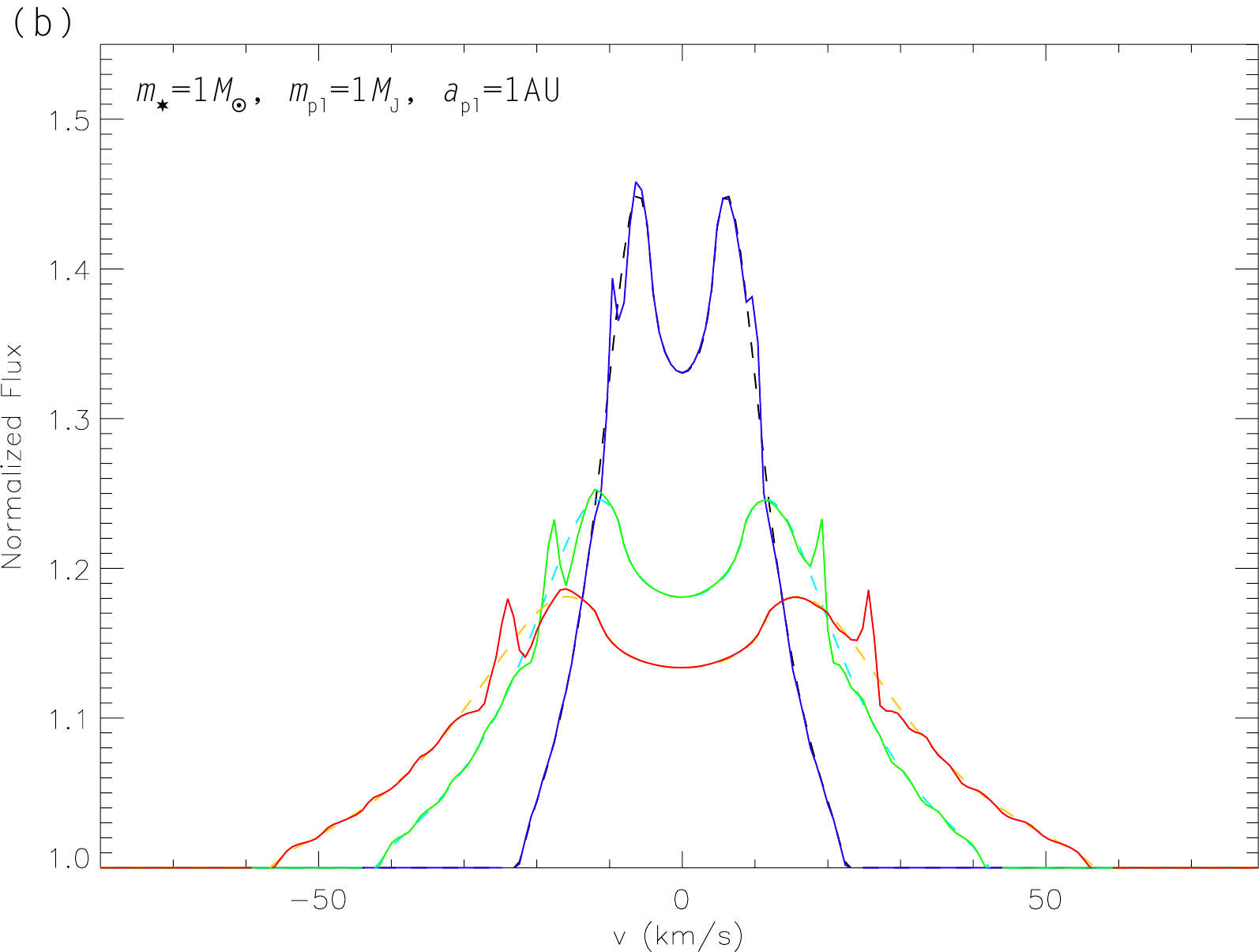}\\
	\includegraphics[width=8cm]{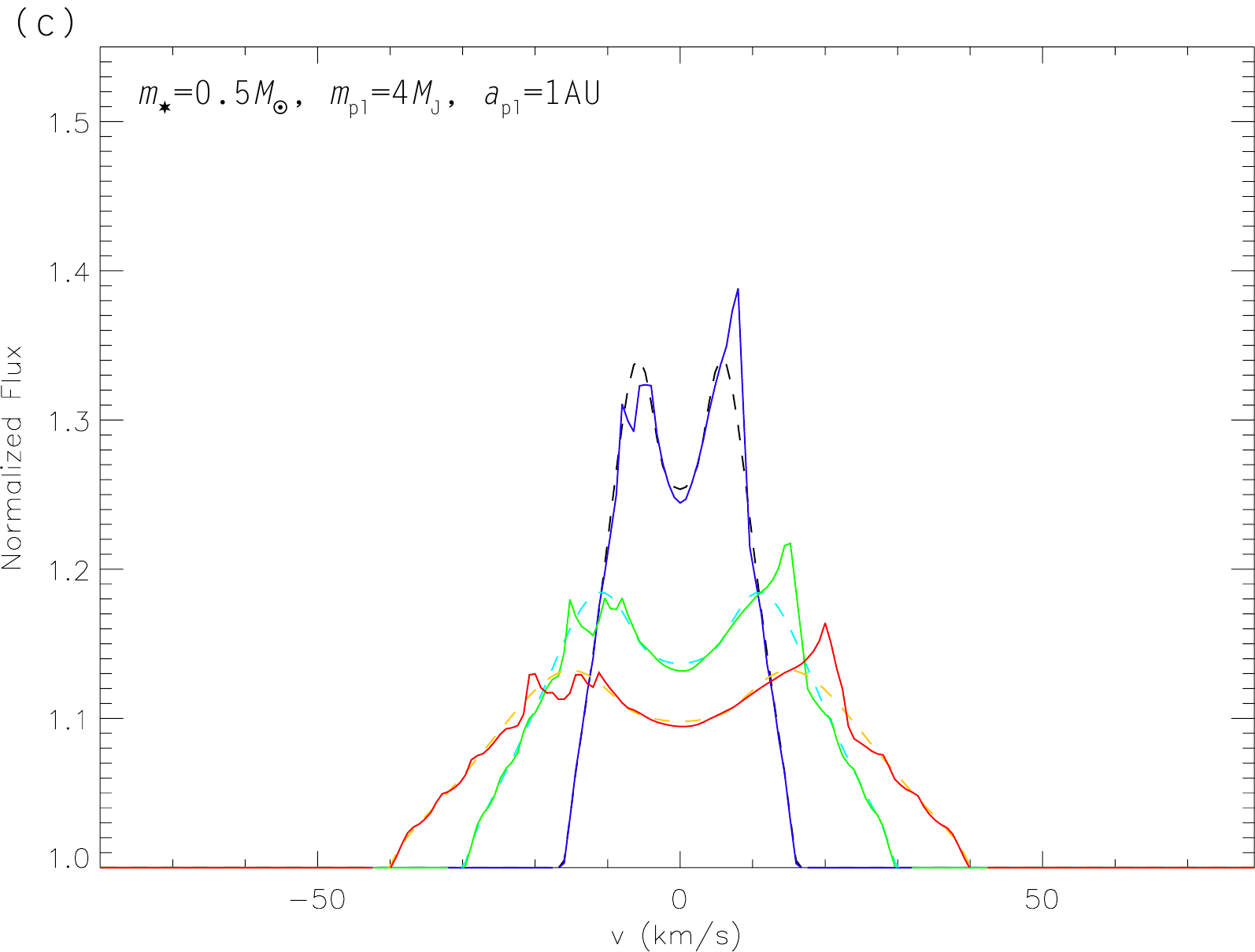}
	\includegraphics[width=8cm]{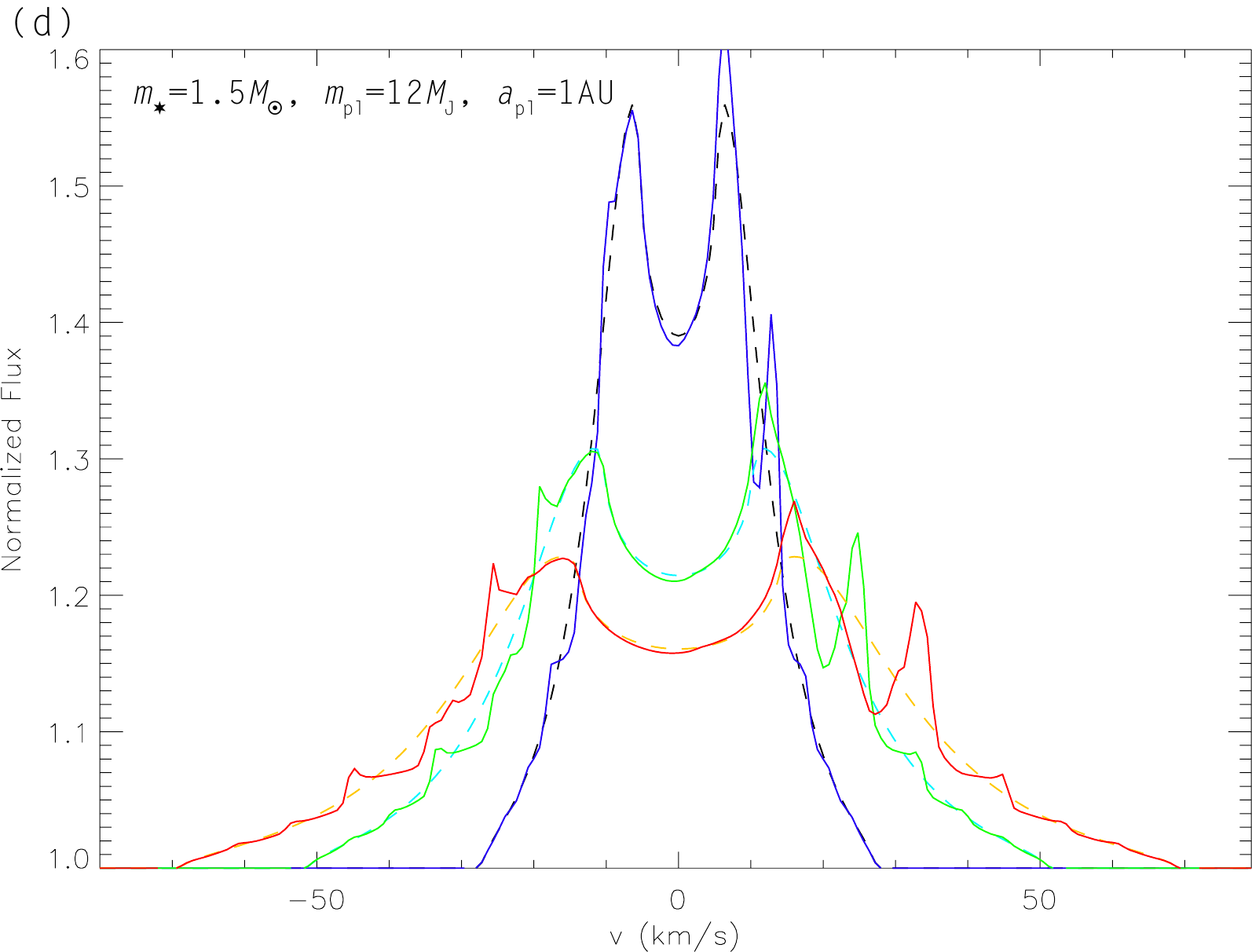}\\
	\includegraphics[width=8cm]{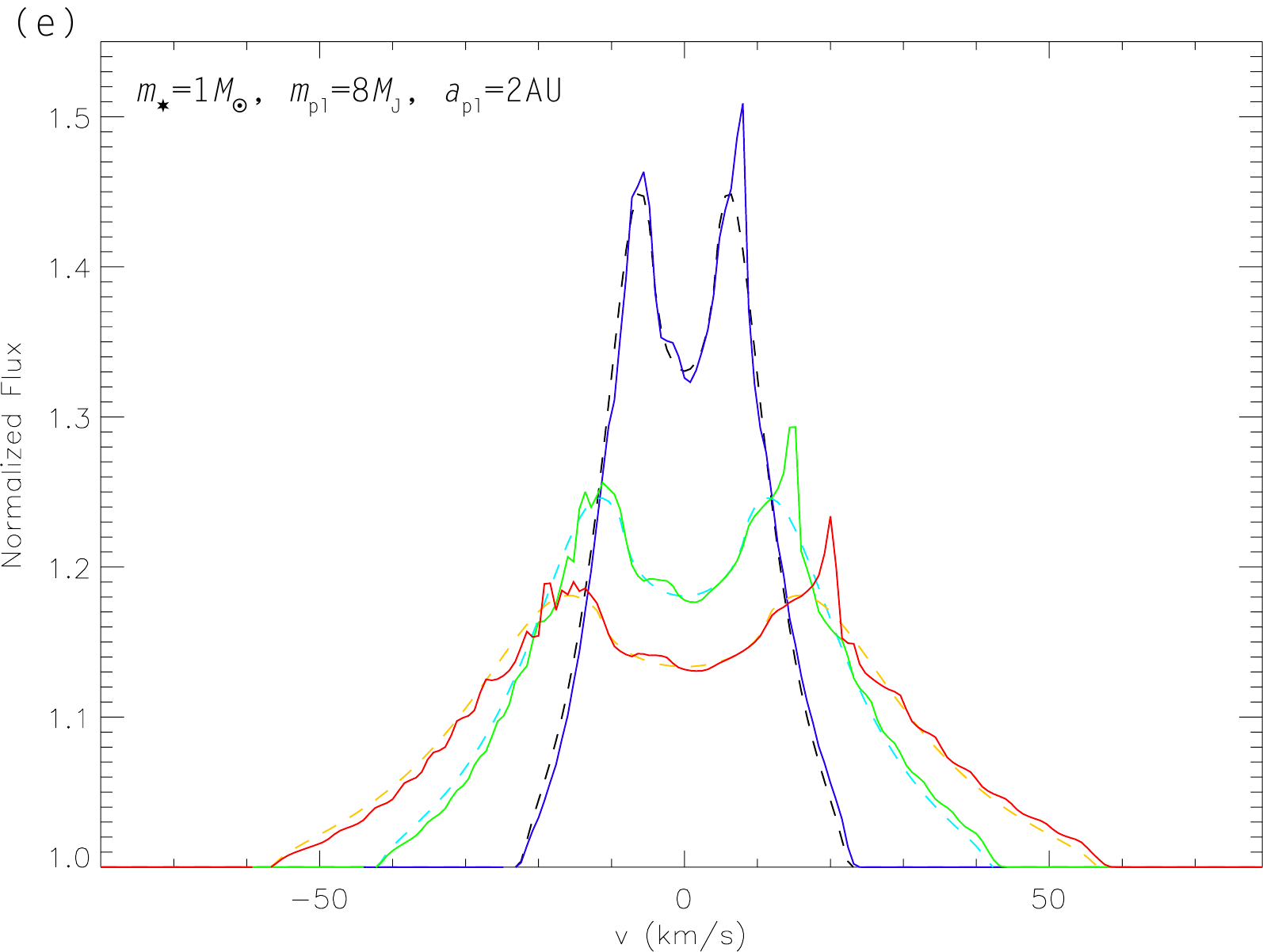}
	\includegraphics[width=8cm]{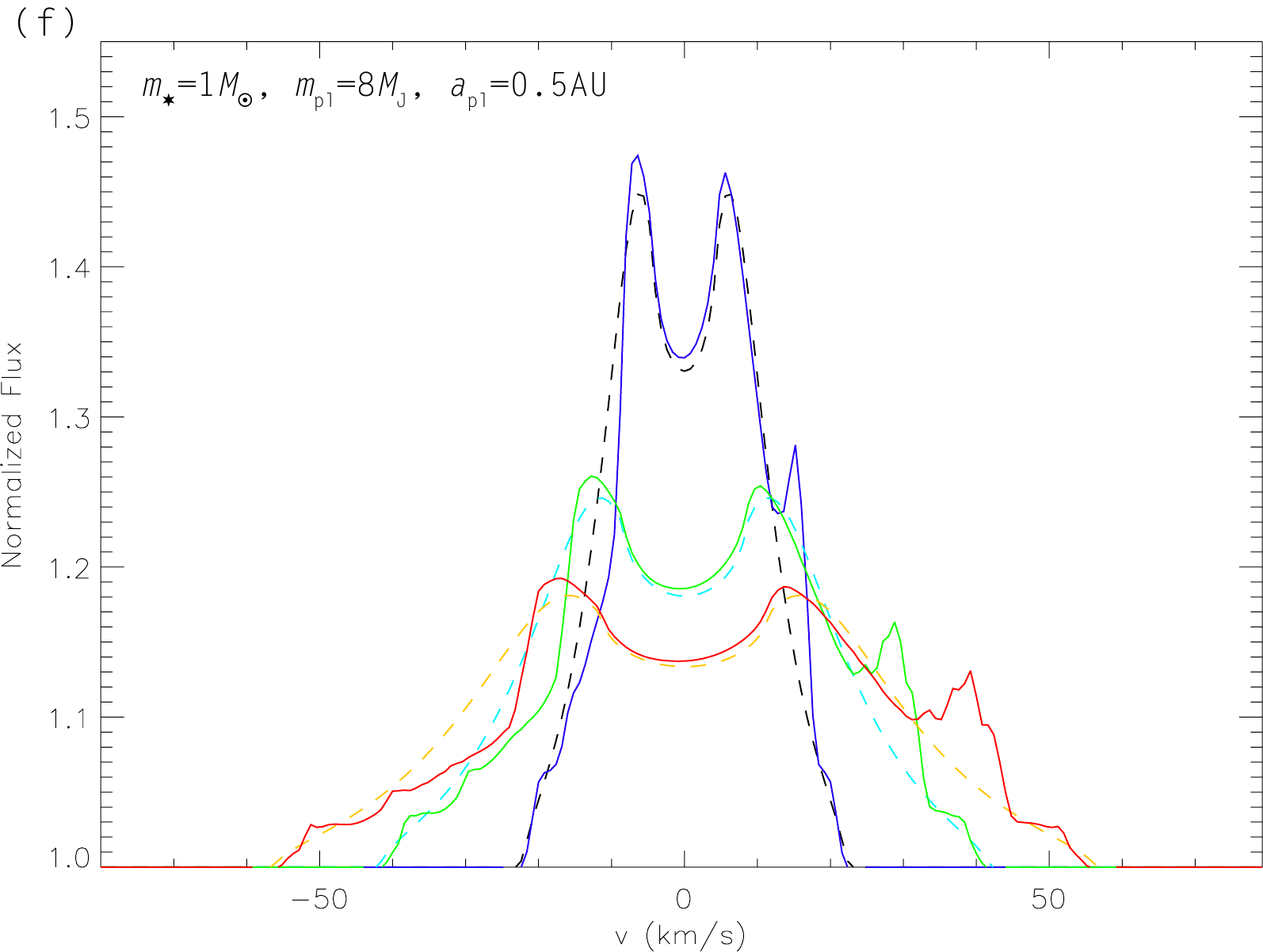}\\
	\includegraphics[width=8cm]{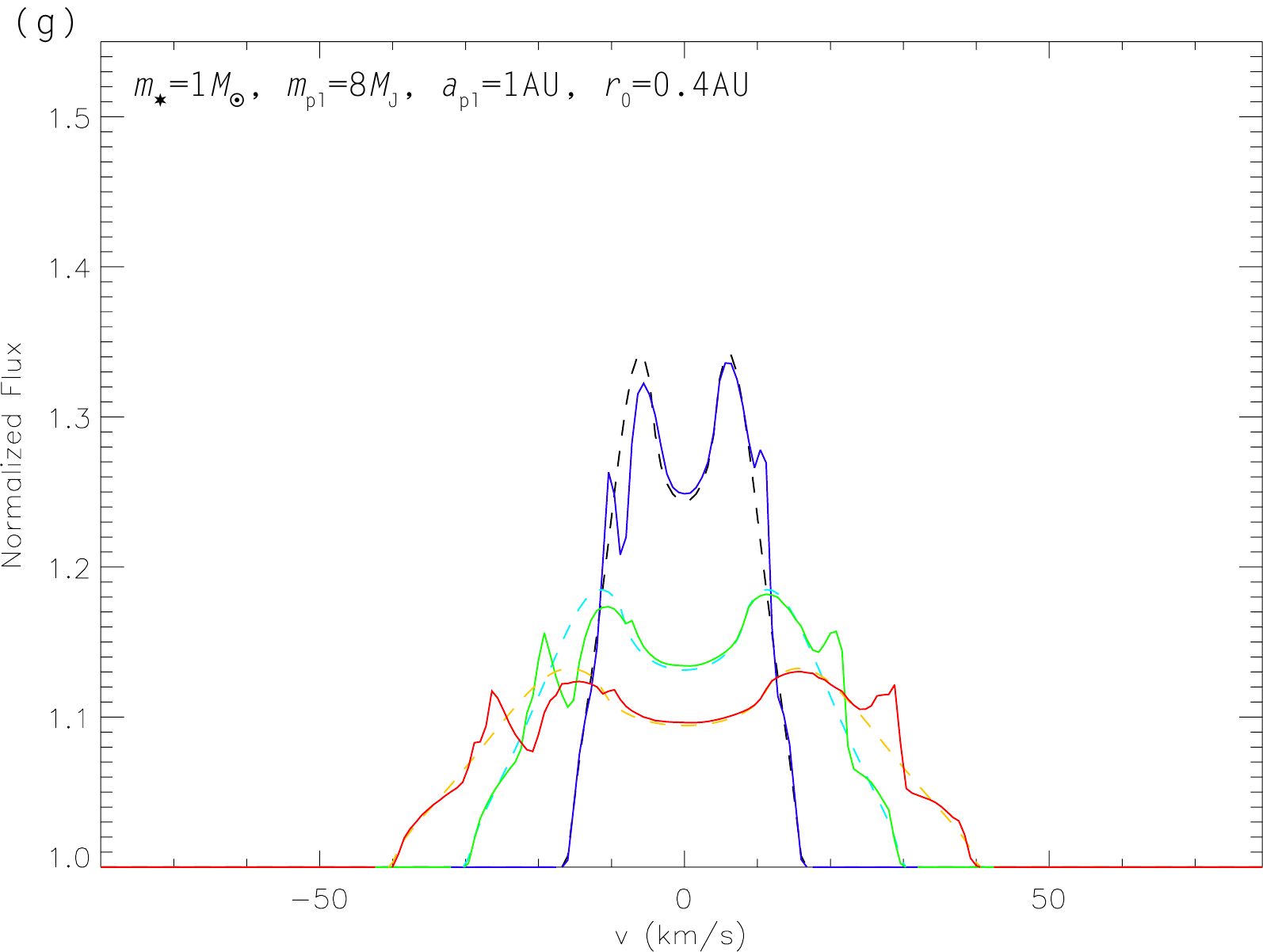}
	\includegraphics[width=8cm]{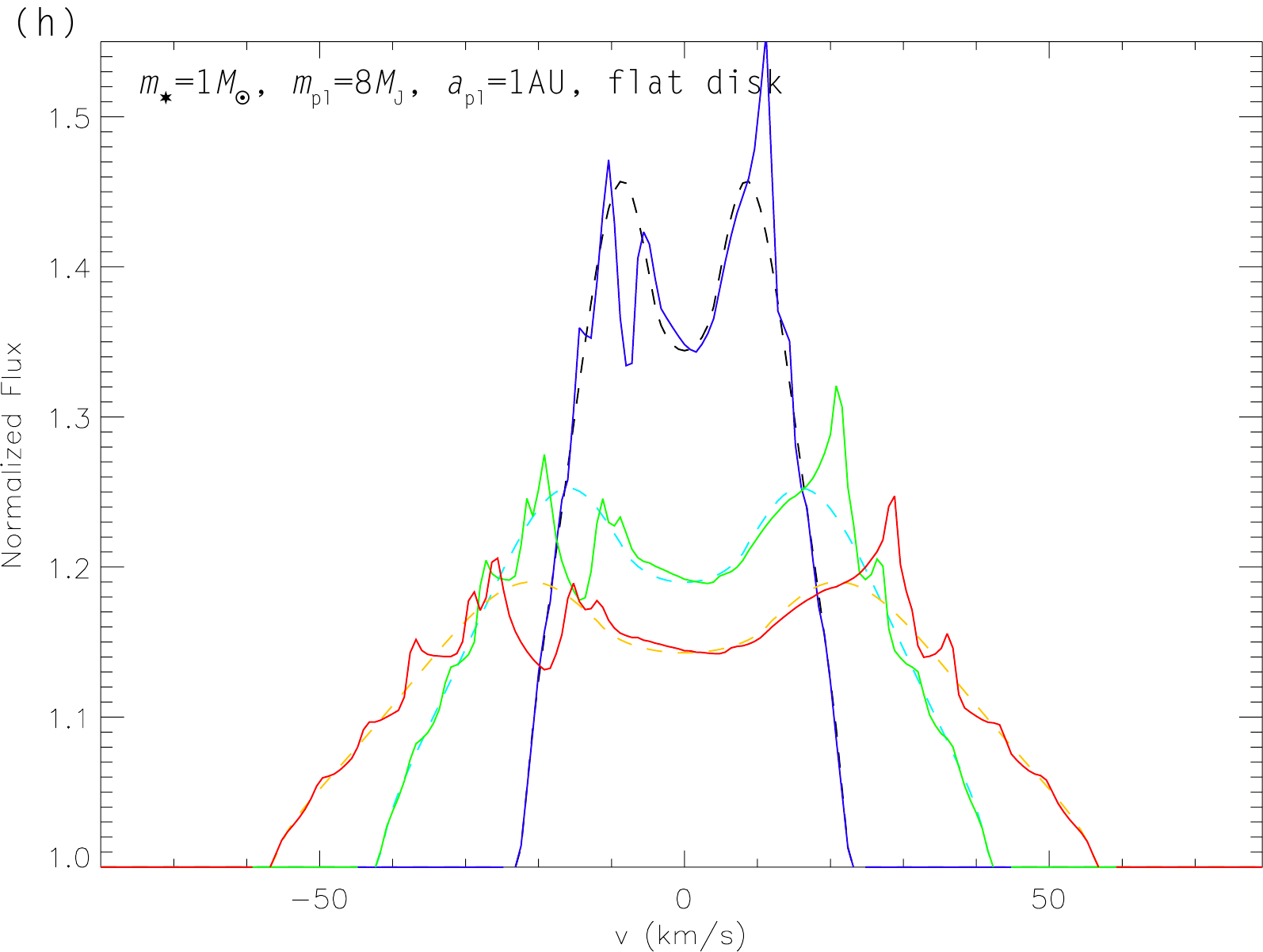}


	\caption{Distorted CO line profiles emerging in different models of planet-bearing disks. The disk inclinations were $20^{\circ}$, $40^{\circ}$  and $60^{\circ}$ shown in blue, green and red colors. For comparison, we display the symmetric double-peaked line profiles (dashed lines) emerging from a planet-free disk as well. Line profiles shown in panel (a) and (b) show the dependence of the distortion strength on the planetary mass. In panels (c) and (d) the effect of stellar mass, in panels (e) and (f) the effect of the orbital distance is presented. The effect of the inner cavity size on the strength of the line profile distortions is presented in panel (g). The impact of the disk geometry is shown in panel (h), where the line profiles emerging in model \#8 were calculated in flat-disk geometry. Note that the line profiles emerging in distinct models is calculated at orbital phases where the line profiles asymmetry is the most prominent, i.e., the orbiting planet is not always at the same position angle in each figure.}
	\label{fig:V1-0P10-profiles}
\end{figure*}

\begin{figure*}
	\centering
	
	\includegraphics[width=5.4cm,height=3.4cm]{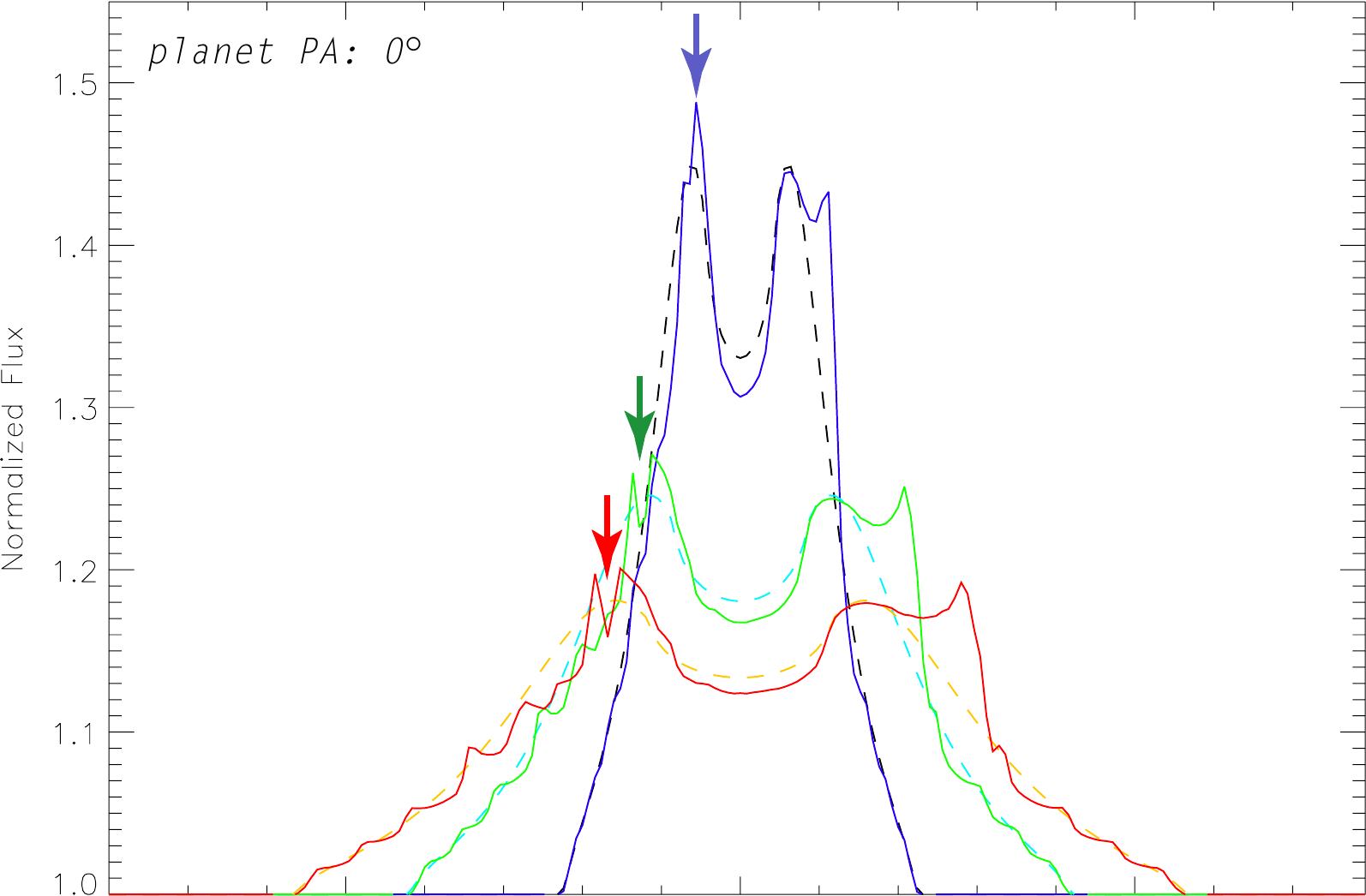}
	\includegraphics[width=5.0cm,height=3.4cm]{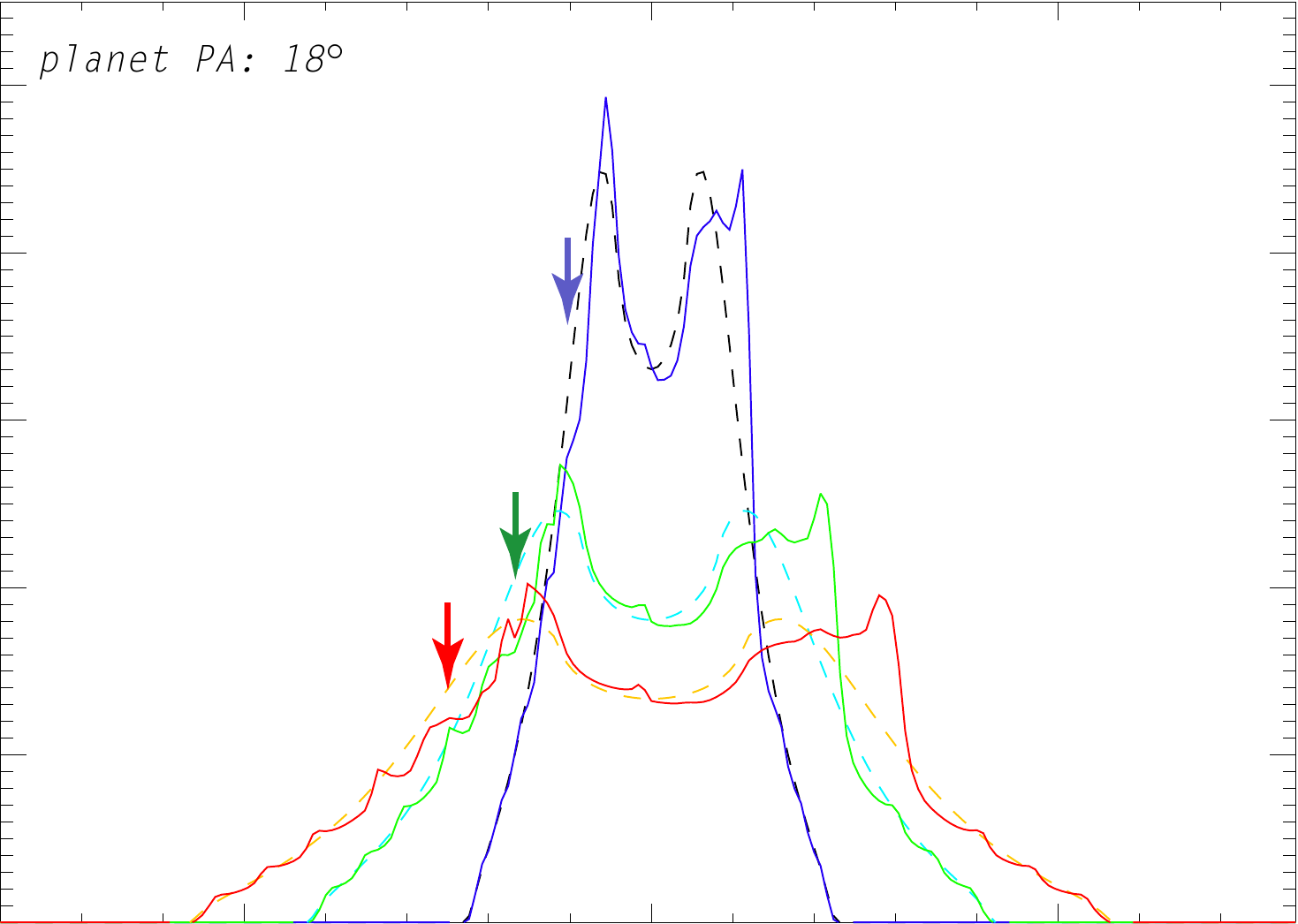}
	\includegraphics[width=5.0cm,height=3.4cm]{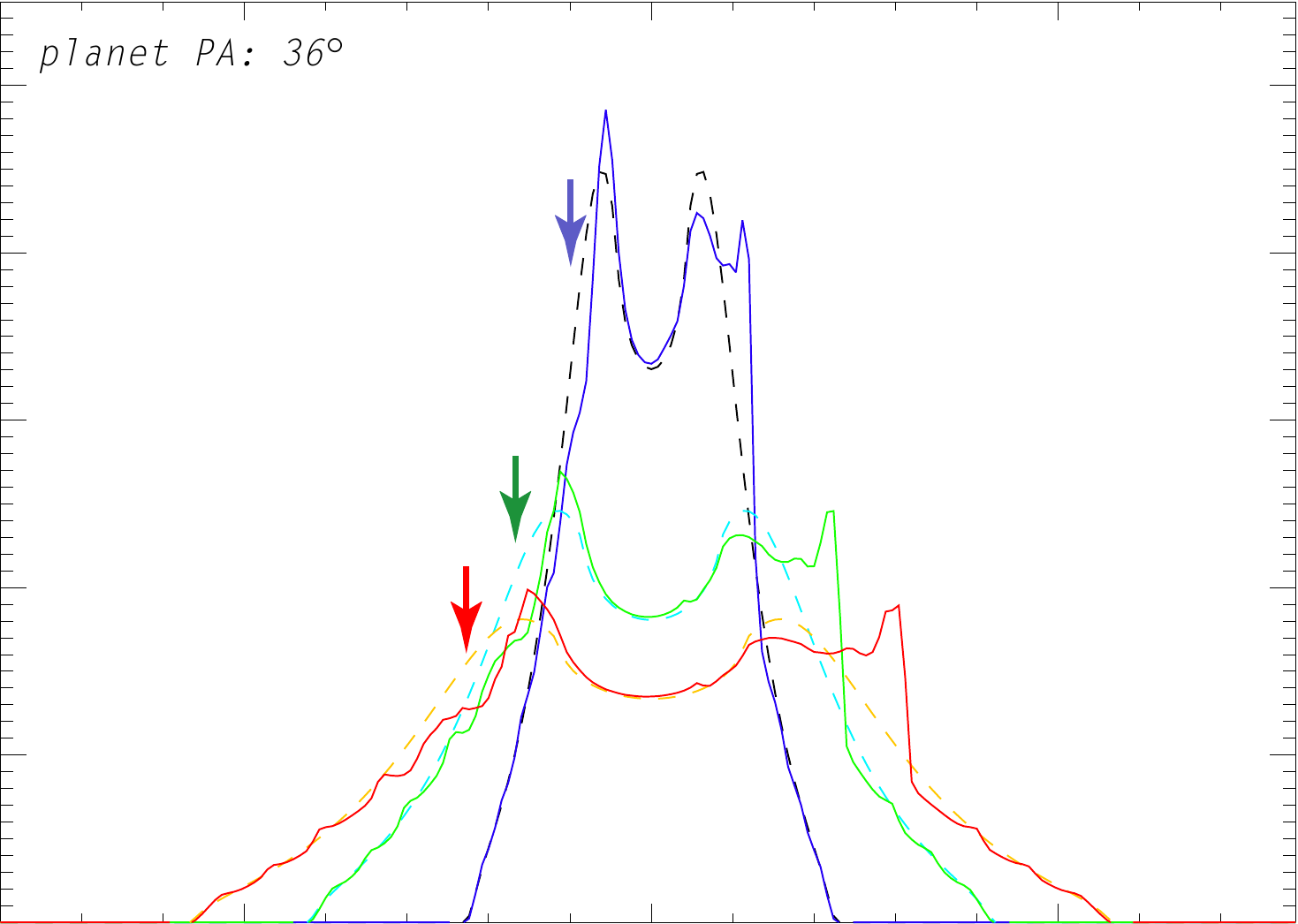}\\
	\includegraphics[width=5.4cm,height=3.4cm]{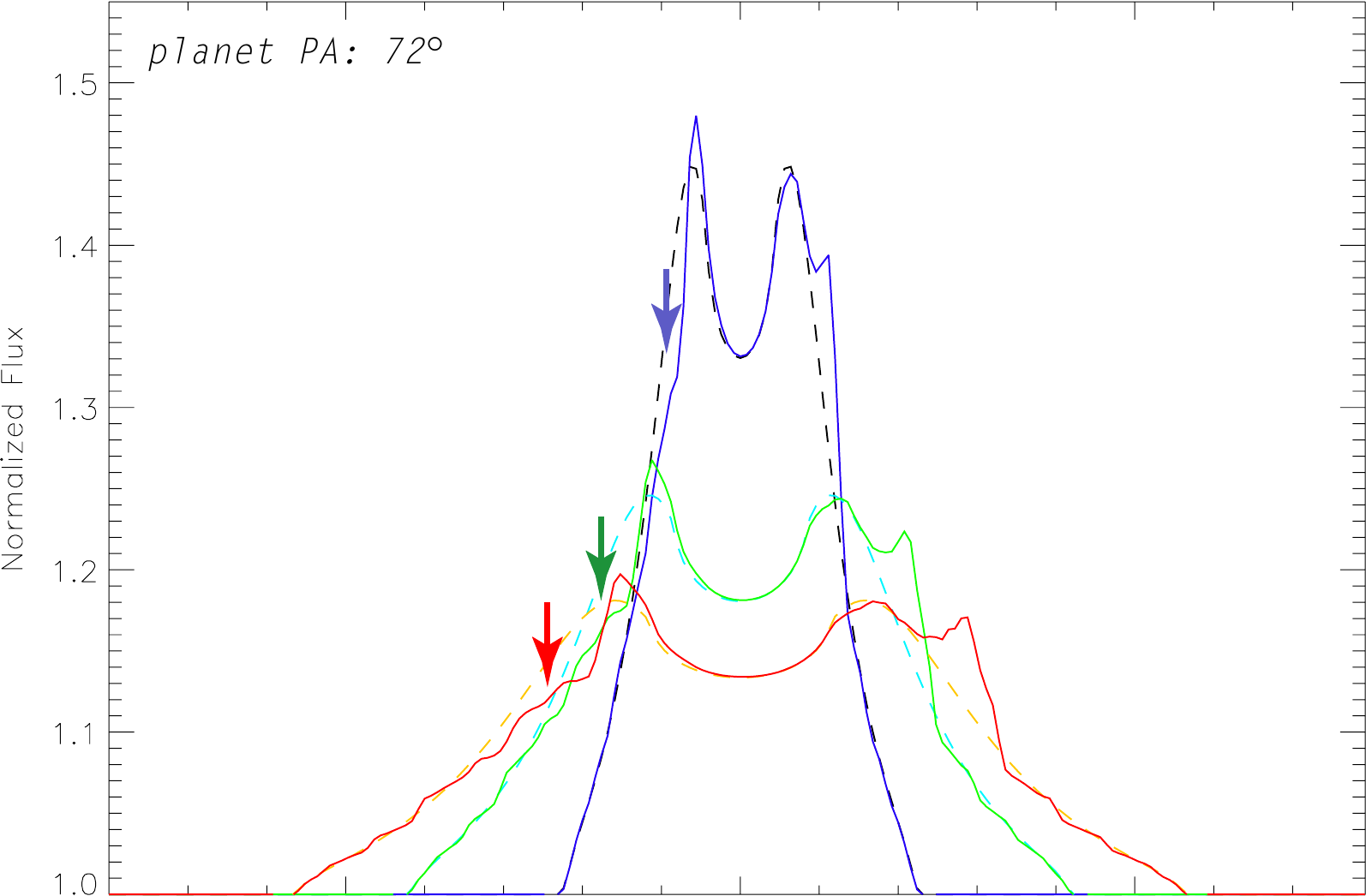}
	\includegraphics[width=5.0cm,height=3.4cm]{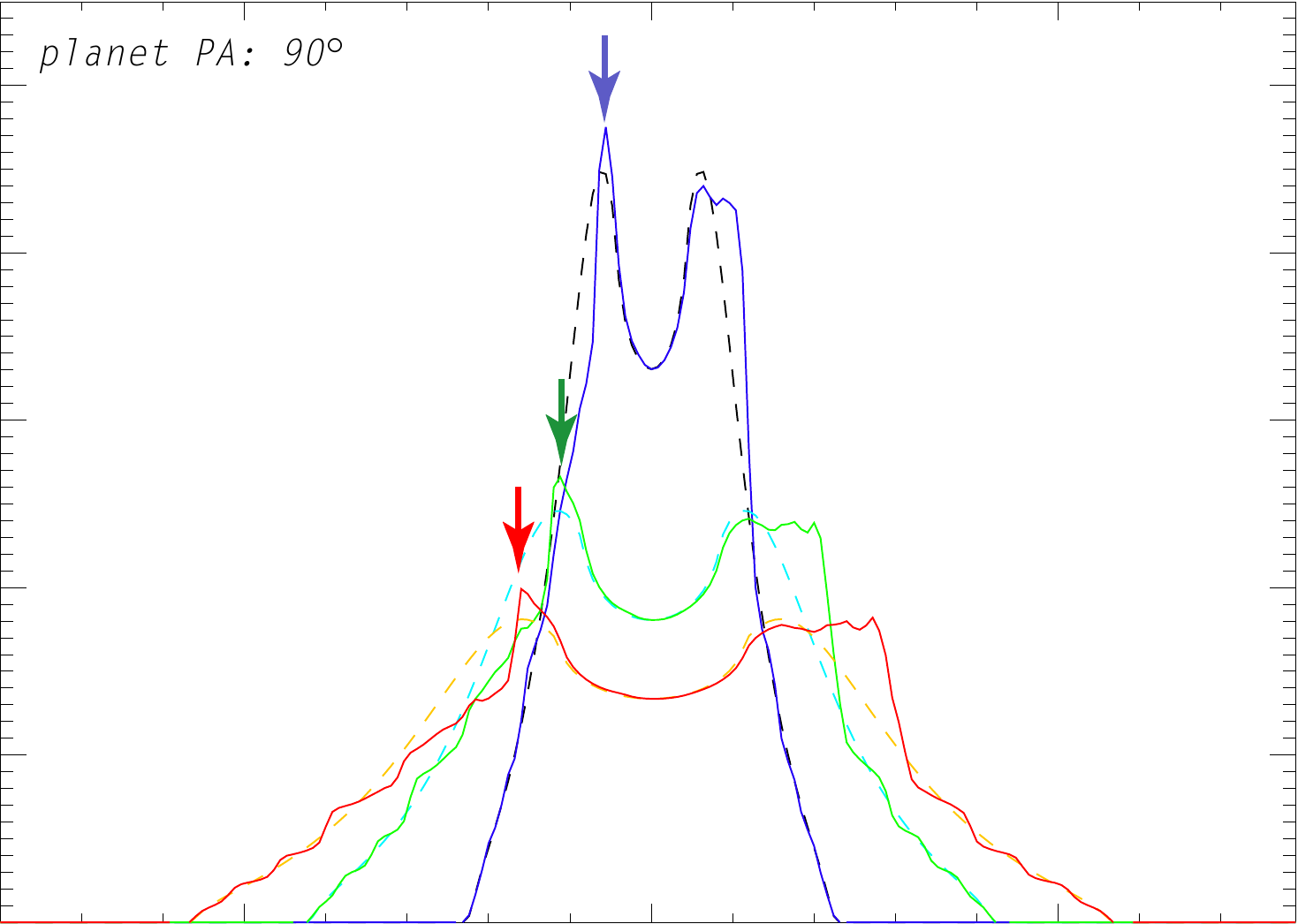}
	\includegraphics[width=5.0cm,height=3.4cm]{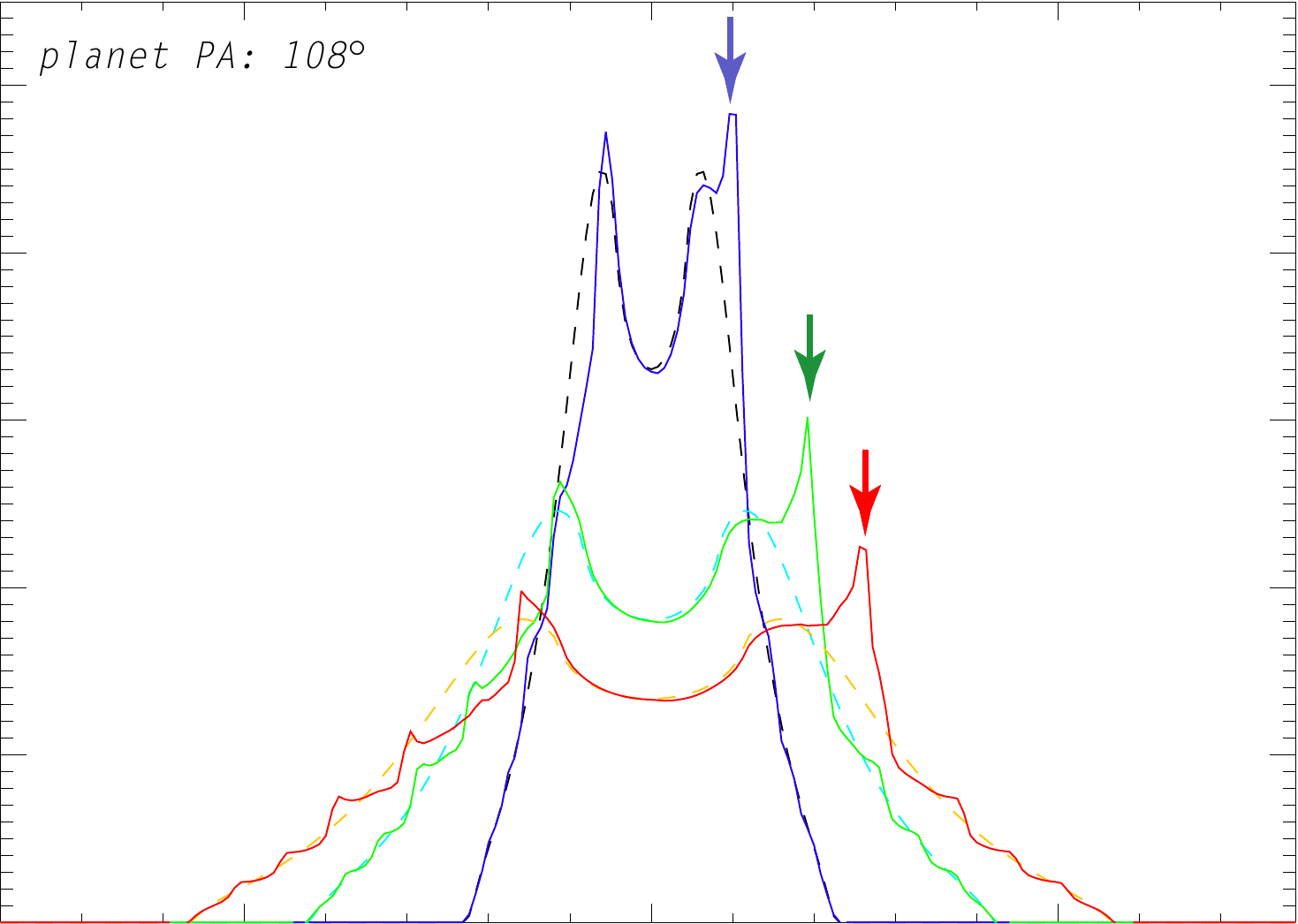}\\
	\includegraphics[width=5.4cm,height=3.4cm]{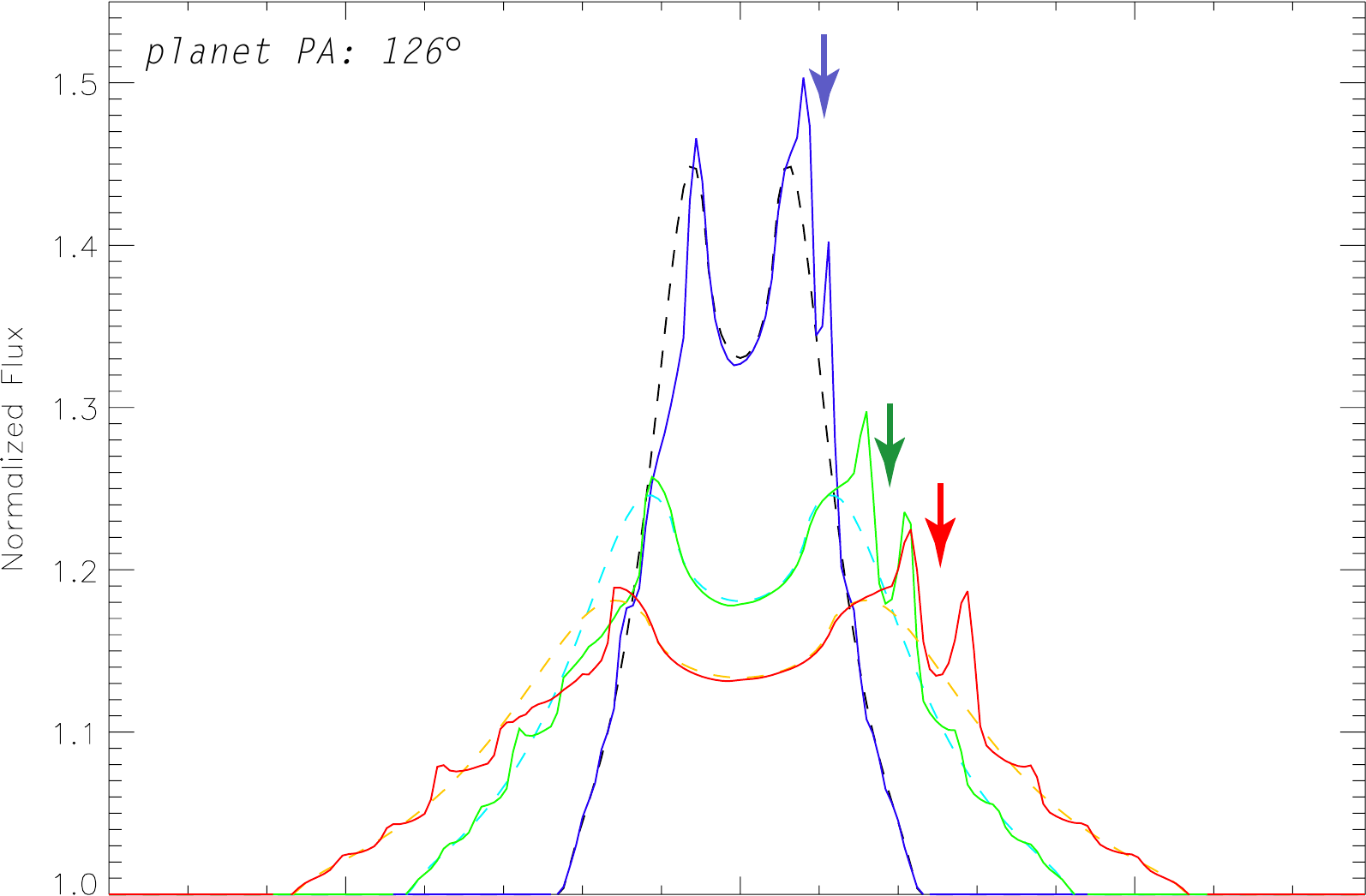}
	\includegraphics[width=5.0cm,height=3.4cm]{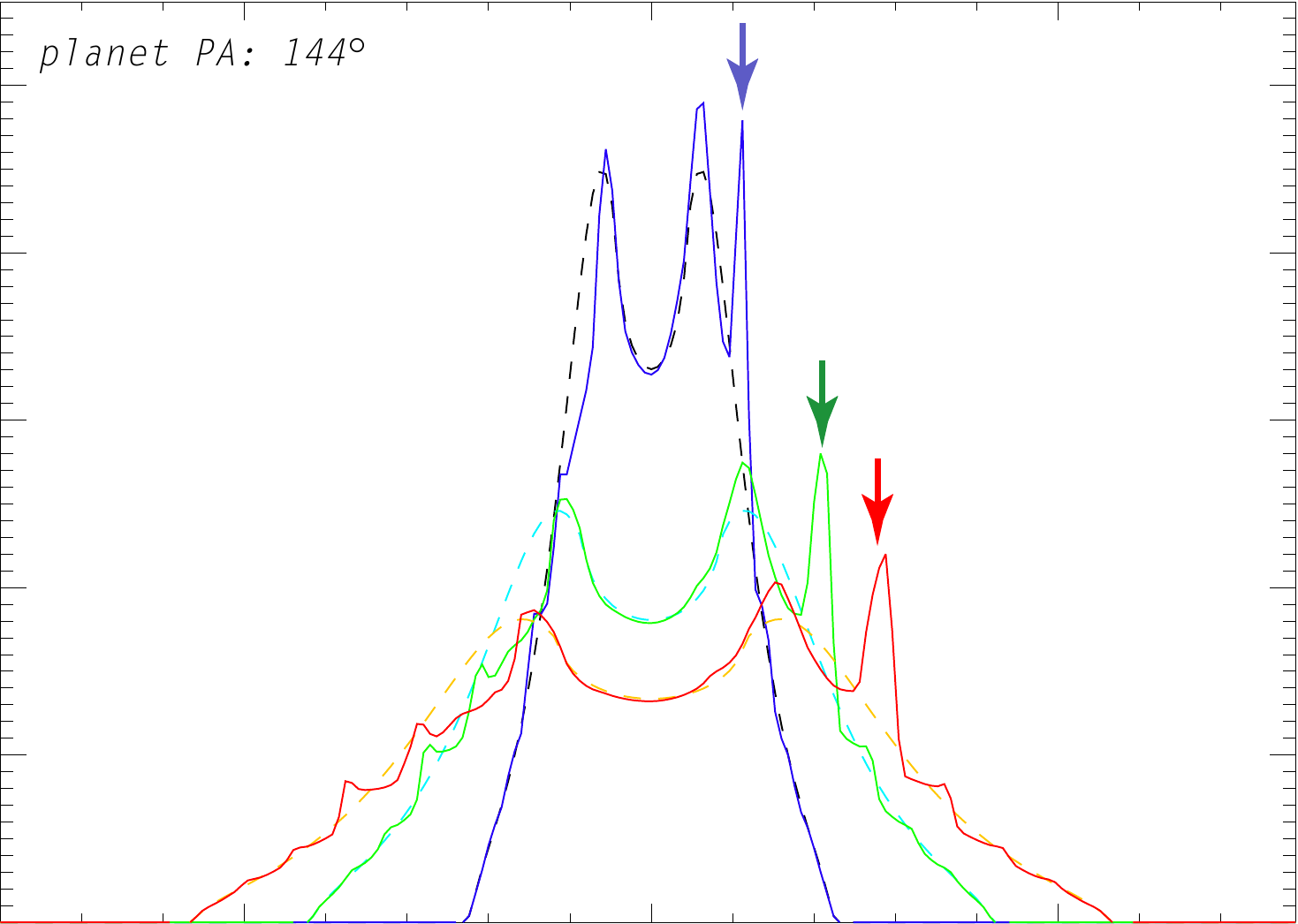}
	\includegraphics[width=5.0cm,height=3.4cm]{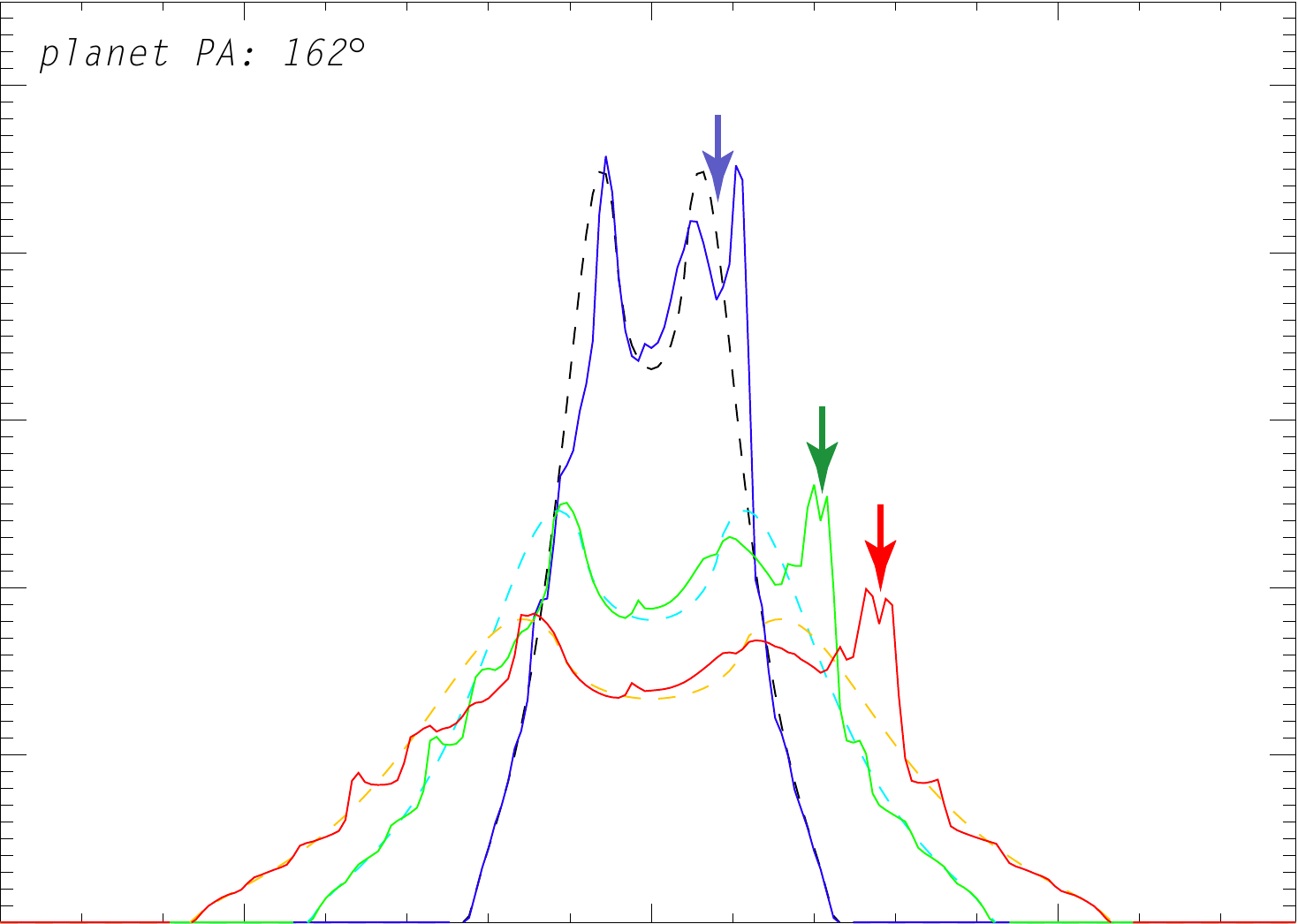}\\
	\includegraphics[width=5.4cm,height=3.4cm]{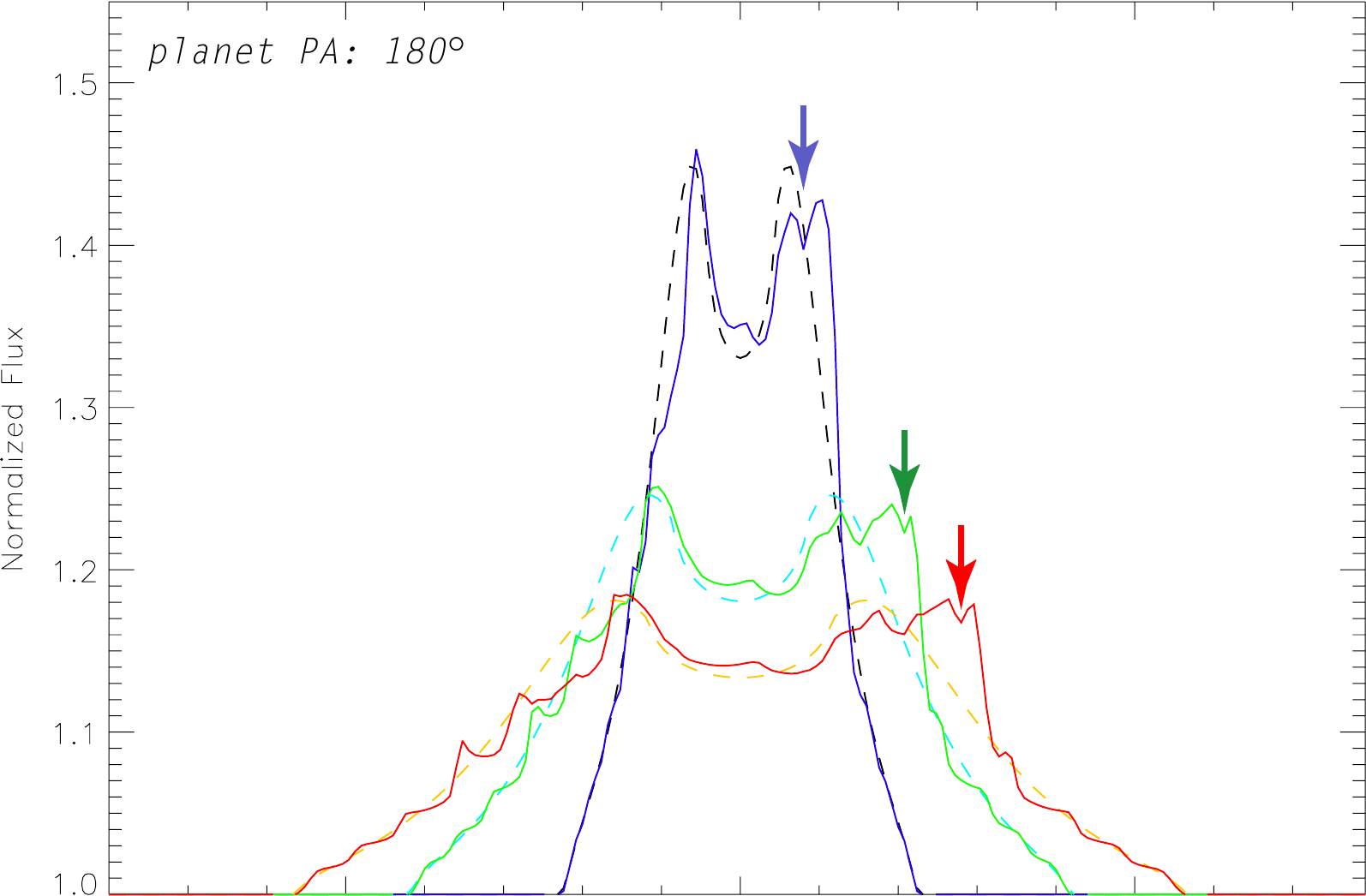}
	\includegraphics[width=5.0cm,height=3.4cm]{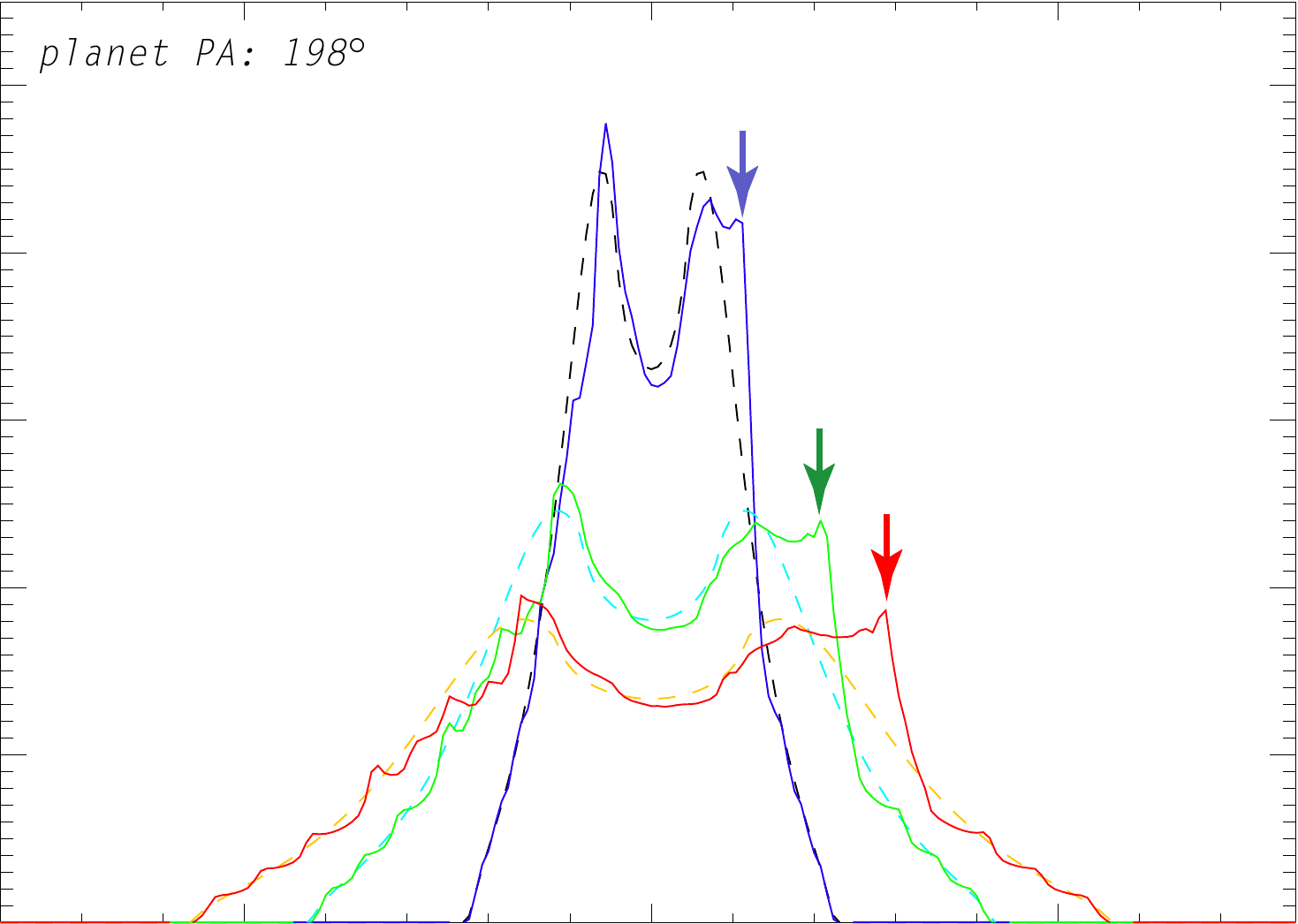}
	\includegraphics[width=5.0cm,height=3.4cm]{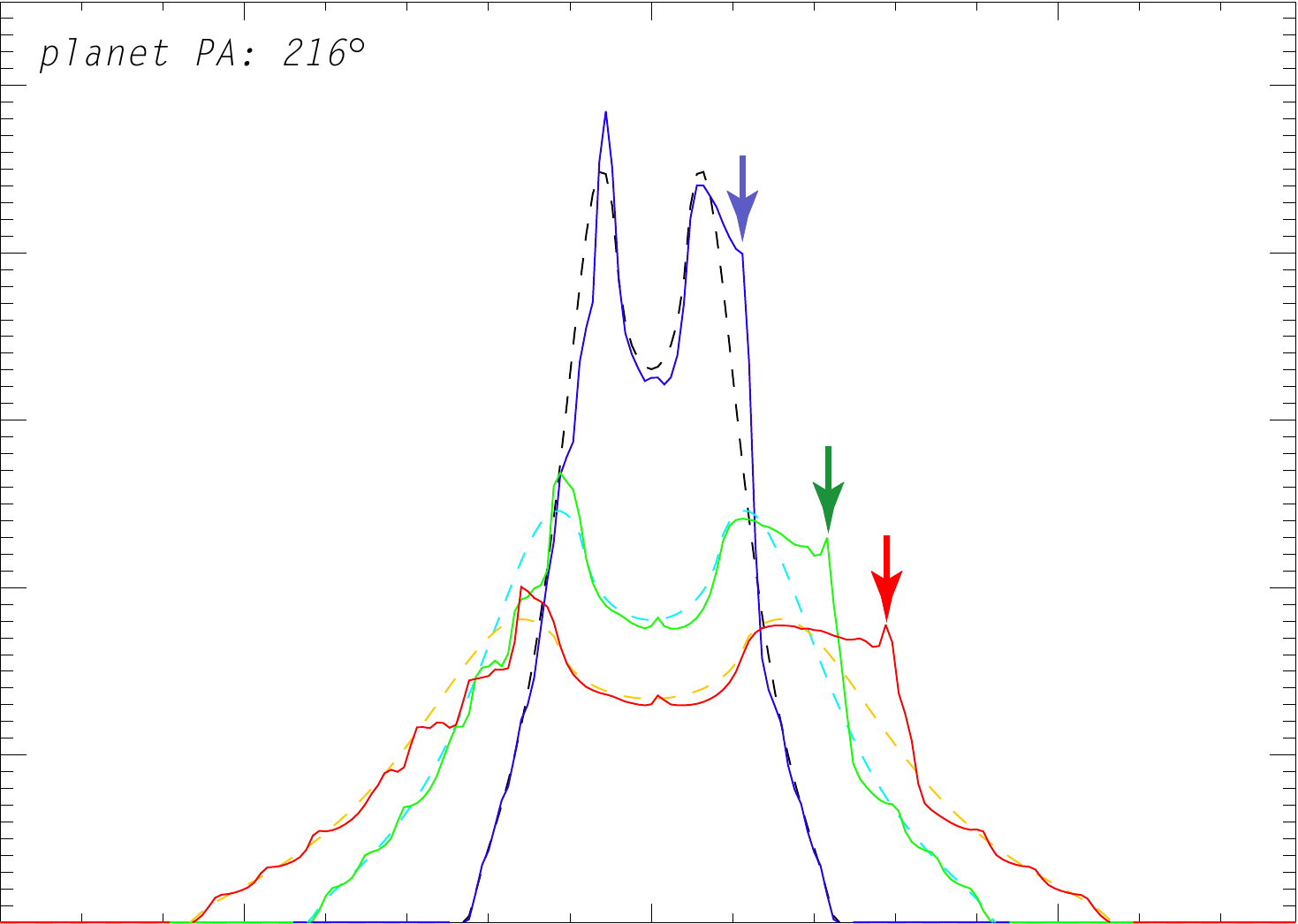}\\
	\includegraphics[width=5.4cm,height=3.4cm]{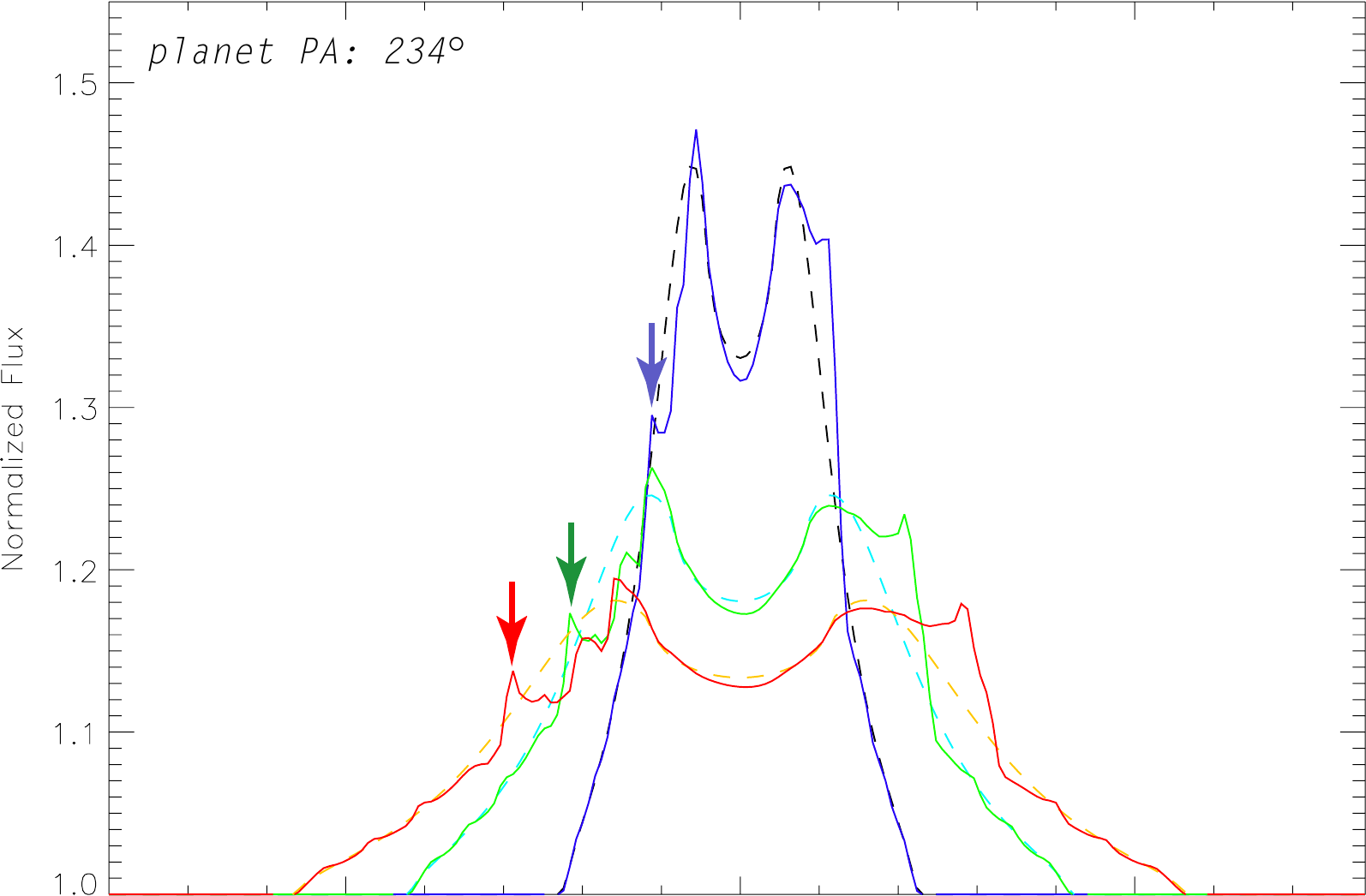}
	\includegraphics[width=5.0cm,height=3.4cm]{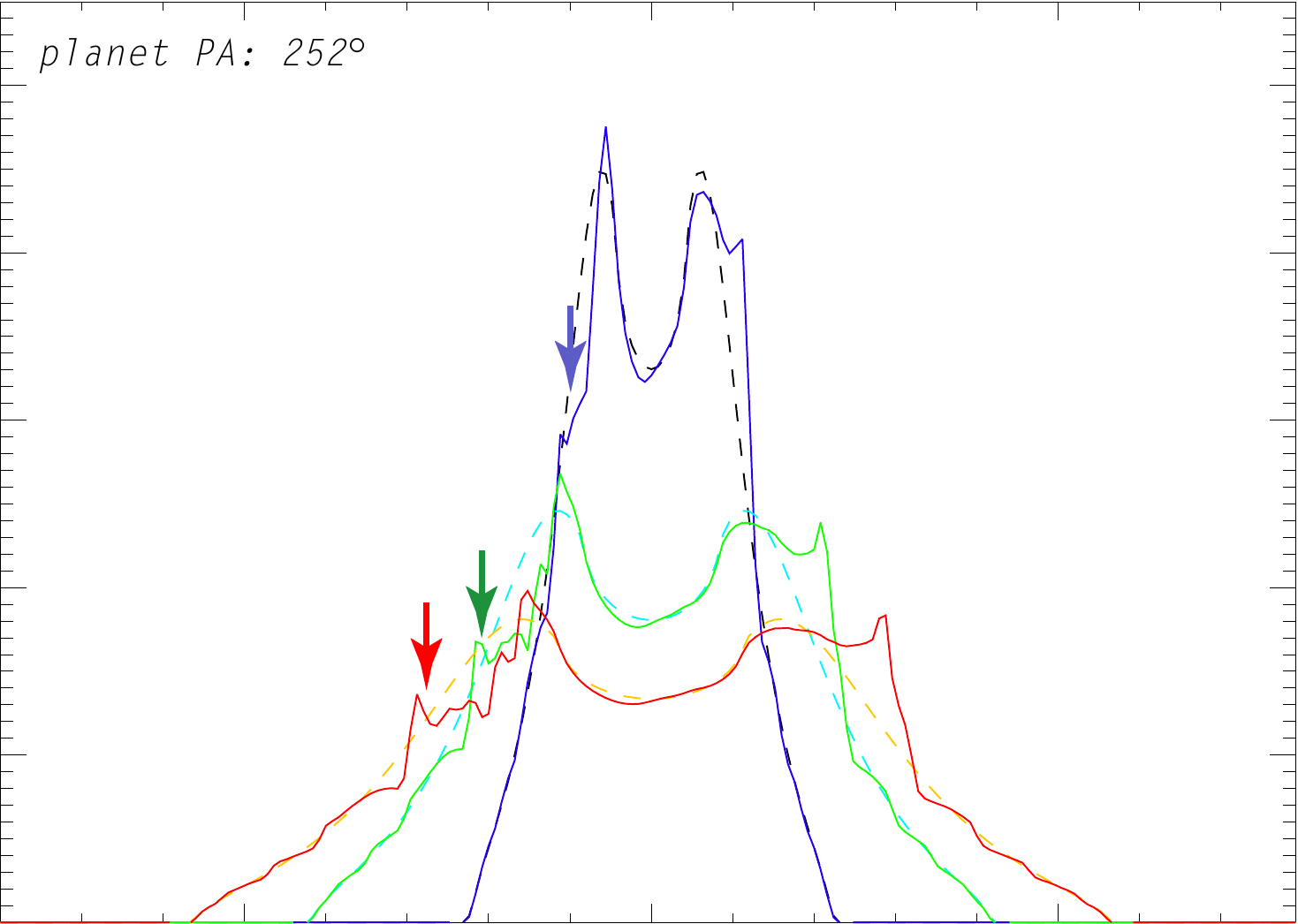}
	\includegraphics[width=5.0cm,height=3.4cm]{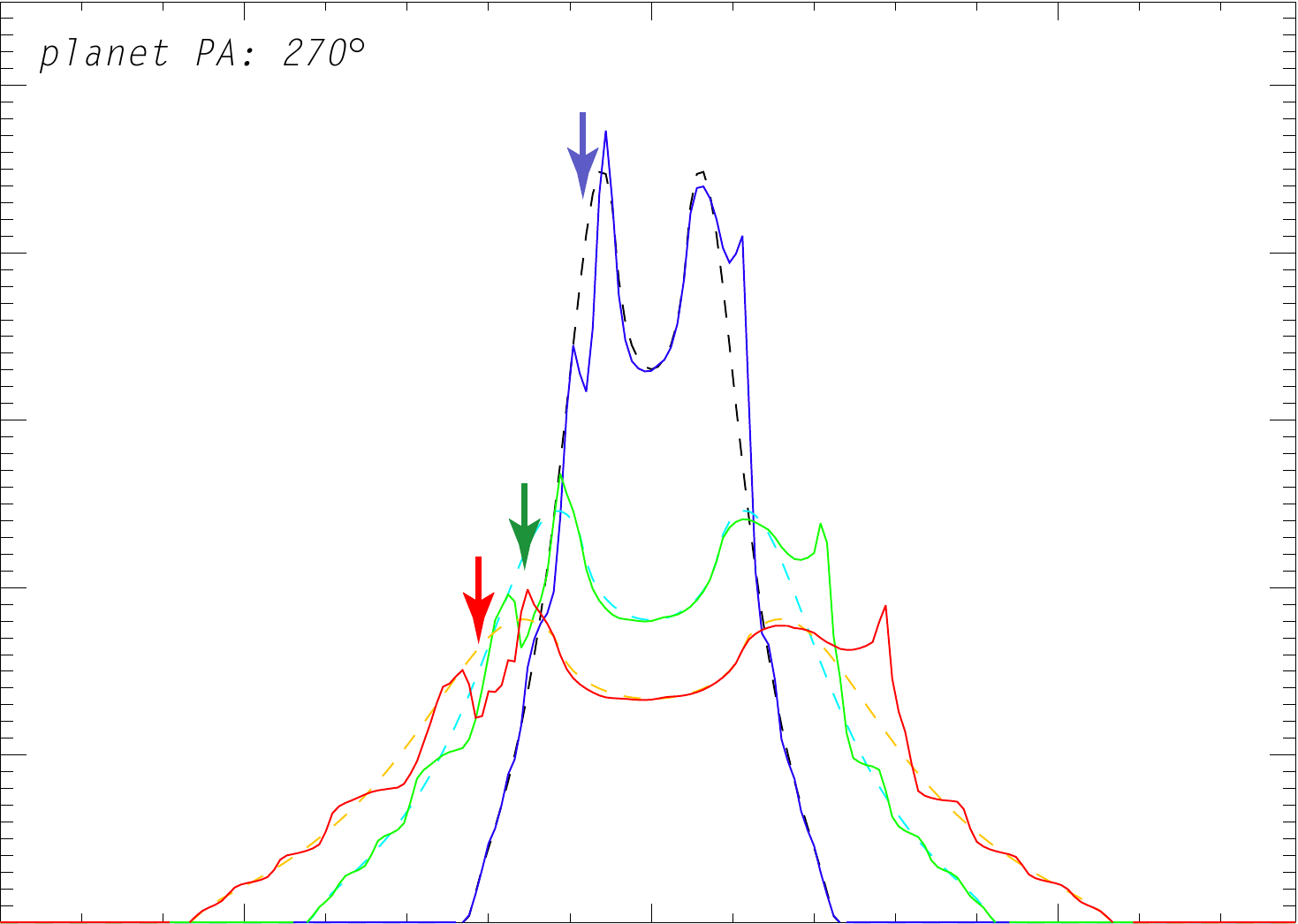}\\
	\includegraphics[width=5.4cm,height=3.4cm]{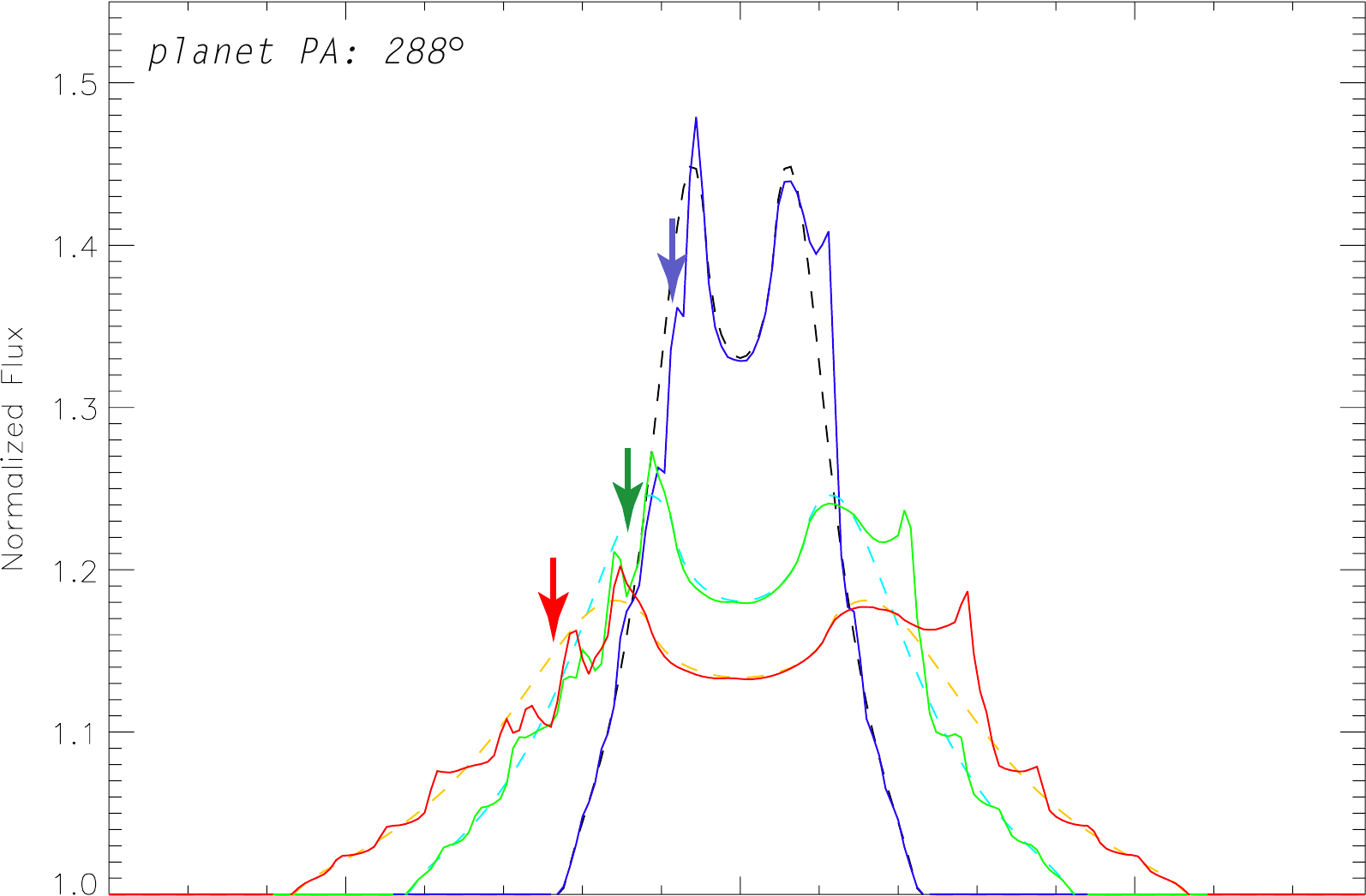}
	\includegraphics[width=5.0cm,height=3.4cm]{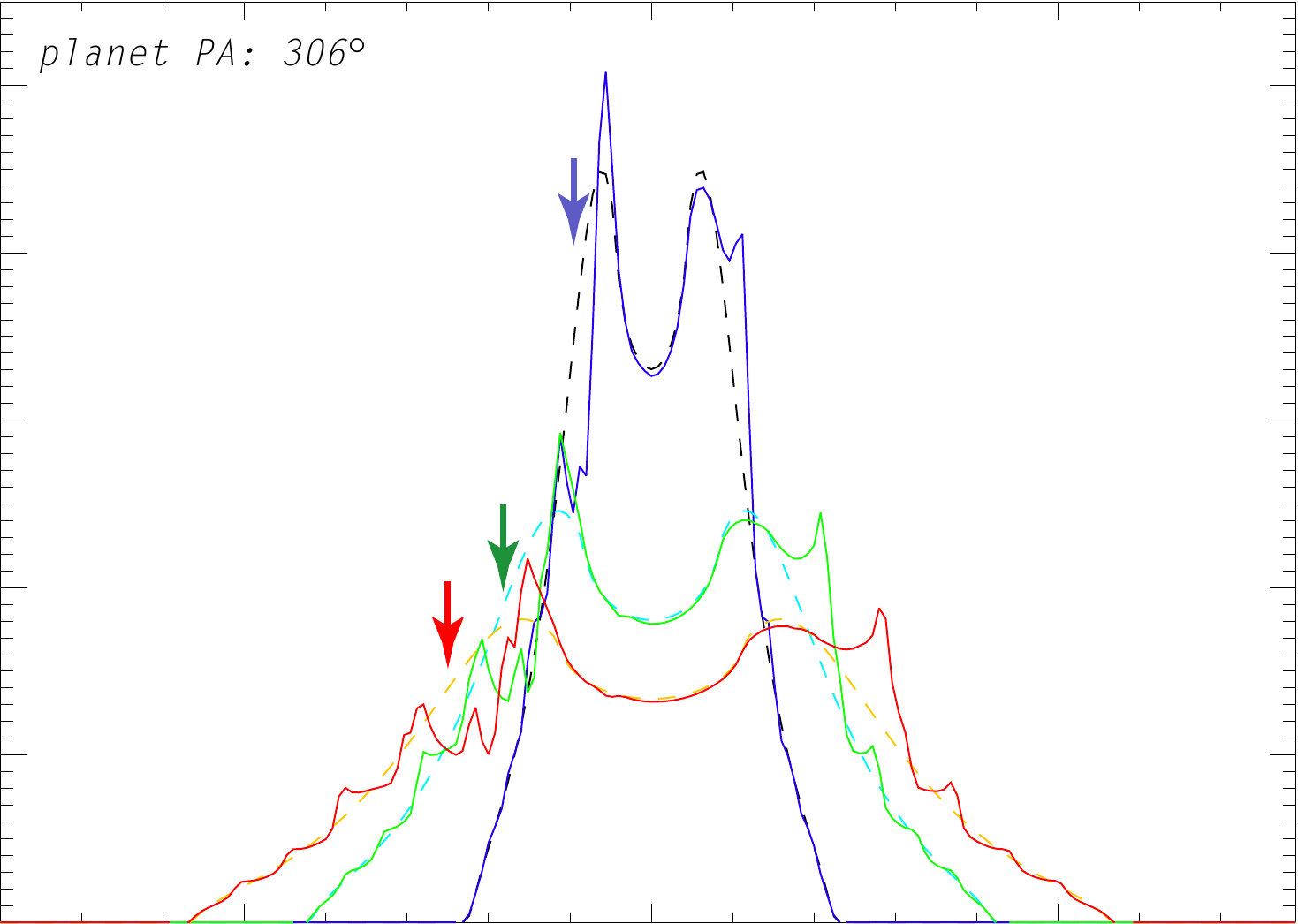}
	\includegraphics[width=5.0cm,height=3.4cm]{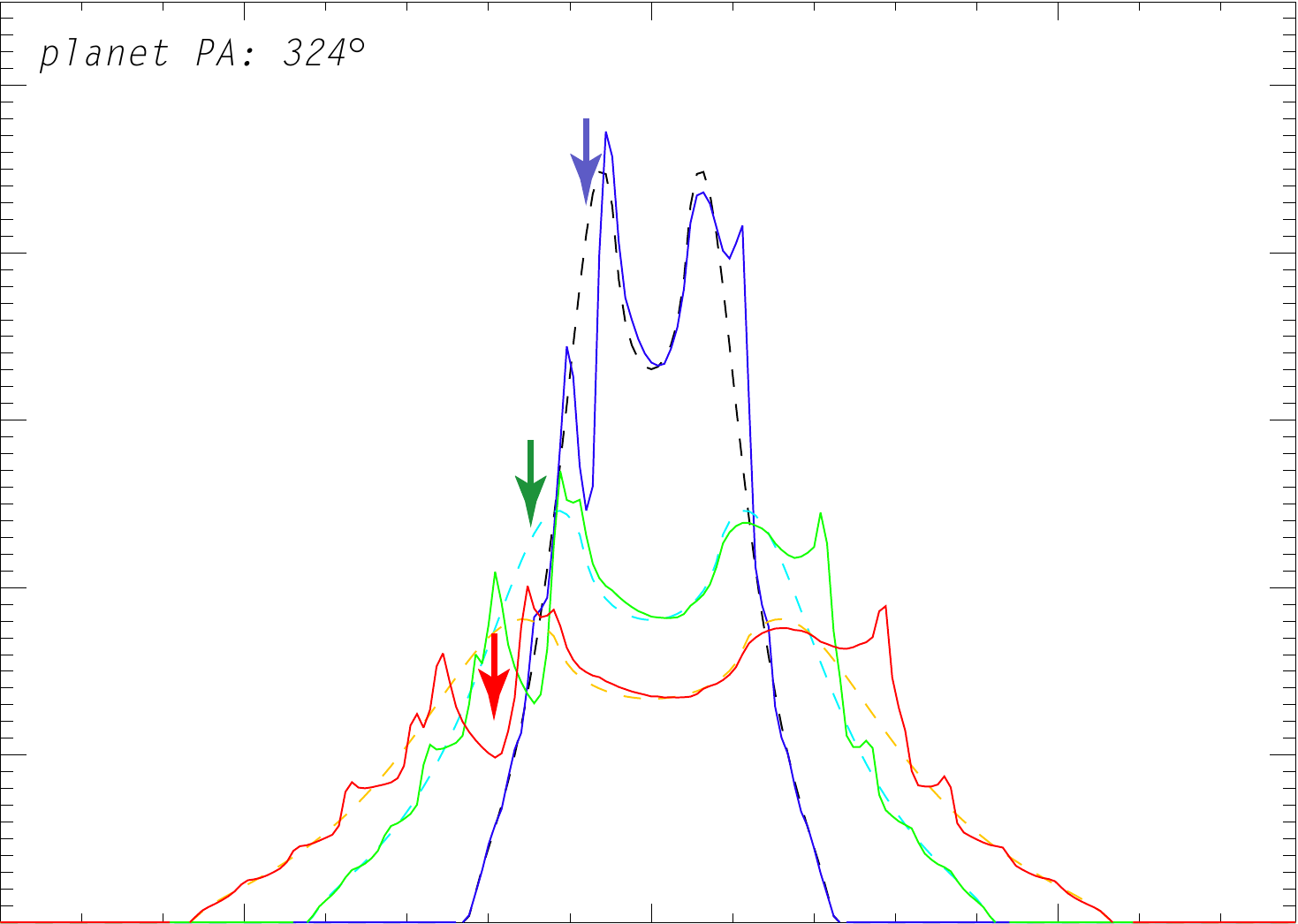}\\
	\includegraphics[width=5.4cm,height=3.8cm]{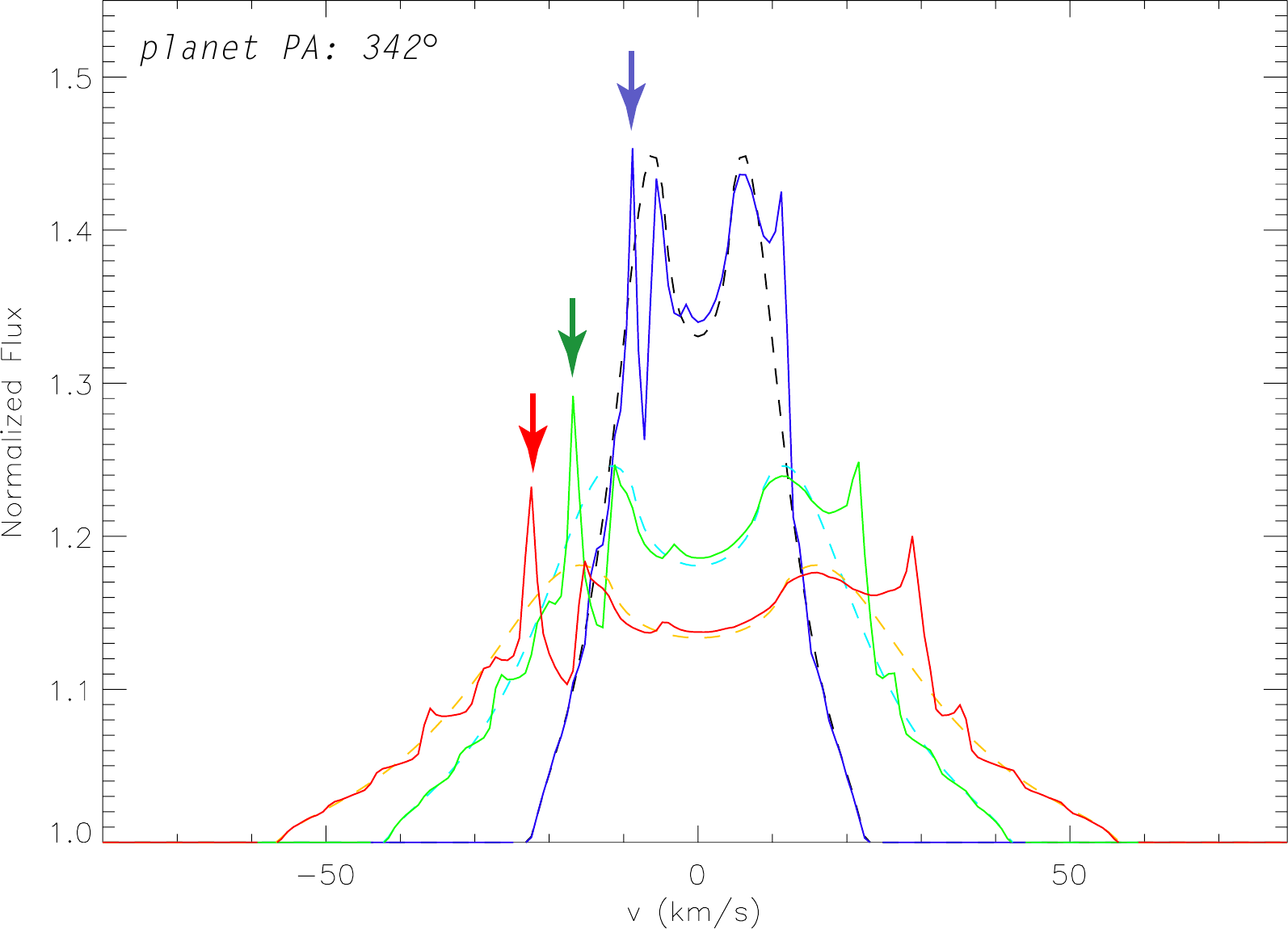}
	\includegraphics[width=5.0cm,height=3.8cm]{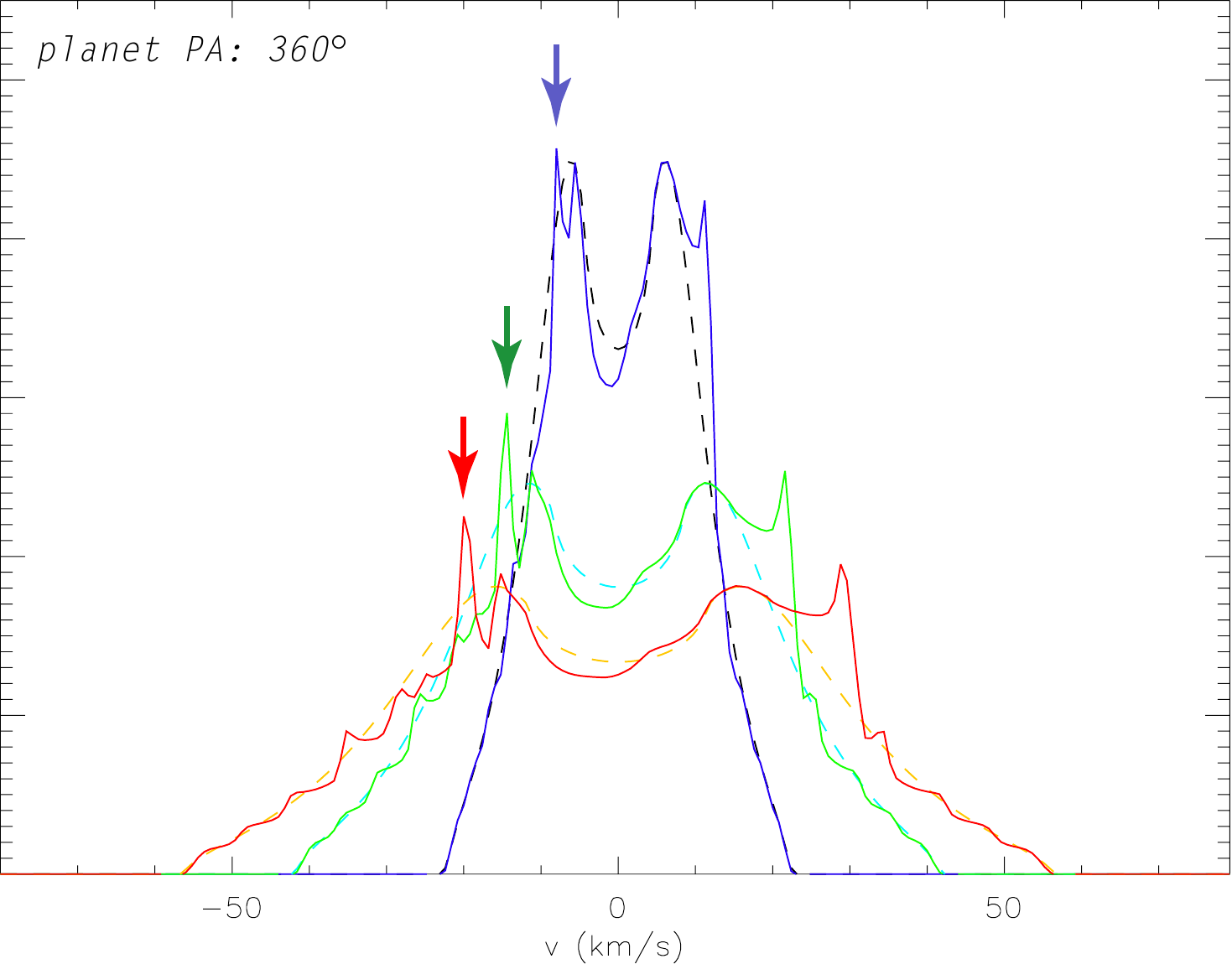}
	\includegraphics[width=5.0cm,height=3.8cm]{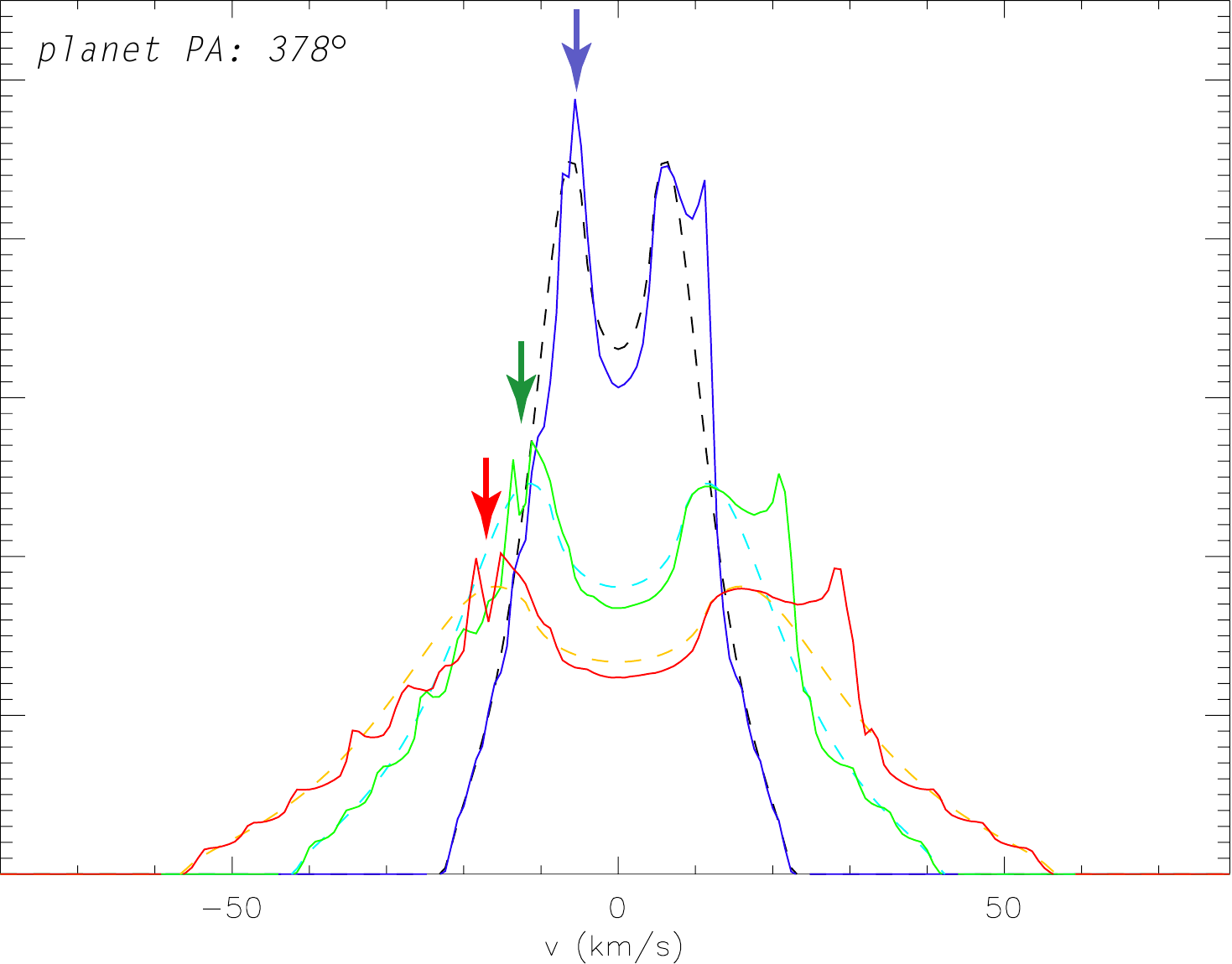}


	\caption{Permanent asymmetry and variations in the distorted CO ro-vibrational line profiles in a disk with an $8\,M_\mathrm{J}$ planet orbiting an $1\,M_{\sun}$ star at 1\,AU (model \#8). The subsequent epochs go across and down. The position angle (PA) of the giant planet is shown in each frame. The arrows point to the variable component moving in between $\sim\pm 25\,\mathrm{km/s}$. The normalized line flux is shown with the same scale as in Fig.\ref{fig:V1-0P10-profiles}(a). The snapshots were calculated during one planetary orbit, thus about $2.5$ weeks elapsed between the frames. The disk inclinations were $20^{\circ}$, $40^{\circ}$  and $60^{\circ}$, shown in blue, green and red colors. For comparison, we displayed the symmetric double-peaked line profiles (dashed lines) emerging from a planet-free disk as well. As can be seen, the line profiles have permanent asymmetry and show a significant change in their shapes within weeks.}
	\label{fig:variation-Ms1-Mb8-V1-0P10}
\end{figure*}

In this section we investigate the effect of the orbital distance of the planet to the strength of the spectral line distortion. One would initially expect that the spectral line distortions are stronger/weaker as the orbital distance of the planet decreases/increases. We recalculated the hydrodynamical simulations in ``tight'' and ``wide'' models, where the planets orbit at 0.5 and 2\,AU distances from the central star, respectively. In good agreement with the expectations, we found that the planet signature in the spectral line was strongly suppressed in the ``wide'' versions of models \#1-\#8, where $m_*=0.5\,M_{\sun}$ and $m_*=1\,M_{\sun}$. The reason for this is obvious: in wide systems the planet is orbiting in cold regions (the atmosphere temperature is varied between 200-350\,K at 2\,AU, depending on the stellar mass), where the atmospheric CO emission measured to the continuum is weak. As can be seen in Fig. \ref{fig:V1-0P10-profiles}(e), the ``wide'' version of model \#8 ($m_\mathrm{pl}=8\,M_\mathrm{J}$, $m_*=1\,M_{\sun}$, 2\,AU orbital distance) shows considerably weaker planet signatures in the CO line profile than the original 1\,AU model, see Fig. \ref{fig:V1-0P10-profiles}(a) for comparison. The distortion patterns appear at lower velocities owing to lower orbital velocities in $0.5\,M_{\odot}$ models. On the other hand, we found about the same strength of the planet signal in the ``wide'' version of model \#12 ($m_\mathrm{pl}=12\,M_\mathrm{J}$, $m_*=1.5\,M_{\sun}$, 2\,AU orbital distance) as in model \#8. This is expected, since in the latter case the luminosity of the host star is high enough to produce sufficient CO excitation even at 2\,AU.

\begin{figure}
	\centering
	\includegraphics[width=\columnwidth]{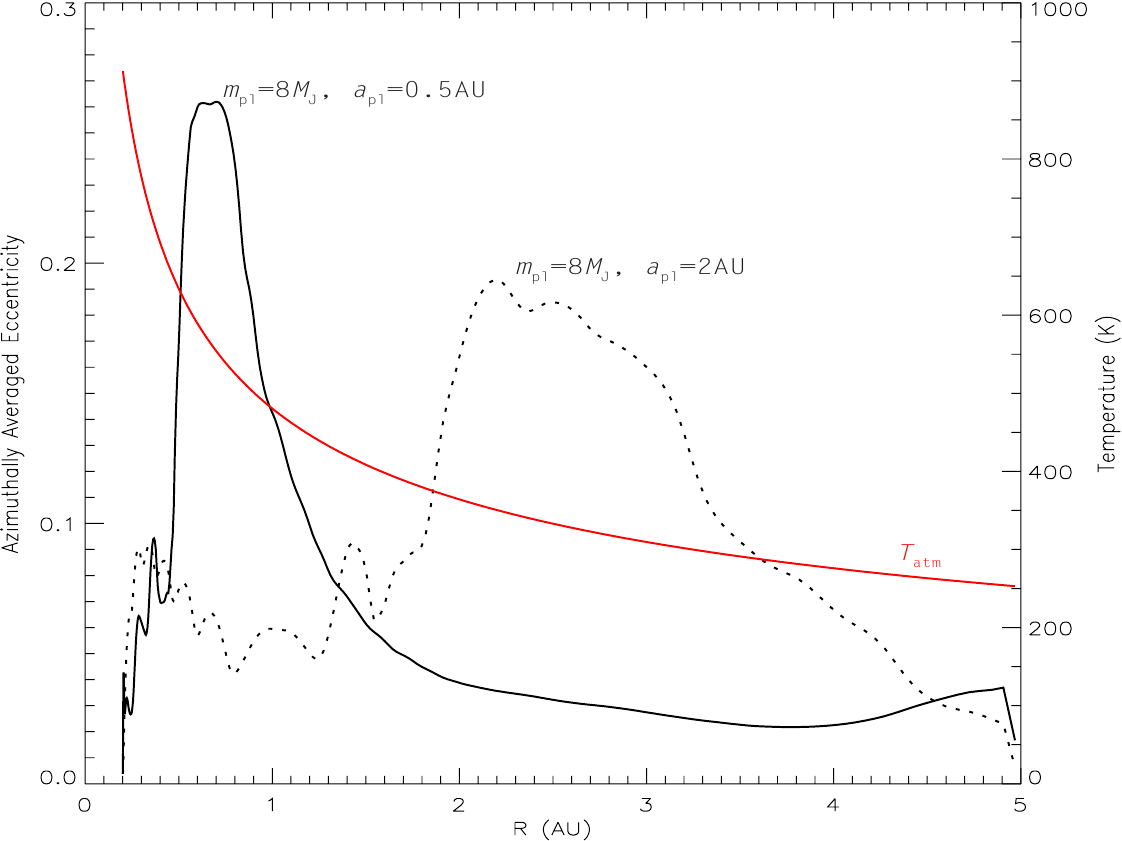}
	\caption{Azimuthally averaged eccentricities in the ``wide'' (dotted black) and ``tight'' (solid black) version of model \#8 ($m_\mathrm{pl}=8\,M_\mathrm{J}$, $m_*=1\,M_{\sun}$) and the atmospheric temperature (red). It is apparent that even though the eccentricity maximum is smaller in the ``wide'' system, the disk is more eccentric on average than in the ``tight'' model. At the same time the planet signal becomes stronger in the ``tight'' (see Fig. \ref{fig:V1-0P10-profiles}(f)) than in the ``wide'' (see Fig. \ref{fig:V1-0P10-profiles}(e)) model, because the disk atmosphere is considerably hotter at 0.5\,AU than 2\,AU, where the disk is dynamically perturbed by the planet.}
	\label{fig:ecc-temp}
\end{figure}

The resulting line profile calculated in the ``tight'' version of the model \#8 ($m_\mathrm{pl}=8\,M_\mathrm{J}$, $m_*=1\,M_{\sun}$, 0.5\,AU orbital distance) is shown in Fig. \ref{fig:V1-0P10-profiles}(f). Because the giant planet orbits closer to the host star, where the disk atmosphere is heated to higher temperatures (the atmosphere temperature varies between 450-700\,K at 0.5\,AU, depending on the stellar mass), the emission emerging from the perturbed regions is stronger than in the original 1\,AU models. It is also evident that the distorted peaks are slightly shifted to higher velocities compared to the original 1\,AU models, because the orbital velocity of the gas parcels in dynamically perturbed orbits is higher in ``tight'' models. The disk is more eccentric on average in the latter case (Fig. \ref{fig:ecc-temp}), but the strength of the line profile distortion is increasing as the planetary orbital distance decreases.

In order to demonstrate that smaller mass planets can also be detected in closer orbits we have calculated the line profile in a ``tight'' version of model \#2, $m_{\mathrm{pl}}=1.5\,M_\mathrm{J}$, $m_*=0.5\,M_{\sun}$, 0.5\,AU orbital distance. We find that one could observe roughly the same strength line profile distortions as in model \#8 ($m_\mathrm{pl}=8\,M_\mathrm{J}$, $m_*=1\,M_{\sun}$, 1\,AU orbital distance).

\subsection{Inner cavity size}

Now we turn our attention to the question under which circumstances planets can  be detected orbiting farther away from the star (farther than 2\,AU for instance). In Sect. 3.2 we argued that there is a region around the star called inner cavity from which the CO is already depleted. In our simulations presented so far, we assumed that the radius of the inner cavity is fixed at 0.2\,AU irrespective of the stellar luminosity. On the other hand, the size of the inner cavity is presumably dependent on the stellar mass, and indeed this is confirmed by observations \citep{Akesonetal2005}. As we mentioned in Sect. 3.2, the size of the inner cavity has a significant influence on the overall line-to-continuum ratio. In disks dynamically perturbed by planets orbiting at 1\,AU and with a significant amount of CO lying inside 0.2\,AU, the majority of CO flux is emerging from the unperturbed innermost regions, which eventually could smear out the planet signatures. On the contrary, if the size of the inner cavity is larger, the contribution of the perturbed regions to the total CO flux becomes stronger. In Fig. \ref{fig:V1-0P10-profiles}(g) we present the line profiles obtained in model \#8, in which the inner cavity is increased to 0.4\,AU in size, while the $8\,M_\mathrm{J}$ mass planet was still orbiting at 1\,AU. As can be seen, the overall line-to-continuum ratio is decreased due to the absence of hot gas, but the level of the asymmetric pattern profile is more apparent than in models with an  inner cavity extending only to 0.2\,AU, see Fig. \ref{fig:V1-0P10-profiles}(a) for comparison. For larger inner cavities the continuum level of the disk interior and inner rim are decreased too, resulting in a moderate weakening of the line-to-continuum ratio and slight strengthening of planet signal. In this case the detection possibility of a planet orbiting at larger distances ($>2\,\mathrm{AU}$) is growing.

\subsection{Disk geometry}

It is well known that many T\,Tauri sources present flatter than $\lambda F_\lambda \sim \lambda^{-4/3}$ SEDs. One attempt to explain this SED flattening was to assume that the disk is flaring \citep{KenyonHartmann1987}, resulting in a geometrically thicker disk atmosphere $H(R)\sim R^{\gamma+1}$, where $\gamma$ is the flaring index. In the approach of \citet{ChiangGoldreich1997} the flaring index is $\gamma=2/7$. In order to investigate the effect of disk geometry on the strength of the line profile distortions we recalculated several models using flat thermal disk approximation (the hydrodynamical inputs were the same). An example of the flat version of model \#8 ($m_\mathrm{pl}=8\,M_\mathrm{J}$, $m_{*}=1\,M_{\sun}$, 1\,AU orbital distance) is shown in Fig. \ref{fig:V1-0P10-profiles}(h). Note that in order to have the same peak line-to-continuum ratio in flat as in flared models, the amount of emitting CO is arbitrary increased by decreasing the dust-to-gas ratio to $1.2\times 10^{-3}$ and keeping the disk mass unchanged. The reason for this is that although the atmospheric temperature does not differ in flared and flat disk models (see the atmospheric temperature given by Eq. (\ref{eq:temp-surf})), the atmospheric surface density of CO is substantially smaller in flat than flared disk models (see the dependence of surface density given by Eq. (\ref{eq:dens-surf}) on the grazing angle, given by the first term of Eq. (\ref{eq:grazing-angle-flaring}) in flat disk models). It is evident that the total line-to-continuum ratio is higher in the flat than in the flared disk model, but the line width at the foot of the line profile does not change. According to our calculations the line profile distortions in flat models compared to the flared one are less significant in ``wide'' systems where planets orbit at larger distances. This can be explained by the effect of flaring getting stronger with increasing distance to the host star.

\subsection{Observational considerations}

\begin{figure}
	\centering
	\includegraphics[width=\columnwidth]{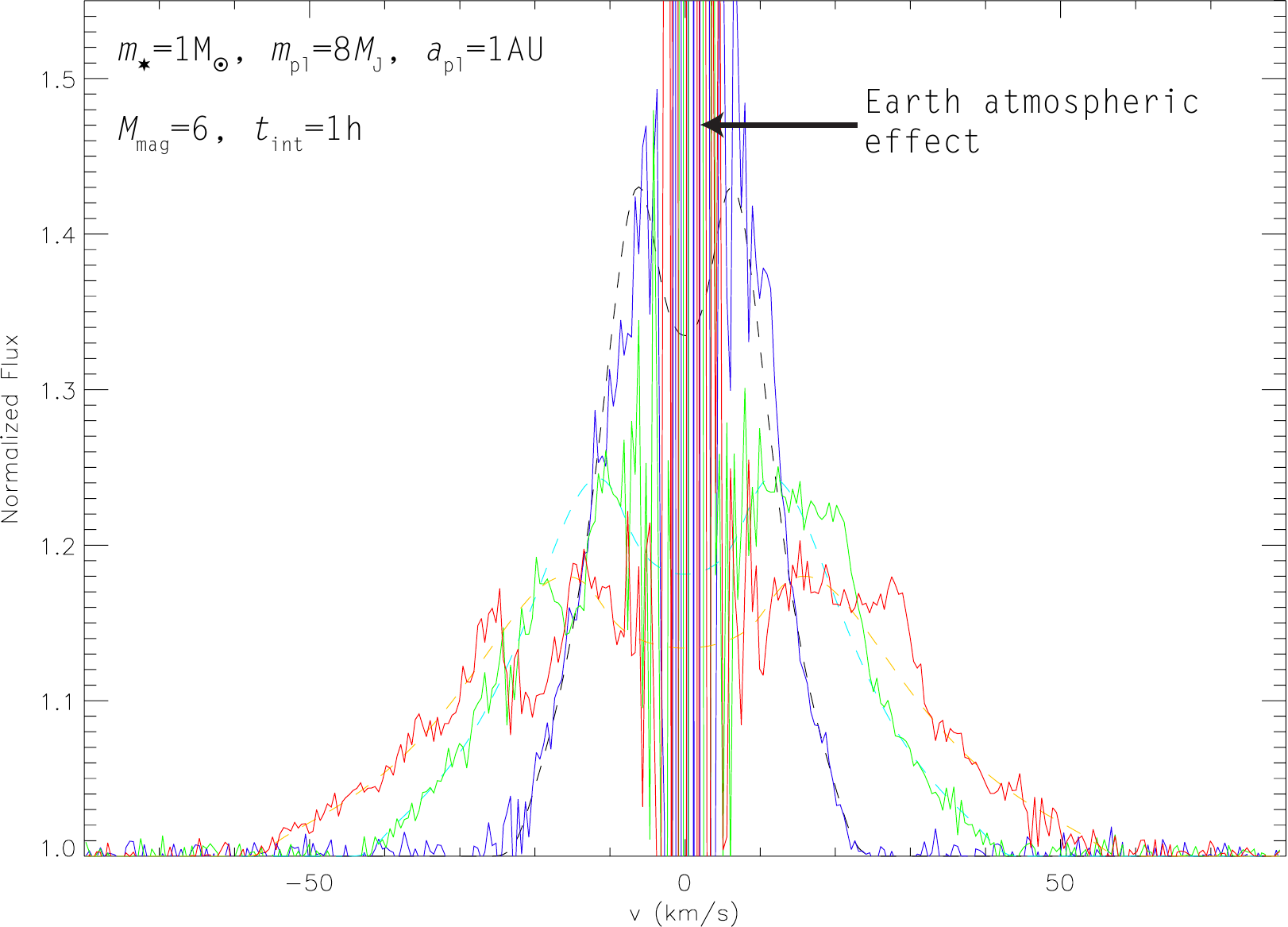}\\
	\includegraphics[width=\columnwidth]{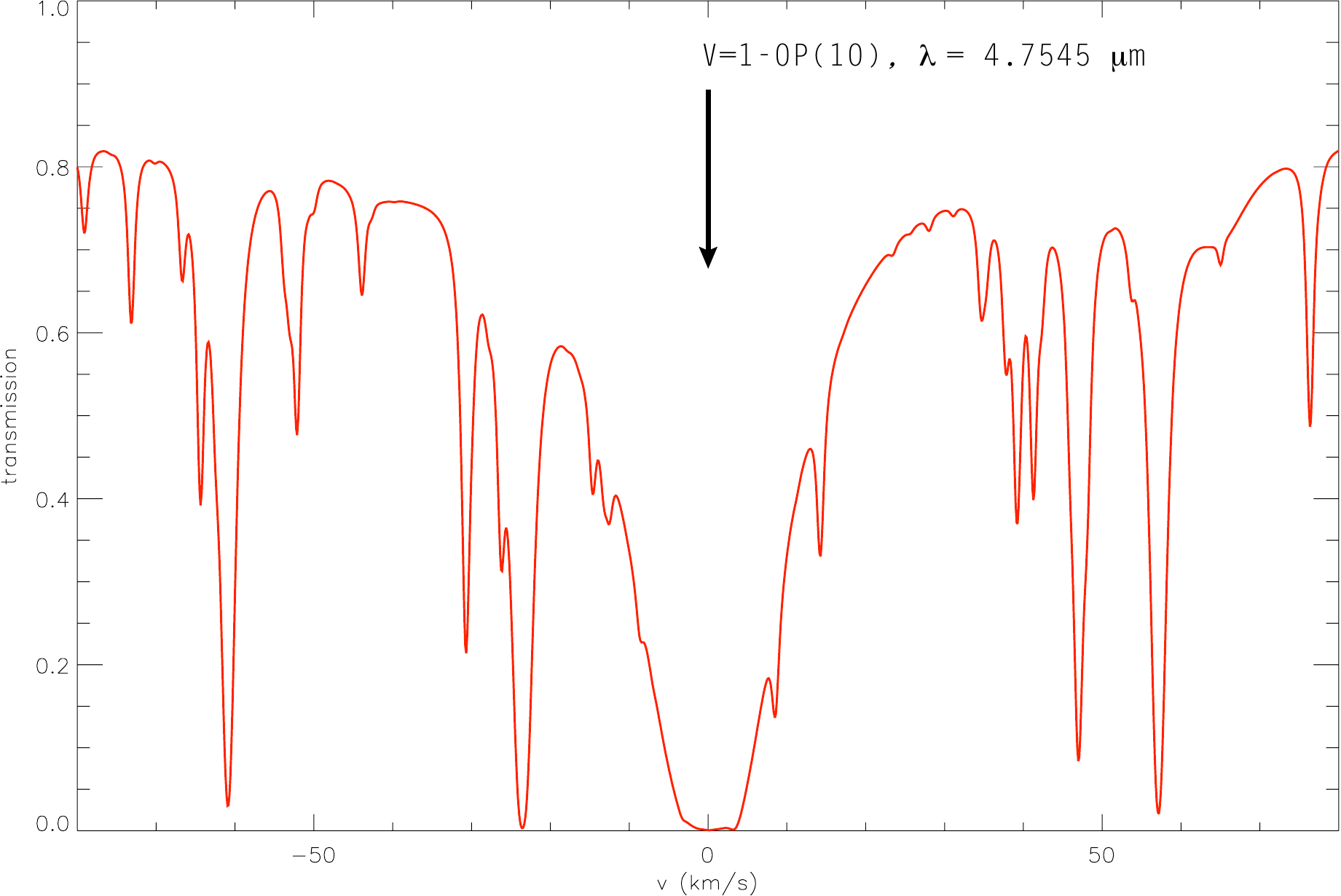}
	\caption{Effect of instrumental noise and atmospheric absorption on the observability of line profile distortions (\emph{top}) and the Earth atmospheric transmission at $4.7545 \mathrm{\mu m}$ (\emph{bottom}). The line profiles were calculated in model \#8, shown already in Fig. \ref{fig:V1-0P10-profiles}(a), but superimposed with an artificial noise expected for one hour of integration on $M_\mathrm{mag}=6$ brightness source with VLT/CRIRES.}
	\label{fig:noise-Ms1-Mb8-V1-0P10}
\end{figure}

\begin{figure}
	\centering
	\includegraphics[width=\columnwidth]{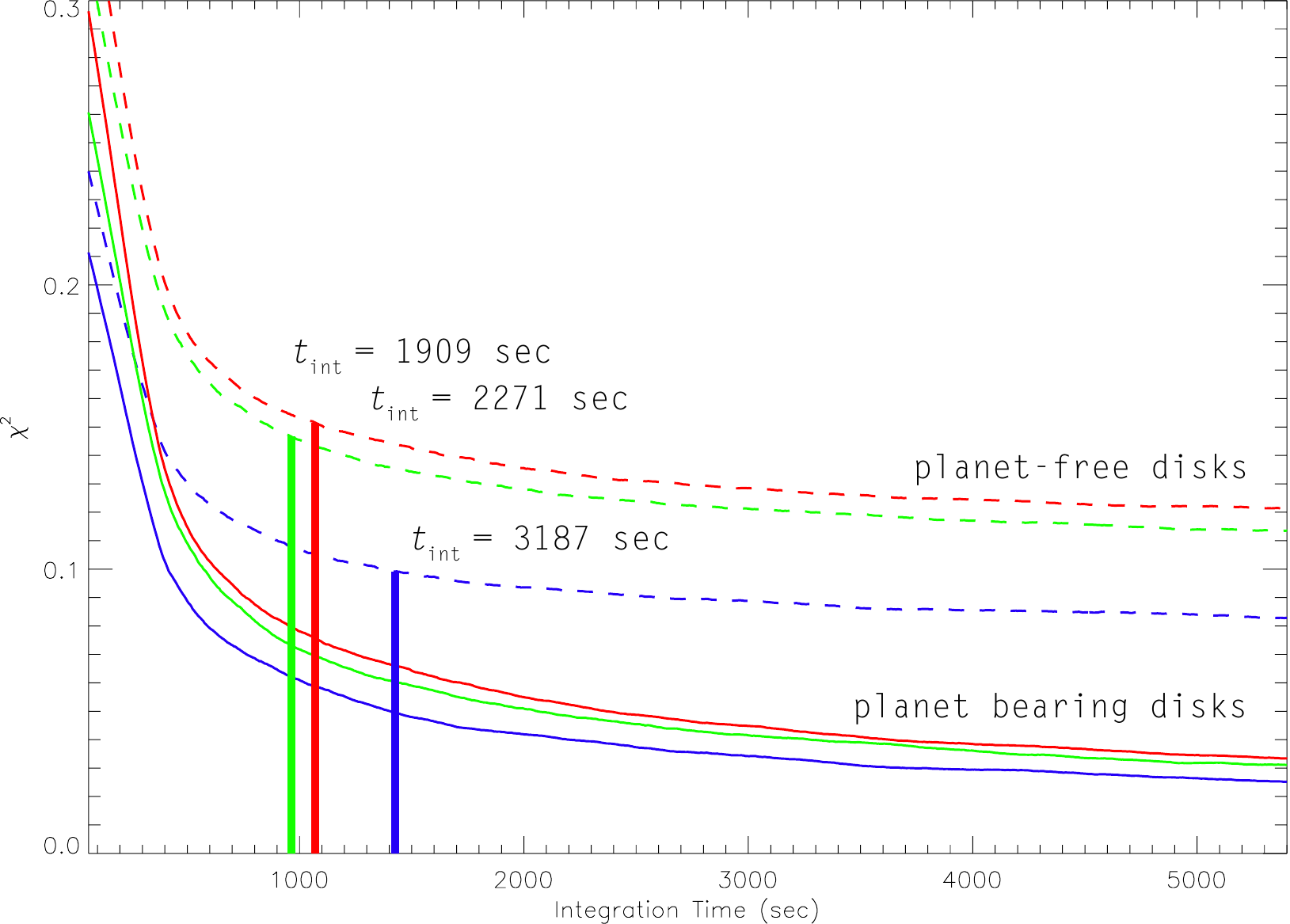}
	\caption{Goodness-of-fits of distorted line profiles contaminated with artificial noise obtained in planet-free (dashed lines) and giant planet-bearing disk (solid lines) models versus the integration time for $20^{\circ}$, $40^{\circ}$ and $60^{\circ}$ inclinations, represented by blue, green and red colors, respectively. It is visible that for a reasonable signal-to-noise ratio, the planet-free models provide always fits with less confidence than the giant planet models, i.e the $\chi^2$ is always larger. Moreover, the smaller the inclination angle, the larger the required minimal integration time ($t_\mathrm{int}>3187\,\mathrm{sec}$, $1909\,\mathrm{sec}$ and $2271\,\mathrm{sec}$ for the $20^{\circ}$, $40^{\circ}$ and $60^{\circ}$ inclinations, respectively) to distinguish the planet-free and planet-bearing disk models.}
	\label{fig:noise-goodness}
\end{figure}

To observe the planet-induced distortions of the CO line profiles that we simulated, a spectroscopic facility is required that meets two basic demands: it needs sufficient spectral resolution to properly resolve the sub-structure in the line profile and sufficient sensitivity to bring out the low-contrast line profile distortions. In this section, we investigate the observability of the modeled effect in the context of contemporary facilities, and give a brief outlook into the ELT era. We present a quantitative example for the \emph{Cryogenic High Resolution Echelle Spectrograph} (CRIRES) at the VLT (Kaeufl et al. 2004), which is currently the most powerful high-resolution spectrograph covering the M-band around $4.8\,\mathrm{\mu m}$, capable of observing CO ro-vibrational spectra.

The main limiting factor for ground-based thermal infrared observations is the Earth atmosphere, which strongly absorbs the radiation from astronomical sources and causes a high thermal background. Particularly strong telluric absorption and emission is present at the wavelengths of low-excitation CO transitions. Within $\sim 10\,\mathrm{km/s}$ from the line center the telluric transmission is so low and the background emission so high that these spectral regions are not accessible. However, depending on the location of the source with respect to the ecliptic, the CO features are Doppler shifted by up to $\pm 30\,\mathrm{km/s}$ due to the Earth's orbital motion, making the whole CO line accessible over the course of several months, in principle. Note though that we can never measure the entire CO line profile simultaneously, there will always be an approximately $20\,\mathrm{km/s}$ wide ``gap'' in our spectral coverage at any observing epoch.

In order to test whether we would be able to detect the planet-induced permanent profile asymmetries with currently available instrumentation, we have simulated a VLT/CRIRES observation of model \#8. We have assumed the source to have a brightness of $M=6\,\mathrm{mag}$, and calculated the achieved $SNR$ as a function of wavelength using the VLT/CRIRES exposure-time calculator\footnote{We used version 3.2.8 of the ETC, assumed a water vapor column of 2.3\,mm, an airmass of 1.4, a seeing of  $1\farcs0$, a slit width of $0\farcs2$, and adaptive optics correction using a guide star of $R=10.0\,\mathrm{mag}$ and spectral type of M0V.}. This calculation takes into account the telluric absorption and emission, the quality of the adaptive optics correction, the system throughput and detector characteristics. We note that our simulation only considers the signal-to-noise ratio, and assumes that systematic effects due to, e.g., time-variable telluric absorption, can be well calibrated. The achieved $SNR$ can be scaled according to
\begin{equation}
SNR \propto 10^{−0.4M} \sqrt{T_{\rm{int}}},
\end{equation}
where $M$ denotes the M-band magnitude and $T_{\rm{int}}$ the integration time. In Fig. \ref{fig:noise-Ms1-Mb8-V1-0P10} we show the emerging V=1-0P(10) line profiles for model \#8 (upper figure) and the atmospheric transmission (lower figure) for an assumed integration time of 1 hour on a source of $M=6\,\mathrm{mag}$. The model spectrum has been convolved with the 3\,km/s CRIRES instrumental resolution. We have assumed zero relative velocity between the source and the Earth at the time of the observation. The effects of the telluric CO line are clearly seen around the line center, where no meaningful measurement can be obtained. The line wings, where the effects of the planet are most prominent, can be well measured.

In order to quantify the significance with which a planetary signal (the most prominent permanent asymmetric pattern) may be detected, we have tried to fit the modeled observation with both planet-free and planet-bearing models. In Fig. \ref{fig:noise-goodness} we show the goodness-of-fits (reduced $\chi^2$) as a function of integration time for both cases (with dashed line for planet-free and solid line for planet-bearing disk models) for $20^\circ$, $40^\circ$, and $60^\circ$ inclination angles. Above a certain integration time (presented by vertical lines) the planet-bearing models yield a more than 50\% better reduced $\chi^2$ fit to the simulated data than those of the planet-free models. The  asymmetric pattern induced by an embedded planet increase and thus become easier to detect with disk inclination.

Because the fundamental band ro-vibration R(0), R(1), R(12), R(13), R(19)-(21), R(25), R(26) and R(30)-(33), P(6-12) and P(22-32) CO lines are not blended by the higher vibrational lines\footnote{Assuming that the abundance of $\mathrm{^{12}C^{16}O}$ isotopologues are low, thus  their fundamental ${\mathrm{V=1\rightarrow 0}}$ contribution to lines is negligible.}, the line profiles can be averaged, using appropriate scaling, resulting in an increased signal-to-noise ratio. Note that the unblended lines with higher $J > 25$ or lower $J < 2$ rotational quantum numbers are too weak for averaging: using them would introduce only additional noise to the averaged profiles. Also note that, since the higher vibrational levels are not excited at all in disk atmospheres similar to that used in our models, the blended lines can also be used for averaging, resulting in an even more increased signal-to-noise ratio. 

We conclude that in the confines of our model, the permanent asymmetric signals produced by an $8\,M_\mathrm{J}$ giant planet orbiting at $\sim 1\,\mathrm{AU}$ embedded in the disk of a T\,Tauri star of brightness $M=6\,\mathrm{mag}$ would be detectable with VLT/CRIRES with an integration time of 1\,hour. Moreover, because the variable component at the top of the permanent asymmetry has a width of $\sim 5-10\,\mathrm{km/s}$, exceeding the CRIRES $\sim 3\,\mathrm{km/s}$ resolution, and $\sim 10$\% magnitude, there is a certain possibility to strengthen its planetary origin.

With the next generation of giant, ground-base telescopes our sensitivity to low-contrast features in infrared spectra will increase greatly. We have used the ESO exposure-time calculator for the ELT (version 2.14) to simulate an observation with the proposed first generation infrared instrument \emph{Mid-infrared E-ELT Imager and Spectrograph} (METIS)  \citep{Brandletal2008}. The exposure-time calculator necessarily incorporates planned telescope and instrument specifications, because this facility is yet to be built. We find that ELT/METIS yields an increase in sensitivity of a factor of $\sim$30 compared to VLT/CRIRES. Thus also sources that are substantially fainter or have weaker spectral features than those modeled here in the context of VLT/CRIRES are within reach of the observational facilities of the next decade.

\section{Conclusions}

We have shown that our semi-analytical 2D double-layer thermal disk model (in which the disk layers are heated mainly by stellar irradiation, and the emission of the disk inner rim is accounted for) is capable to reproduce the  double-peaked Keplerian line profiles in CO fundamental ro-vibrational band in young (about $2.5\,\mathrm{Myr}$ old) T\,Tauri-type protoplanetary disks assuming canonical dust and gas properties. However, note that several T\,Tauri stars show centrally peaked symmetric profiles instead, which can be attributed to an enhanced CO flux formed in distant regions (up to 10\,AU), presumably above the disk atmosphere. This can be explained by non-LTE heating, such as UV fluorescence \citep{Krotkovetal1980}, or dust gas temperature decoupling \citep{Glassgoldetal2004,KampDullemond2004}, resulting in highly excited CO emissions. These effects were not taken into account in our calculations. Thus the gas temperature at disk atmosphere provided by our thermal model is presumably somewhat lower than the realistic one. Our predictions, however, demonstrate a ``the worst case'' because the planet signal is getting stronger with increasing atmospheric temperature. It is revealed that significant line profile distortions appear in the CO fundamental ro-vibrational band caused by a giant planet. As we have demonstrated, a giant planet with a mass of at least $1\,M_\mathrm{J}$ can perturb the disk sufficiently to be detected by permanent asymmetry in line profiles and its short timescale variations. The permanent asymmetry in line profiles can be interpreted by the disk being in an eccentric state only in the gap, but the short timescale variability is certainly connected to the local dynamical perturbations of the orbiting planet. Depending on the signal-to-noise ratio and the resolution of the spectra, even lower-mass giant planets can be detected revolving on close orbits in disks observed with high-inclination angles and with large size inner cavities. The summary of our findings is:
\begin{enumerate}
	\item{The dynamical perturbation induced by a giant planet significantly distorts the CO ro-vibrational fundamental emission line profiles.}
	\item{The line profile distortions are more apparent for larger inclination angles.}
	\item{The main component of line profile distortions is a permanent asymmetry. Its position in the line profile depends on the stellar mass and orbital distance of the planet. The permanent asymmetry shape depends on the orientation of the gas elliptic orbit with respect to the line of sight.}
	\item{The line profiles are changing with time, correlated to the orbital phase of the giant planet. The timescale of the variation is on the order of weeks, depending on the orbital period of the planet, i.e, the orbital distance and the mass of the host star. The magnitude of variable component is $\sim10\%$, its width is $\sim 10\,\mathrm{km/s}$, depending on the inclination angle.}
	\item{The planet signal becomes stronger with increasing planetary mass.}
	\item{The planet signal strengthens/weakens with increasing/decreasing mass of the host star, neglecting the effect of the stellar luminosity on the size of the inner cavity}
	\item{The planet signal strengthens/weakens with decreasing/increasing orbital distance of the planet.}
	\item{The size of the inner cavity has a strong influence on the giant planet observability. If the size of the inner cavity is substantially smaller than 0.2\,AU, we can detect a giant planet only at close (about 0.5\,AU) orbit. On the contrary, if the cavity is larger (0.4\,AU), a distant planet (in 2\,AU orbital distance) can also be detected.}
	\item{The influence of disk geometry on the planet signal is modest. With a decreasing flaring index the strength of planet signal does not change although the line profiles are strengthened, except in ``wide'' flat models where the planet signal substantially suppressed.}
	\item{The lowest mass of giant planet orbiting at 1\,AU that still can be detected is $0.5\,M_\mathrm{J}$ for a $0.5\,M_{\sun}$ star in the confines of our models.}
	\item{By high-resolution near-IR spectroscopic monitoring with VLT/CRIRES, giant planets with $\geq 1\,M_\mathrm{J}$ mass orbiting within 0.2-3\,AU in a young disks may be detectable, if other phenomena do not confuse, mimic, or obscure the planet signature. The level of confidence is growing with increasing inclination angle.}
\end{enumerate}

In the light of our findings, we propose to observe T\,Tauri stars that shows an appreciable amount of CO in emission with high-resolution near-IR spectrograph to search for giant planet companions. The asymmetric and time-varying double-peaked line profiles can be explained by strongly non-circularly Keplerian gas flow in the disk caused by the giant planets, as was presented. Although higher $V\geq 2$ vibrational states of CO are not excited in our models, overlaps can occur, resulting in asymmetric line profile. To avoid this, one should consider only non-blended lines for profile averaging such as P(6)-P(12), and R(0), R(1), R(12), R(13), R(19)-(21), R(25), R(26) and R(30)-(33). Another possible source of permanent line profile distortions could be a large companion deforming the circumprimary disk to fully eccentric in mid-separation binaries (e.g. \citet{Kleyetal2008}; Regaly et al. in prep.). Last but not least, the distorted PSF, caused by tracking errors of the telescope or unstable active optics during an exposure, can induce artificial signals. A real signal from a bipolar structure will change sign when observed using antiparallel slit position angles whereas the artificial 
signal will not. Thus artificial signatures can be successfully identified with the comparison of the spectra of disks obtained at two antiparallel slit position angles, e.g. $0^\circ$ and $180^\circ$ \citep{Branniganetal2006}. By detecting short time-scale (weeks to months) variations of the asymmetric profiles with similar patterns as presented in Fig. \ref{fig:variation-Ms1-Mb8-V1-0P10}, we could infer the presence of a giant planet in formation. 
 
The proposed instrument METIS for the ELT \citep{Brandletal2008}, providing high-resolution ($\lambda/\Delta\lambda=10^5$) infrared spectroscopy with an increase in sensitivity of a factor of $\sim 30$ to that of VLT/CRIRES will give the opportunity to even decrease more the mass and brightness limits of possible giant planets. If we revealed for instance the distance of the observed planets to the star as a function of the age of the star-disk system, we would be able to constrain planetary migration models. Moreover, determining the youngest star host planet-bearing disks could also provide constraints on planet-formation models.

Although our assumptions are reasonable from the point of view of finding a simple model of the signal of dynamical perturbations, it can be developed further of course. For example,  if the gap is cleared by the planet to a fraction of the surrounding density, the disk is not directly irradiated, but is shadowed by the inner wall of the gap \citep{Varniereetal2006}. This mild heating causes a temperature drop at the inner gap wall, which results in a depressed CO flux. On the other hand, the outer edge of the gap is directly illuminated by the star, resulting in a slight temperature increase, which produces stronger CO emission from the regions where the eccentricity reaches its maximum (see e.g. Fig.\ref{fig:FARGO-ecc}). According to our results the gap is strongly non-axisymmetric, thus 3D radiative transfer using 3D density structure needed to calculate the real temperature profile in perturbed disks. Moreover, as the gap is depleted in large grains because of the trapping of dust grains larger than $\sim 10\,\mathrm{\mu m}$ \citep{Riceetal2006}, the CO flux normalized to the continuum is strengthened. The gas and dust can be thermally uncoupled (see e.g., \citet{Glassgoldetal2004,KampDullemond2004,Woitkeetal2009}) in the tenuous gap, which may result in CO being overheated to the dust. These effects could cause additional permanent line profile distortions due to the elliptic shape of the gap seen in density distribution (Fig.\,\ref{fig:FARGO-output}). If we consider an excess emission originating from the planetary accretion flow suggested by \citet{ClarkeArmitage2003}, or the density waves in disk surface caused by the orbiting planet, it would cause a varying planet signal. To these caveats, one might also consider the disk turbulence (e.g., due to MRI effects), which may add random asymmetries and ``blobbiness'' to the line profiles. We will investigate these effects in a forthcoming paper using a more sophisticated thermal disk model.

CO exists inside the dust sublimation radius in some T\,Tauri stars \citep{Carr2007} or Herbig\,Ae/Be stars \citep{Brittainetal2009}, where the disk is optically thin in the absence of dust. One possibility to excite CO in the optically thin inner disk close to the stellar surface is the UV fluorescence, which is far from LTE \citep{Krotkovetal1980} applied in our calculations. Depending on the efficiency of X-ray heating the gas temperature can be as high as $\sim 4000-5000\,\mathrm{K}$. Somewhat below this temperature, where CO still exists, the high-excitation vibrational levels of CO such as $V\geq 2$ are populated, resulting in significant $V=2\rightarrow 1$, $V=3\rightarrow 2$, etc. ro-vibrational lines. According to \citet{Skrutskieetal1990}, the inner cavity in transitional disk may be formed by a close giant planet. Indeed, \citet{LubowDAngelo2006} found that an $1-5M_\mathrm{J}$ mass planet significantly lowers the accretion of dust and gas to 10\%-25\% of the accretion rate outside the gap. We expect that the planetary signal forming in the optically thin inner disk is also detectable, which is strongly supported by the revealed line profile distortion strengthening with decreasing orbital distance of the planet. The UV pumping requires considerable UV flux from the central star, which is a characteristic of Herbig Ae/Be stars rather than smaller T\,Tauri stars. Considering that the CO may be excited up to to 10\,AU by UV fluorescence \citep{Brittainetal2007,Brittainetal2009} in Hebig\,Ae/Be stars, planet orbiting at larger distances than 2\,AU may be also detectable by CO ro-vibrational line profile distortions.

Spectroscopy of T\,Tauri stars shows emission of molecules such as $\mathrm{H_2O}$, $\mathrm{OH}$, $\mathrm{HCN}$, $\mathrm{C_2H_2}$, and $\mathrm{CO_2}$, as well. Nevertheless CO is more abundant than these molecules by $\sim 10$ times, only $\mathrm{H_2O}$ could reach the abundance of CO, predicted by recent models that calculate the vertical chemical structure of the gas in disk atmosphere (e.g. \citet{Glassgoldetal2004}, \citet{KampDullemond2004}, and \citet{Woitkeetal2009}). Indeed in some cases, like AA\,Tau \citep{CarrNajita2008}, and AS\,205A and DR\,Tau \citep{Salyketal2008}, the  rotational transitions of $\mathrm{H_2O}$ dominate the mid-infrared ($10-20\,\mathrm{\mu m}$) spectra, suggesting that $\mathrm{H_2O}$ is abundant in disk atmospheres. The strong water emission could be the consequence of turbulent mixing that carries molecules from the disk midplane, where they are abundant, to the disk atmosphere \citep{CarrNajita2008}, or the effects of an enhanced mechanical heating of the atmosphere \citep{Glassgoldetal2009}. While the $\mathrm{H_2O}$ ro-vibrational lines probe the inner regions of disk out to radii $\geq 2\,\mathrm{AU}$, the rotational lines are produced between radii 10-100\,AU \citep{Meijerinketal2008}.  The rotational and ro-vibrational line profiles of water are also subject to distortions caused by the dynamical perturbations of a giant planet. Thus it is reasonable to search for signature of giant planets in the high-resolution spectra of $\mathrm{H_2O}$ in the mid-infrared band (for planets orbiting int the inner disk), and rotational spectra of $\mathrm{H_2O}$ in the far-infrared band (for planets orbiting in the outer disk), as well. Because $\mathrm{H_2O}$ is heated by stellar X-rays and sub-thermally populated beyond ~0.3\,AU, X-ray-heating and non-LTE level population treatment is needed to calculate water lines (e.g., \citet{Meijerinketal2008,Kampetal2010}).

Finally, we point out some noticeable resemblance to our findings revealed in observed CO line profiles. The V836\,Tau transitional disk  shows strongly distorted $4.7\,\mathrm{\mu m}$ CO ro-vibrational line profile presented by \citet{Najitaetal2008}. The averaged CO line profiles shows very similar features to our results. According to \citet{Najitaetal2008} the possibility of a massive planet is also restricted by current limits on the stellar radial velocity of V836\,Tau, which constrain the mass of a companion within $0.4-1\,\mathrm{AU}$ to $5-10\,M_\mathrm{J}$. Because the disk extents to a limited range of radii ($\sim 0.05-0.4\,\mathrm{AU}$), the dynamical perturbation of a close orbiting planet with a mass smaller than $5\,M_\mathrm{J}$ presumably could cause the observed line profile distortions. Note that radial velocity measurements indicate that no planet is present larger than $1-2\,M_\mathrm{J}$ closer than $\sim 0.5\,\mathrm{AU}$ \citep{Pratoetal2008}. The transitional disks that encompass the young sources SR\,21 and HD\,135344\,B have been observed by VLT/CRIRES  \citep{Pontoppidanetal2008} and they already show clear line profile asymmetries in CO fundamental ro-vibrational band. Our scenario of a planet is a reasonable explanation of the asymmetry. The presence of a planet was also proposed by \citet{Gradyetal2009} for HD\,135344\,B and \citet{Eisneretal2009} for SR\,21, based on SED and visibility data, respectively. EXLupi (prototype of EXor type young variable stars) showed remarkable line profile variations during its 2008 outburst (Goto et al. 2010, in prep.). Goto et al. revealed that the CO emission has a spatially multiple origin. The quiescent component forms in the outer optically thick disk, while the outburst component in the inner optically thin gas disk, where the higher vibrational levels of CO are presumably excited by UV pumping. These high-excitation $V=2\rightarrow 1$ and $V=3\rightarrow 2$ lines are double peaked, strongly asymmetric and variable on a short (weekly) time scale. Thus for example, by measuring short timescale (on the order of weeks or less) periodic variations in the CO ro-vibrational spectra of these sources, by mid-IR monitoring observations would eventually indirectly detect an embedded Jupiter-like planet in birth.

\begin{acknowledgements}
It is a pleasure to acknowledge helpful discussions with Ewine van Dischoeck, Klaus Pontoppidan and Attila Juh\'asz. This research has bee supported in part by DAAD-PPP mobility grant P-M\"OB/841/ and ``Lend\"ulet'' Young Researcher Program of the HAS. We are grateful to L. Kiss who helped us to substantially clarify the text. We also thank the anonymous referee for thoughtful comments that helped to significantly improve the quality of the paper.
\end{acknowledgements}

\appendix

\section{Temperature distribution in the disk}
\label{apx:temp-dist}

For the sake of completeness we present here the derivation of the dust temperature distributions in the disk interior and atmosphere according to \citet{ChiangGoldreich1997}. Below we use the local thermodynamical equilibrium (LTE) assumption everywhere. The stellar flux in the disk atmosphere at a distance $R$ from the stellar surface is
\begin{equation}
	F_{*}(R)\simeq\frac{\delta(R)}{2}\left(\frac{R_{*}}{R}\right)^{2}\sigma T_{*}^4,
\end{equation}
where $\delta(R)$ is the grazing angle of the incident irradiation, $R_{*}$ and $T_{*}$ are the stellar radius and surface temperature, respectively, and $\sigma$ is the Stefan--Boltzmann constant. Here we assumed that only half of the stellar surface is visible from a given point at a distance $R$ from the star in the disk atmosphere. According to \citet{ChiangGoldreich1997} the grazing angle can be given by
\begin{equation}
	\label{eq:grazing-angle-flaring}
	\delta(R)=\frac{2}{5}\left(\frac{R_{*}}{R}\right)+\frac{8}{7}\left(\frac{T_*}{T_\mathrm{g}}\right)^{4/7}\left(\frac{R_*}{R}\right)^{-2/7},
\end{equation}
in a flared disk assumed to be in hydrostatic equilibrium. In Eq. (\ref{eq:grazing-angle-flaring}) $T_\mathrm{g}$ is the gravitational temperature at which the thermal energy of gas parcel balances the gravitational energy at the stellar surface, which can be given by
\begin{equation}
	T_\mathrm{g}=\frac{GM_*m_\mathrm{H}}{kR_*},
\end{equation}
where $m_\mathrm{H}$ is the mean molecular weight of the gas, $G$ and $k$ is the gravitational constant and Boltzmann constant, respectively. Note that for a  non-flared, i.e flat disk the grazing angle is defined by only the first term of Eq. (\ref{eq:grazing-angle-flaring}).

By definition the optical depth of the disk atmosphere along the incident stellar irradiation is $\tau_\mathrm{V}=1$ in the optical wavelengths. The optical depth of the atmosphere at near-IR heated to $T_\mathrm{atm,irr}(R)$ perpendicular to the disk plane is $\tau_\mathrm{V}\delta(R)\epsilon_\mathrm{atm}(R)$, where $\epsilon_\mathrm{atm}(R)=(T_\mathrm{atm,irr}(R)/T_\mathrm{*})^\beta$ is the dust emissivity, and $\beta$ is the power law index of the absorption coefficient of the dust, see Appendix \ref{apx:emissivity} for details. The total flux emitted inward and upward by the optical thin atmosphere can be given by
\begin{equation}
	F_\mathrm{atm}(R)=2\delta(R)\epsilon_\mathrm{atm}(R)\sigma T_\mathrm{atm,irr}(R)^4.
\end{equation}
In LTE the absorbed flux of atmosphere equals to the emitted one ($F_{*}(R)=F_\mathrm{atm,irr}(R)$), thus the disk atmosphere is heated to the temperature
\begin{equation}
	\label{eq:temp-surf}
	T_\mathrm{atm,irr}(R)=\left(\frac{1}{4}\right)^{1/(4+\beta)}\left(\frac{R_{*}}{R}\right)^{2/(4+\beta)}T_{*}
\end{equation}
by stellar irradiation.

Assuming that the disk interior with a temperature $T_\mathrm{int,irr}(R)$ is optically thick to its radiation everywhere in our computational domain,\footnote{As mentioned before, in the computational domain, the disk interior remains optically thick to its own radiation.} the flux $F_\mathrm{int,irr}(R)$ emitted by the disk interior is
\begin{equation}
	F_\mathrm{int,irr}(R)=\sigma T_\mathrm{int,irr}(R)^{4}.
\end{equation}
Considering that the absorbed stellar flux will be re-emitted by the disk atmosphere upward and downward only half of this flux heats the interior, i.e  $0.5F_\mathrm{atm}(R)=0.5F_{*}(R)=F_\mathrm{int,irr}(R)$. Accordingly, the temperature of the disk interior set by the stellar irradiation can be given by
\begin{equation}
	\label{eq:temp-int-irr}
	T_\mathrm{int,irr}(R)=\left(\frac{\delta(R)}{4}\right)^{1/4}\left(\frac{R_{*}}{R}\right)^{1/2}T_{*}.
\end{equation}

Owing to the accretion process a significant amount of gravitational potential energy has to be dissipated by a viscous processes. According to \citet{Lynden-BellPringle1974}, in a steady state disk\footnote{The disk being in steady state means that the density distribution corresponding to the surface density distribution does not considerably change in time.} with a constant accretion rate $\dot M$, the flux $F_\mathrm{acc}(R)$ released by the disk mid-plane due to the change of potential energy can be given by
\begin{equation}
	\label{eq:acc-flux}
	F_\mathrm{acc}(R)=\frac{3GM_{*}\dot{M}}{8\pi}\left(1-\left(\frac{R_\mathrm{*}}{R}\right)^{1/2}\right) R^{-3}.
\end{equation}
In this way the disk mid-plane is heated to the temperature 
\begin{equation}
	\label{eq:acc-temp}
	T_\mathrm{acc}(R)=\left[\frac{3GM_{*}\dot{M}}{8\pi\sigma}\left(1-\left(\frac{R_\mathrm{*}}{R}\right)^{1/2}\right)\right]^{1/4}R^{-3/4}.
\end{equation}
Here we have to note that only half of the total accretion power is involved in Eq. (\ref{eq:acc-flux}), the remaining is stored in the kinetic energy of orbiting gas. This energy should be radiated away by the disk boundary layer \citep{Pophametal1993,Pophametal1995} or by the accreting material in the funnel flow formed along the magnetic field lines in magnetospheric accretion model, see \citet{Hartmannetal1994,Bouvieretal2007}. Because the disk boundary layer and the funnel flows are confined into such small volumes that the gas temperature  reaches about 10000K emitting in the UV band, its radiation is not taken into account.

Now let us take into account the heating due to viscous dissipation in disk interior using the superposition principle, i.e the radiation at a given location is the sum of the radiations corresponding to different heating sources. To determine the disk interior temperature $T_\mathrm{irr,acc}(R)$ caused by the viscous dissipation first assume that the flux emitted by the optical thick disk interior is 
\begin{equation}
	F_\mathrm{int,acc}(R)=\sigma T_\mathrm{int,acc}(R)^4.
\end{equation}
Because the disk mid-plane radiates the accretion flux into two directions $0.5F_\mathrm{acc}(R)=F_\mathrm{int,acc}(R)$, the disk interior is heated to 
\begin{equation}
	\label{eq:temp-acc}
	T_\mathrm{int,acc}(R)=(1/2)^{1/4}T_\mathrm{acc}(R)
\end{equation}
by accretion. Taking into account the heating of the disk interior by irradiation of atmosphere and viscous dissipation together (simply summing the radiation fluxes), the resulting temperature of disk interior is
\begin{equation}
	\label{eq:temp-int}
	T_\mathrm{int}(R)=\left(T_\mathrm{int,irr}(R)^4+T_\mathrm{int,acc}(R)^4\right)^{1/4}.
\end{equation}

\section{Emission from the disk inner rim}
\label{apx:sinner-rim}

To determine the temperature profile of the disk interior close to the disk inner edge, where the simple double layer assumption cannot be applied, we first assume that the rim is a perfect vertical wall. Any given optically thick vertical slab with a thickness of $dR$ at $R+dR$ distance from the central star is irradiated by the neighboring hotter slab located at $R$. Assuming that half of its radiation is received by the one at $R+dR$, the emitted and irradiated fluxes are in equilibrium in LTE, i.e.
\begin{equation}
	\frac{1}{2}\sigma T_\mathrm{rim}(R)^4=\sigma\epsilon(R) T_\mathrm{rim}(R+dR)^4,
\end{equation}
where $\epsilon(R)=(T_\mathrm{rim}(R+dR)/T_\mathrm{rim}(R))^\beta$ is the emissivity of the slab at $R+dR$, see in Appendix \ref{apx:emissivity}. Note that here we neglect the irradiation of the neighboring slab at $R+2dR$ with a lower temperature than the slab at $R+dR$. To calculate $T_\mathrm{rim}(R)$, we first approximate $T_\mathrm{rim}(R+dR)$ with $T_\mathrm{rim}(R)+T_\mathrm{rim}^\prime (R)dR$. This results in an ODE for $T(R)$, which has the solution
\begin{equation}
	\label{eq:temp-rim}
	T_\mathrm{rim}(R)=T_\mathrm{rim}(R_0)\exp\left[-\left(1-\left(\frac{1}{2}\right)^{1/(4+\beta)}\right)\left (R-R_0 \right)\right],
\end{equation}
where $R_0$ is the radius of the disk inner edge. To take into consideration the additional irradiation of the rim at the opposite side, we assume that the temperature at the innermost slab is $T_\mathrm{rim}(R_0)=qT_\mathrm{int,irr}(R_0)$, where $q\simeq1.2$ according to results of our simulations done by 2D RADMC \citep{DullemondDominik2004}. Thus incorporating the additional heating by the disk rim, the disk interior temperature, previously given by Eq. (\ref{eq:temp-int}), can be given by
\begin{equation}
	\label{eq:temp-int-mod}
	T_\mathrm{int}(R)=\left(T_\mathrm{int,irr}(R)^4+T_\mathrm{int,acc}(R)^4+T_\mathrm{rim}(R)^4\right)^{1/4}.
\end{equation}
where the superposition rule is applied.

\section{Dust emissivity}
\label{apx:emissivity}

In this section the derivation of dust emissivity at a specific temperature is given, assuming optically thick environment. Let us define the dust emissivity at a temperature $T_\mathrm{dust}$ heated by a blackbody of the temperature $T_\mathrm{irr}$ as the ratio of the Rosseland mean opacities
\begin{equation}
	\label{eq:emissivity}
	\epsilon_\mathrm{dust}=\frac{\left<\kappa(T_\mathrm{dust})\right>_\mathrm{R}}{\left<\kappa(T_\mathrm{irr})\right>_\mathrm{R}}.
\end{equation}
The Rosseland mean opacity is defined by
\begin{equation}
	\label{eq:R-mean-opac}
	\left<\kappa(T)\right>_\mathrm{R}=\frac{\int_0^\infty dB(\nu,T)/dTd\nu}{\int_0^\infty dB(\nu,T)/dT\kappa(\nu,T)^{-1}d\nu},
\end{equation}
where $B(\nu,T)$ is the Planck function. The dust used in our model generates the $\kappa(\nu)=\kappa_0 (\nu/\nu_0)^{\beta}$ opacity law, where $\beta>0$. Note that wee use $\beta=1$ throughout all our models \citep{Rodmannetal2006}. Following the calculations of \citet{StahlerPalla2005} in Appendix G, we assume that the dust absorption coefficient can be given by
\begin{equation}
	\kappa(\nu,T)=\kappa_0\left(\frac{kT}{h\nu_0}\right)^{\beta} x^{\beta},
\end{equation}
where $k$ and $h$ are Boltzmann and Planck constants, respectively, and  $x=h\nu/kT$. Supposing that
\begin{equation}
	\partial B(\nu,T)/\partial Td\nu=Af(x)dx,
\end{equation}
where A is an appropriate dimensional constant, and substituting this into the definition of Rosseland mean opacity, Eq. (\ref{eq:R-mean-opac}), leads to
\begin{equation}
	\label{eq:Rosseland-mo}
	\left<\kappa(T)\right>_\mathrm{R}=\kappa_0\left(\frac{kT}{h\nu_0}\right)^{\beta}\frac{\int f(x)dx}{\int x^{-\beta}f(x)dx}.
\end{equation}
Because the quotient of integrals is a pure number, the Rosseland mean opacity is proportional to $T^\beta$. Applying \ref{eq:emissivity} and \ref{eq:Rosseland-mo} we find that the dust emissivity can be given by
\begin{equation}
	\epsilon_\mathrm{dust}=\left(\frac{T_\mathrm{dust}}{T_\mathrm{irr}}\right)^\beta.
\end{equation}

\section{Optical depth of the disk atmosphere}

Assuming that the disk is observed with an inclination angle $i$, the monochromatic optical depth of the disk atmosphere is the sum of optical depths of dust and gas along the line of sight, i.e.:
\begin{equation}
	\label{eq:tau_nu_1}
	\tau({\nu,R,\phi},i)=\frac{1}{\cos(i)}\left( \kappa_\mathrm{d}(\nu)\Sigma_\mathrm{d}(R)+\kappa_\mathrm{g}(\nu,R,\phi,i)\Sigma_{g}(R)\right),
\end{equation}
where $\kappa_\mathrm{d}(\nu)$ and $\kappa_\mathrm{g}(\nu,R,\phi,i)$ are the dust and gas opacity at frequency $\nu$, while $\Sigma_\mathrm{d}(R)$ and $\Sigma_\mathrm{g}(R)$ are the dust and gas surface densities in the disk atmosphere. Note that because the dust opacity ($\kappa_\mathrm{d}$) is taken to be uniform throughout the disk, the optical depth of the disk atmosphere depends on $R$ and $\phi$ via the gas opacity characterized by the temperature distribution and the line-of-sight component of the orbital velocity of gas parcels. By definition the optical depth at the bottom of the disk atmosphere is unity at optical wavelengths along the stellar irradiation. Assuming that the mere opacity source at visual wavelengths is the dust, the dust surface-density of disk atmosphere is
\begin{equation}
	\label{eq:dens-surf}
	\Sigma_\mathrm{d}(R)=\frac{\delta(R)}{\kappa_\mathrm{V}},
\end{equation}
where $\kappa_\mathrm{V}$ is the overall opacity of dust at visual wavelengths. Assuming that the gas- and dust-mass ratio to the total mass ($X_\mathrm{g}$ and $X_\mathrm{d}$, respectively) is constant throughout the disk (i.e. there are no vertical or radial variations in the mass ratios), the surface density of gas in the disk atmosphere is
\begin{equation}
	\label{eq:dens-gas}
	\Sigma_\mathrm{g}(R)=X_\mathrm{g}\Sigma(R)=\frac{X_\mathrm{g}}{X_\mathrm{d}}\Sigma_\mathrm{d}(R),
\end{equation}
using the dust surface-density Eq. (\ref{eq:dens-surf}). Thus, the optical depth in the disk atmosphere along the line of sight can be given by
\begin{equation}
	\label{eq:tau_nu_2}
	\tau(\nu,R,\phi,i)=\frac{1}{\cos(i)}\left(\kappa_\mathrm{d}(\nu)\frac{\delta(R)}{\kappa_\mathrm{V}}+\frac{X_\mathrm{g}}{X_\mathrm{d}}\frac{\delta(R)}{\kappa_\mathrm{V}}\kappa_\mathrm{g}(\nu,R,\phi,i)\right),
\end{equation}
where we used Eqs. (\ref{eq:tau_nu_1}-\ref{eq:dens-gas}).

\section{Monochromatic opacities}

In LTE the monochromatic opacity of the emitting gas at frequency $\nu$ with a molecular mass of $m_\mathrm{CO}$ due to transitions between states u$\rightarrow$l can be given by
\begin{eqnarray}
	\label{eq:gas-kappa}
	\kappa_\mathrm{g}({\nu},R,\phi,i)&=&\frac{1}{m_\mathrm{CO}8\pi}\frac{1}{Q(T_\mathrm{atm}(R))}\left(\frac{c}{\nu_{0}}\right)^{2}A_\mathrm{ul} g_\mathrm{u}  \nonumber \\
	&\times&\left( \exp\left[-\frac{E_\mathrm{l}}{k T_\mathrm{atm}(R)}\right] - \exp\left[-\frac{E_\mathrm{u}}{k T_\mathrm{atm}(R)}\right]\right)\\
	&\times&\Phi(\nu,R,\phi,i)
\end{eqnarray}
where $\nu_{0}$ is the fundamental frequency of the transition, $Q(T_\mathrm{atm}(R))$ is the partition sum at the gas temperature $T_\mathrm{atm}(R)$, $A_\mathrm{ul}$ and $g_\mathrm{u}$ are the probability of transition (i.e the Einstein $A$ coefficient of the given transition) and the statistical weight of the upper state, respectively, while $c$ is the the light speed. The partition sum can be given by
\begin{equation}
	Q(T_\mathrm{atm}(R))=\sum_\mathrm{i}g_\mathrm{i}\exp\left[-\frac{E_\mathrm{i}}{k T_\mathrm{atm}(R)}\right],
\end{equation}
where $g_\mathrm{i}$ and $E_\mathrm{i}$ are the statistical weight and  energy level of the $i$th excitation state. In Eq. (\ref{eq:gas-kappa}) the $\Phi(\nu,R,\phi,i)$ is the local intrinsic line profile originating by the natural thermal and the local turbulent broadening acting together. If the pressure of gas is negligible, the intrinsic line profile can be represented by a normalized Gauss function
\begin{equation}
	\label{eq:ilp}
	\Phi(\nu,R,\phi,i)=\frac{1}{\sigma(R)\sqrt{\pi}}\exp\left[-\left(\frac{\nu-\nu_{0}+\Delta\nu(R,\phi,i)}{\sigma(R)}\right)^2\right],
\end{equation}
where $\sigma(R)$ is the line width, $\Delta\nu(R,\phi,i)$ is the line center shift due to Doppler shift caused by the apparent motion of the gas parcels along the line of sight. The line width is determined by the natural thermal broadening 
\begin{equation}
	\label{eq:natural-width}
	\sigma_\mathrm{therm}(R)=\frac{\nu_{0}}{c}\sqrt{\frac{2 k T_\mathrm{atm}(R)}{m_\mathrm{CO}}},
\end{equation}
and the local turbulent broadening
\begin{equation}
	\label{eq:turbulent-width}
	\sigma_\mathrm{turb}(R)=\frac{\nu_{0}}{c}\chi\sqrt{\frac{\gamma k T_\mathrm{atm}(R)}{m_\mathrm{H}}},
\end{equation}
assuming that the speed of turbulent motions is $\chi$ times the local sound speed. In Eqs. (\ref{eq:natural-width}) and (\ref{eq:turbulent-width}) $m_\mathrm{CO}$ and $m_\mathrm{H}$ are the molecular masses of H and CO and $\gamma$ is the adiabatic index of the main constituent of the gas, i.e that of hydrogen. Considering the thermal and local turbulent broadening, the resulting profile will be the convolution of the two Gaussian line profiles, i.e
\begin{equation}
	\sigma(R)=\sqrt{\sigma_\mathrm{therm}(R)^2+\sigma_\mathrm{turb}(R)^2}.
\end{equation}

In a planet-free disk, in which the gas parcels are moving on circularly Keplerian orbits, the line center at $\nu_0$ fundamental frequency shifts due to the Doppler shift is
\begin{equation}
	\label{eq:Doppler-shift}
	\Delta\nu(R,\phi,i)=\frac{\nu_0}{c}\sqrt{\frac{GM_{*}}{R}}\cos(\phi)\sin(i).
\end{equation}
Here we neglect that the massive planet and the host star are orbiting the common center of mass, instead the center of mass is set to the center of host star. Moreover, the influence of the gas pressure on the angular velocity of gas parcels, which causes slightly sub-Keplerian orbital velocities due to radial pressure support, is also not taken into account.

\bibliographystyle{aa}
\bibliography{regaly}

\end{document}